Hiu Yung Wong

# Quantum Computing Architecture and Hardware for Engineers - Step by Step - Volume II

# Preface

After publishing my book “Quantum Computing Architecture and Hardware for Engineers: Step by Step” [1] (now I call it Volume I), in which spin qubit and superconducting qubit quantum computers were covered, I decided to continue to write the second volume to cover the trapped ion qubit quantum computer, which was also taught in my EE274 class. I follow the same structure as in Volume I by discussing the physics, mathematics, and their connection to laser pulses and electronics based on how they fulfill the five DiVincenzo’s criteria. I also think it would be a good idea to share the second volume on arXiv so that more people can read it for free, and I can continue to update the contents. As of July 2025, I have finished the trapped ion quantum computer part. In the future, I plan to write more critical topics in a step-by-step manner to bridge engineers who did not receive rigorous training in Physics to the quantum computing world.

For any comments, please send an email to intro.qc.wong@gmail.com.

California<br>June 2025

*Hiu Yung Wong*

# Preface

**Volume I Preface**

The purpose of this book is to teach quantum computing hardware from an engineer's perspective. Engineers play an important role in quantum computers. However, college and graduate engineering students usually do not have the required physics and mathematics training to understand how quantum computer hardware works. This book provides step-by-step guidance to connect engineers to the quantum world. It is based on the teaching materials I created at San José State University for EE274 Quantum Computing Architectures in Spring 2023.

In this book, quantum computers built on silicon spin qubits and superconducting qubits are discussed in detail. The physics, mathematics, and their connection to microwave electronics are presented based on how they fulfill the five DiVincenzo's criteria. Readers will be able to understand the nuts and bolts of qubit initialization, readout, and manipulation. At the end, I also present a superconducting qubit integrated circuit design example, which is not commonly found elsewhere.

The characteristics of this book are three. Firstly, I try to write the equations and explain the process step-by-step to avoid ambiguity. As an engineer, I believe that this is very useful for serious engineers who want to connect their knowledge to quantum computing hardware. Such a connection is difficult to find in the literature. Secondly, I try to connect and contrast the spin qubits with superconducting qubits so that the readers can understand them in a unified theory. Lastly, simulation programs and design examples, along with teaching videos, are provided.

I learned quantum computing hardware by myself, and I believe I have written the book in a way that the confusions and ambiguities an engineer usually faces are highlighted and answered. I also need to thank Dr. Yaniv Jacob Rosen and Dr. Kristin Beck from the Lawrence Livermore National Laboratory for the collaborations since 2021, through which I learned a lot from both of them and also had the opportunity to practice my knowledge in superconducting qubit hardware.

This book is organized into four parts. Part I gives an overview of quantum computers and the basics of linear algebra, i.e., the Schrödinger equation, the density matrix, and the Bloch sphere. These represent the minimal knowledge required to understand quantum computing hardware. Readers are also encouraged to read my other book, “Introduction to Quantum Computing: From a Layperson to a Programmer in 30 Steps” if one is interested in learning the basics of quantum algorithms.

In Part II, we discuss the physics of spin magnetic moment and magnetic field interaction which is the basics of spin qubits. Besides Larmor precession, we go through the physics and mathematics of Rabi oscillation under a linearly oscillating magnetic field and rotating magnetic field. In the process, we build the mathematical tools and concepts, such as rotating frame and rotating wave approximation which are required to understand qubit operations in many technologies. With this knowledge, we show how a silicon electron qubit can be implemented on a Metal-Oxide-Semiconductor (MOS) technology.

In Part III, we first introduce Lagrangian and Hamiltonian mechanics, which are usually not covered in engineering classes. Then we show how to connect classical mechanics to quantum mechanics through operator promotion. We will discuss the details of a mechanical simple harmonic oscillator and reuse their equations in an LC tank. Then we show that an LC tank is not a suitable qubit due to the lack of anharmonicity and introduce the physics and fabrication of the Josephson junction due to its ability to provide nonlinear inductance for anharmonicity in a superconducting qubit. We go through circuit quantization and discuss the characteristics of the Cooper pair box in its charge qubit and transmon qubit regimes. We highlight the similarities between the superconducting qubit and spin qubit Hamiltonian and show that we can reuse the solutions from the spin qubit after careful treatments. In this process, we share scripts for solving the superconducting qubits.

In Part IV, we discuss the roles of microwave electronics in quantum computers. Then we show the design parameters and methodologies of superconducting integrated quantum chip. Finally, we touch upon errors and decoherence time measurements. After reading this book, readers will be more ready to study more advanced topics such as open quantum systems.

This book is accompanied by many teaching videos and GitHub codes. Readers can find them in Appendix A.

For any comments, please send an email to intro.qc.wong@gmail.com.

California *Hiu Yung Wong*
August 2024

# Contents

# Part V
# Volume I Addenda and Errata

# Chapter 26
# Volume I Addenda and Errata

## 26.1 Introduction

In this Chapter, we will fix some of the typos in Volume I [1] and add clarifications. They are ordered according to subsection numbers.

## 26.2 Addenda and Errata

**Section: 12.3.2.2**

*Location: Page 163, Right before Eq. (12.12). Highlight*: Clarification of tensor product.

Therefore, the Hamiltonian of the system, which can be obtained by a tensor product based on the individual Hamiltonians, $\boldsymbol{H_1} \otimes \boldsymbol{I_2} + \boldsymbol{I_1} \otimes \boldsymbol{H_2}$, is

$$\boldsymbol{H} = \begin{pmatrix} -E_z & 0 & 0 & 0 \\ 0 & \frac{-dE_z}{2} & 0 & 0 \\ 0 & 0 & \frac{dE_z}{2} & 0 \\ 0 & 0 & 0 & E_z \end{pmatrix}, \qquad (12.12)$$

**Section: 17.3**

*Location: Page 241, Fig. 17.2. Highlight*: Typos in Steps 3 and 5 in Fig. 17.2.

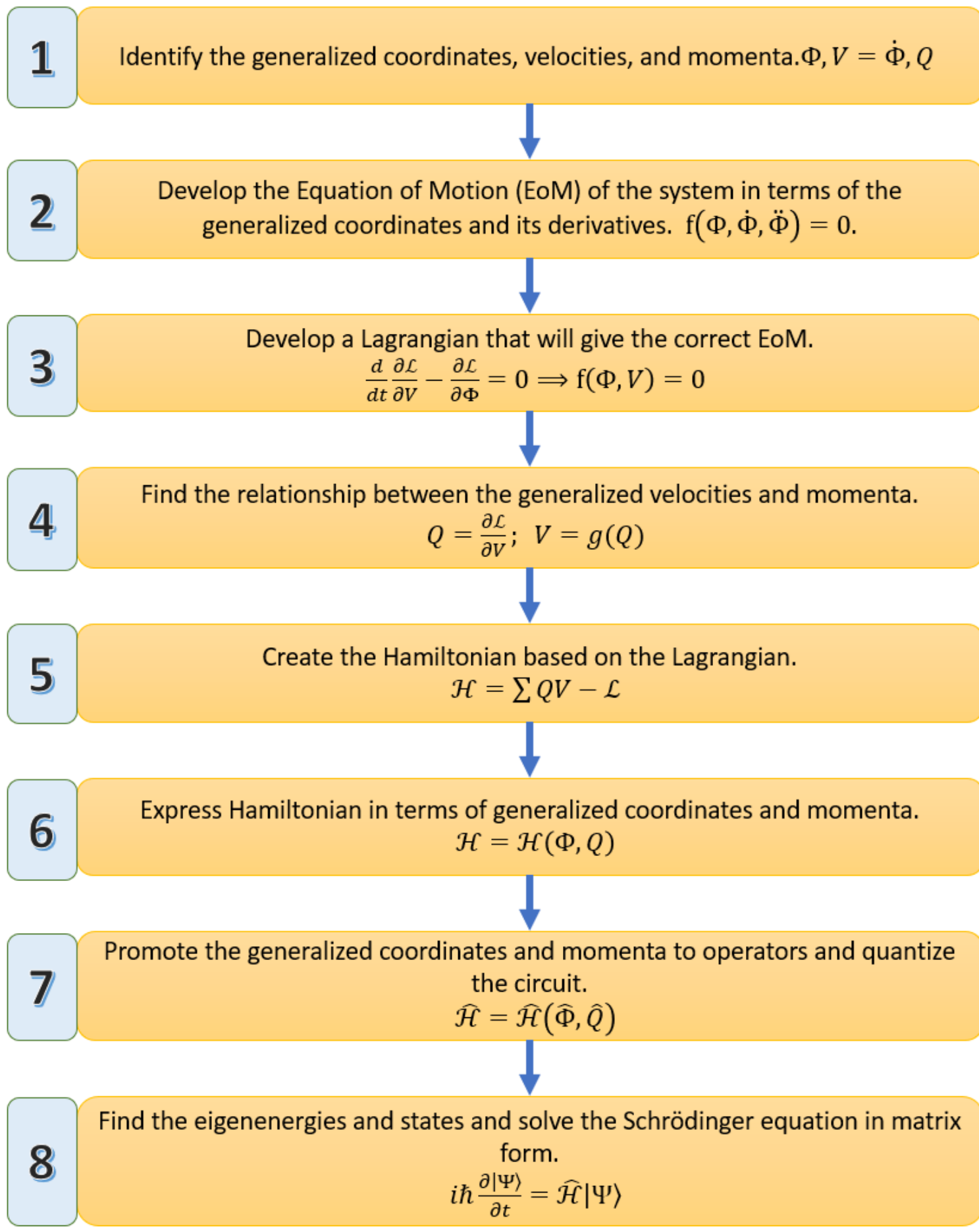

Fig. 17.2 Quantization flow using the electrical circuit as an example

**Section: 17.3**

*Location: Page 246, Steps 5 and 6, Right before Eq. (17.35). Highlight*: Integration should be used since the Hamiltonian is time-dependent.

It, thus, adds an additional total phase shift to the unitary matrix constructed from $e^{-\frac{i}{\hbar}\int_0^t \mathcal{H}(t')dt'}$.

**Section: 23.3.1**

*Location: Pages 329-331 Highlight*: The initial signals and LO signal need to be changed so that the low-pass filter will filter the higher frequency component. The text and Fig. 23.2 are updated accordingly.

Firstly, an "in-phase" (or $I$) component and an "out-of-phase" (or $Q$ or **quadrature** component) are generated by AWGs (Fig. 23.2). AWGs can generate any (arbitrary) waveform as long as the required frequency is not too high. The $I$ component is

$$-2s(t)I\sin\omega_{AWG}t = -2s(t)\cos\phi\sin\omega_{AWG}t, \tag{23.5}$$

and the $Q$ component is

$$-2s(t)Q\sin\omega_{AWG}t = 2s(t)\sin\phi\sin\omega_{AWG}t, \tag{23.6}$$

where $\omega_{AWG}$ is the frequency of the pulses and $s(t)$ is the envolope function. Note that $s(t)$ is scaled by $-2I$ and $-2Q$, respectively, as the final envelope functions of the pulses generated by the AWG in this particular example. Since $\omega_{AWG}$ is in the order of tens of $MHz$ (e.g., $50MHz$), it can be generated digitally by the AWGs easily. Note that $\phi$ is embedded as the amplitude of the $I$ and $Q$ signals.

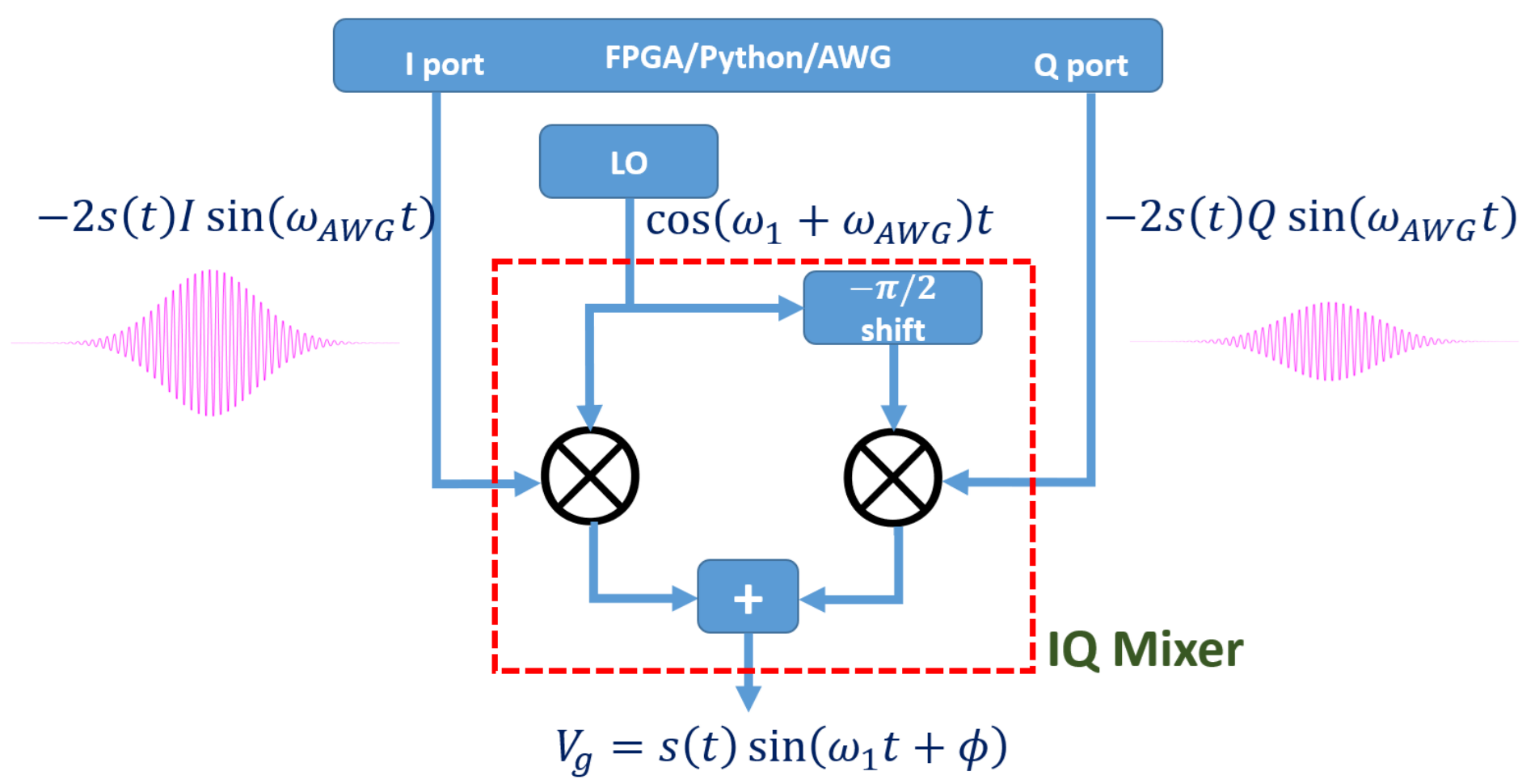

Fig. 23.2 Schematic showing how the $I$ and $Q$ signals generated by AWG are mixed with the LO signal in an $IQ$ mixer to achieve the desired wavefunction for qubit manipulations

However, the pulses we need to interact with the qubits are in $GHz$. An $I-Q$ mixer is then used to multiply the AWG signals with the signal from a local os-cillator to bring it to a high enough frequency. This is called **up-conversion**. The LO generates a high-frequency sinusoidal wave at the $GHz$ range, $\cos(\omega_1 + \omega_{AWG})t$ (e.g., 5 GHz). We assume the amplitude is one for simplicity. Its frequency is chosen to be $\omega_1 + \omega_{AWG}$ because the goal is to achieve a signal with $\omega_1$ at the mixer output. Note that, in an $I-Q$ mixer, the $I$ part is multiplied by the LO signal (in-phase component) directly. The $Q$ component is multiplied by the LO signal phase-shifted by $-\pi/2$ (quadrature component). Note that we make it $-\pi/2$ instead of the commonly used $\pi/2$ because we want to make the final equation in the desired form for instructional purposes. They are then added together as the output.

Therefore, the signal after the $I-Q$ mixer, $V_g$, is given by

$$\begin{aligned} V_g &= -2s(t)I\sin\omega_{AWG}t\cos(\omega_1+\omega_{AWG})t \\ &\quad -2s(t)Q\sin\omega_{AWG}t\cos\left((\omega_1+\omega_{AWG})t-\frac{\pi}{2}\right), \\ &= -2s(t)\cos\phi\sin\omega_{AWG}t\cos(\omega_1+\omega_{AWG})t \\ &\quad +2s(t)\sin\phi\sin\omega_{AWG}t\sin(\omega_1+\omega_{AWG})t, \end{aligned} \tag{23.7}$$

where we have used the identity, $\cos(\theta-\pi/2)=\sin\theta$ in the second line. Then, we will use the product formulae, $\sin\alpha\cos\beta=\frac{1}{2}(\sin(\alpha+\beta)+\sin(\alpha-\beta))$ for the first term and $\sin\alpha\sin\beta=\frac{1}{2}(\cos(\alpha-\beta)-\cos(\alpha+\beta))$ for the second term. We get:

$$\begin{aligned} V_g = &-2s(t)\cos\phi\frac{1}{2}(\sin(\omega_1 t+2\omega_{AWG}t)+\sin(-\omega_1 t)) \\ &+2s(t)\sin\phi\frac{1}{2}(\cos(-\omega_1 t)-\cos(\omega_1 t+2\omega_{AWG}t)), \end{aligned} \tag{23.8}$$

where the $(2\omega_{AWG}t+\omega_1 t)$ terms will be filtered by a **low-pass filter (LPF)** and the signal becomes:

$$\begin{aligned} V_g &= s(t)\cos\phi\sin\omega_1 t+s(t)\sin\phi\cos\omega_1 t, \\ &= s(t)\sin(\omega_1 t+\phi)., \end{aligned} \tag{23.9}$$

Therefore, we have achieved the pulse in Fig. 21.1 for qubit manipulation. More specifically, we only need to set the amplitude of the pulse from the $I$ ($Q$) port to control the amount of rotation about the $x$-axis ($y$-axis) on the Bloch sphere. We can also control the rotation speed $\Omega_R(t)$ by applying an appropriate envelope function $s(t)$ using AWG.

It should be noted that if $\omega_{AWG}$ is not large enough, filtering of the $(2\omega_{AWG}t+\omega_1 t)$ terms can be difficult as it is approximately the same as $\omega_1 t$. Then a **single-side band (SSB)** mixer is required, which can achieve the same purpose of keeping only one of the components.

**Section: 24.3.2.3**

*Location: Page 348. Highlight*: *Wrong* definition of $Q$ was used. $Q$ is the average number of cycles for the stored energy to be dissipated **scaled by 2$\pi$**. The factor $2\pi$ was missing in the definition of $Q$. Therefore, it should be

The quality factor of a resonator, $Q_r$, can be thought of as the number of oscillations, $N$, that occur before the energy is dissipated (in this case, through the coupling with the feedline to the outside world) *and is scaled by 2$\pi$*. Therefore, $Q_r = 2\pi N$. Usually, $Q_r$ is between $5,000$ and $10,000$. Therefore, the time, $T_r$, it takes to lose its energy is

$$\begin{aligned} T_r &= \frac{2\pi}{\omega_r} N, \\ &= \frac{Q_r}{\omega_r}, \\ &= \frac{5000}{2\pi \times 7 \times 10^9}\,\mathrm{s}, \\ &= 113.7\,\mathrm{ns}. \end{aligned} \tag{24.5}$$

**Example: 24.1**

*Location: Page 351. Highlight*: *Wrong* definition of $Q$ was used. $Q$ is the average number of cycles for the stored energy to be dissipated *scaled by 2$\pi$*. The factor $2\pi$ was missing in the definition of $Q$. Therefore, it should be

**Example 24.1** Find the energy relaxation time due to the coupling of the qubit to the outside world for $Q_q = 10^6$ and $Q_q = 10^7$.

Based on Eq.(24,5)

$$\begin{aligned} T_q &= \frac{2\pi}{\omega_q} N, \\ &= \frac{Q_q}{\omega_q}, \\ &= \frac{10^6}{2\pi \times 4.5 \times 10^9}\,\mathrm{s}, \\ &= 35.4\,\mu\mathrm{s}. \end{aligned} \tag{24.12}$$

This is still too small because it is similar to the qubit decoherence time (about 50µs–200µ). If $Q_r = 10^7$, it will increase to $354\mu$s.

---

**Section: 24.3.4**

*Location: Page 348. Highlight*: *Wrong* definition of $Q$ was used. $Q$ is the average number of cycles for the stored energy to be dissipated **scaled by 2$\pi$**. The factor $2\pi$ was missing in the definition of $Q$. *Also*, $\Delta_f = \frac{1}{2\pi T_r}$. Therefore, the equations should be changed to

This can be estimated using

$$\begin{aligned} \Delta_f &= \frac{1}{2\pi T_r}, \\ &= \frac{\omega_r}{2\pi Q_r}, \\ &= \frac{2\pi \times 7 \times 10^9}{2\pi \times 5000}\,\text{Hz}, \\ &= 1.4\,\text{MHz}. \end{aligned} \tag{24.13}$$

# Part VI
# Trapped Ion Qubit Architecture and Hardware

# Chapter 27
# Trapped Ion Qubit - Preliminary Physics

## 27.1 Introduction

In this chapter, we will introduce a few important concepts to help us appreciate the nature and operations of trapped ion qubits in the following chapters. We will review the electronic structure of atoms based on the hydrogen atom model. We will appreciate that further energy splitting is possible in a multi-electron atom. We then discuss the physics of **fine structures**, which is one of the important energy splittings and is caused by **spin-orbit coupling**. Then we will introduce **term symbols** in atomic physics. Finally, we apply these concepts to the Ytterbium atom and use the term symbols to explain the ionization process to create a Ytterbium ion to be used in a quantum computer.

### *27.1.1 Learning Outcomes*

Able to describe the cause of fine structures; be comfortable in using term symbols and understand that they are just labels of energy levels; understand some of the term symbols in the Ytterbium atom.

### *27.1.2 Teaching Videos*

- `https://youtu.be/nQAujUTRtps`
- `https://youtu.be/mymwdbBBgAQ`

## 27.2 Electronic Structure

From our basic knowledge in atomic physics or chemistry, we know that the electrons in an atom occupy discrete levels (orbitals) [2]. Here we remind ourselves that electron orbitals are the solutions (eigenstates) of the Schrödinger equation of the atom. The simplest one is that of a hydrogen atom. In a hydrogen atom, it turns out that there are four quantum numbers that can be used to describe the eigenstates. They are the **principal quantum number**, $n$, the **azimuthal (orbital angular momentum) quantum number**, $l$, **magnetic quantum number**, $m$, and **spin quantum number**, $m_s$. We label the states using $|n, l, m, m_s\rangle$.

$n$ takes the values of 1, 2, 3, $\cdots$. $l$ takes 0, 1, 2, $\cdots$, $n-1$ for a given $n$. Note that $l$ is also labeled as $s$ ($l=0$), $p$ ($l=1$), $d$ ($l=2$), $f$ ($l=3$), etc. In a hydrogen atom, the energy of the electron (or the eigenenergy of the eigenstate, the orbital) is *only* determined by $n$ and is independent of $l$. $l$ only determines the angular momentum of the electron in that orbital. There is no physical reason why the energy is independent of $l$. This just *accidentally* happens in Coulomb potential, and thus this is called the **accidental degeneracy**. Again, degeneracy means more than one eigenstate (orbital) having the same eigenenergy. $l$ also determines the angular momentum, $\vec{L}$, of the electron (see also Section 7.2 in [1]). In quantum mechanics, since the angular momentum in orthogonal directions cannot be determined precisely at the same time (Heisenberg uncertainty principle), only the size, $|\vec{L}|$, or the square, $\vec{L}^2$, of the angular momentum can be determined. It is given that [3],

$$\vec{L}^2 = l(l+1)\hbar^2, \tag{27.1}$$

where $\hbar$ is the reduced Planck's constant. We say that $l$ is the quantum number for angular momentum. It is related to the eigenvalue of the angular momentum operator ($\hat{L}^2$) through Eq. (27.1). In a hydrogen atom, $\hat{L}^2$ commutes with the Hamiltonian and thus it is a **good quantum number**. A good quantum number means the observable (in this case $\hat{L}^2$) corresponding to that quantum number (in this case $l$) commutes with the Hamiltonian of a system (in this case a hydrogen atom) and, thus, its eigenstates do not change as a function of time (up to a phase factor).

$m$ ranges from $-l$ to $l$ for a given $l$. $m$ is also a good quantum number and it is related to the eigenvalue of the $z$ component of the angular momentum, $\vec{L}_z$. They are degenerated as expected before an external magnetic field is applied. If an external magnetic field is applied, the field sets the $z$ direction and splits them into different energy levels (see also Section 7.4 in [1]). This splitting is due to the interaction between the magnetic moment generated by the electron orbital angular momentum and the external magnetic field. This is called the **Zeeman splitting**. The eigenvalues of the observable $\hat{L}_z$ are given by $m\hbar$.

Furthermore, due to electron spin, $m_s$ can be either $\frac{-1}{2}$ or $\frac{1}{2}$. The spin magnetic moment will interact with the external magnetic field to split the levels further. When the splitting due to spin is included in addition to the orbital angular motion, it is called the **anomalous Zeeman effect**.

We thus write the **electron configuration** of a hydrogen atom as $1s^1$ because it has 1 electron in the state $|n = 1, l = s = 0\rangle$. Strictly speaking, these are *good* quantum numbers only for the hydrogen atom (with only Coulomb potential and a single electron). However, we continue to use these quantum numbers for other atoms, and they are still meaningful in practical use. For example, we write the electron configuration of a silicon atom as $1s^2 2s^2 2p^6 3s^2 3p^2$, which clearly shows how electrons fill various orbitals of different $n$ and $l$ in the absence of an external magnetic field. Fig. 27.1 shows the electronic configuration of silicon. Each line represents an orbital state an electron can occupy. Since electrons may have two different spins ($m_s = \frac{-1}{2}$ or $m_s = \frac{1}{2}$), each line actually corresponds to two possible states. Therefore, it may have 2 electrons when $n = 1$ (thus $l = 0$ ($s$), $m = 0$, and therefore $1s^2$), 8 electrons when $n = 2$ because $l = 0$ ($s$) (thus $2s^2$) or $l = 1$ ($p$) (thus $m = -1$, 0, or $-1$ and each has 2 electrons giving $2p^6$), and 4 electrons when $n = 3$ as silicon only has a total of 14 electrons. Note that in multi-electron atoms, the orbital energy depends on not only $n$ but also $l$. For example, $3s$ has a lower energy than $3p$ because its wavefunction is closer to the nucleus and thus experiences less screening (or experiences a larger effective nuclear charge). Therefore, $3s$ is filled before $3p$. Here, we see that *once more physics is included in multi-electron atoms, energy splitting occurs.*

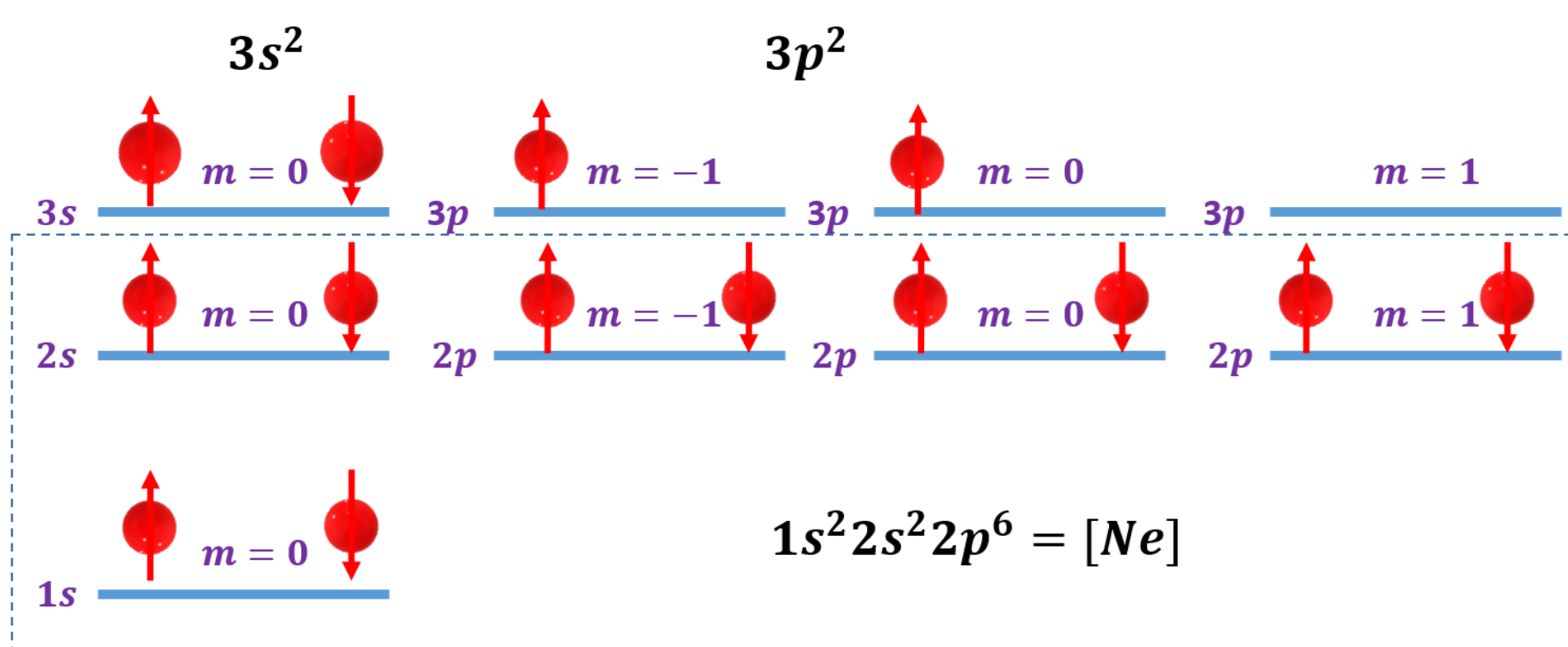

Fig. 27.1: Electron configuration of Si by assuming a hydrogen-like Hamiltonian (potential proportional to $1/r$). It can be regarded as the configuration of neon [Ne]=$1s^2 2s^2 2p^6$ by adding $3s^2 3p^2$. Note that the three $3p$ orbitals are equivalent and indistinguishable. Therefore, the electrons are just assigned randomly in the drawing.

For simplicity, the electron configuration of an element is commonly labeled as that of the nearest smaller noble gas element (with completed shells) with additional electrons. For example, that of silicon can be written as that of neon by adding $3s^2 3p^2$, i.e., [Ne]$3s^2 3p^2$ (Fig. 27.1).

A commonly used trapped ion is from ytterbium-171 atom ($^{171}_{70}$Yb) [4]. This atom has 70 protons, 101 neutrons, and 70 electrons. It has an electron configuration of $1s^2 2s^2 2p^6 3s^2 3p^6 3d^{10} 4s^2 4p^6 4d^{10} 5s^2 5p^6 4f^{14} 6s^2$, which is $[\mathrm{Xe}]4f^{14}6s^2$. That means it has that of xeon (56 electrons) in addition to 14 electrons in $n = 4$, $l = 3$ ($f$) orbitals (with $m = -3$ to $m = 3$) and 2 electrons in $n = 6$, $l = 0$ ($s$) orbitals.

We have reviewed the eigenstates of elements based on the hydrogen-like Hamiltonian. They are pretty accurate in describing many important experimental observations for various elements in the periodic table. The eigenstates $|n,l,m,m_s\rangle$ also form our fundamental understanding of atomic structures.

However, when the atomic size increases, many important effects kick in and cannot be neglected. For example, orbital energies do not depend solely on $n$ anymore when there are multiple electrons (as briefly discussed in the silicon case). For example, ytterbium-171 has its electrons filled in $5s$ and $5p$ first before filling $4f$, while the hydrogen model expects $n = 4$ to have a lower energy than $n = 5$. This is because the screening effect by other electrons is different for different $l$. For $l = 0 = s$, the wavefunction is closer to the nucleus and thus feels less screening than $l = 3 = f$. As a result, a $6s$ electron has a lower energy than a $4f$ electron because a $4f$ electron feels more screening by other electrons and thus is attracted by the nucleus less. The screening effect is also called the **quantum defect** (see section 4.2 in [2]).

## 27.3 Spin-Orbit Coupling and Fine Structure

For heavy atoms, another important phenomenon is the so-called **spin-orbit coupling**. Spin-orbit coupling introduces an additional term in the Hamiltonian. When it is large, the error in the hydrogen model becomes large. *Spin-orbit coupling increases as the atomic number, Z, increases.*

Spin-orbit coupling is a relativistic effect. In the rest frame of the nucleus, the electron experiences an electric field, $\vec{E}$, due to the nucleus. This field can be approximated as a radial field (i.e. depending only on the radial coordinate), $\vec{E}(r)$, due to the Coulomb potential, $U = \frac{Ze}{4\pi\varepsilon_0 r}$, where $r$ is the distance from the nucleus and $\varepsilon_0$ is the vacuum permittivity. Therefore,

$$\begin{aligned}\vec{E} &= |E|\frac{\vec{r}}{r}, \\ &= \left|\frac{\partial U}{\partial r}\right|\frac{\vec{r}}{r}, \\ &= \frac{Ze}{4\pi\varepsilon_0 r^3}\vec{r}.\end{aligned} \tag{27.2}$$

The electron, with mass $m$, may be described as moving at velocity $\vec{v}$ and thus has a linear momentum, $\vec{p} = m\vec{v}$. If we work in the *rest frame of the electron*, according to relativity, the electron will experience a magnetic field instead, which is

$$
\begin{aligned}
\vec{B} &= -\frac{\vec{v}\times\vec{E}}{c^2},\\
&= \frac{\vec{E}\times\vec{p}}{mc^2},\\
&= \frac{\frac{Ze}{4\pi\varepsilon_0 r^3}\vec{r}\times\vec{p}}{mc^2},\\
&= \frac{Ze\vec{r}\times\vec{p}}{4\pi\varepsilon_0 r^3 mc^2},\\
&= \frac{Ze\vec{L}}{4\pi\varepsilon_0 r^3 mc^2},
\end{aligned}
\tag{27.3}
$$

where $c$ is the speed of light in the vacuum. We also used Eq. (27.2) in line 3 and the definition of angular momentum in line 5 (see Eq. (7.1) in [1]).

The electron has a spin angular momentum, $\vec{S}$. It has the corresponding spin magnetic moment of $\vec{\mu}_e = g\mu_B\frac{\vec{S}}{\hbar}$ (see Eq. (7.9) in [1]). Therefore, there will be Zeeman splitting due to this magnetic field as we discussed in Section 7.4 in [1]. Based on Eq. (7.10) in [1] and by including a factor of $1/2$ due to **Thomas precession** (see below), we have

$$
\begin{aligned}
H_{so} &= -\frac{1}{2}\vec{B}\cdot\vec{\mu}_e,\\
&= -\frac{1}{2}\frac{Ze\vec{L}}{4\pi\varepsilon_0 r^3 mc^2}\cdot g\mu_B\frac{\vec{S}}{\hbar},\\
&= -\frac{1}{2}\frac{Zeg\mu_B}{4\pi\varepsilon_0 r^3 mc^2\hbar}\vec{L}\cdot\vec{S},\\
&= \frac{Ze\mu_B}{4\pi\varepsilon_0 mc^2\hbar}\frac{1}{r^3}\vec{L}\cdot\vec{S},
\end{aligned}
\tag{27.4}
$$

where we used the definition of spin magnetic moment and Eq. (27.3) in line 2. In line 4, since $g\approx -2$ for an electron, it is cancelled with the $1/2$ factor due to Thomas precession. Thomas precession is "a relativistic effect that arises because we are calculating the magnetic field in a frame of reference that is not stationary but rotates as the electron moves about the nucleus" [2].

At this point, it might be clearer why it is called the spin-orbit coupling. While the electric field is generated by the nucleus, it is the orbital movement of the electron that causes the effective magnetic field to appear in its rest frame due to relativity. If the electron were not moving, it would not have felt a magnetic field. Therefore, the coupling is due to its spin and its orbital movement. It should be highlighted that while the physics is due to the interaction between the perceived magnetic field and the spin magnetic moment, they are represented by the angular momentum and spin angular momentum of the electron, respectively, in Eq. (27.4).

To further calculate Zeeman splitting, we need to find the expectation value of $H_{so}$, that is

$$\begin{aligned} E_{so} &= \langle H_{so} \rangle \\ &= \frac{Ze\mu_B}{4\pi\varepsilon_0 mc^2\hbar}\left\langle \frac{1}{r^3}\right\rangle \langle \vec{L}\cdot\vec{S}\rangle. \end{aligned} \tag{27.5}$$

From Eq. (2.23) in [2],

$$\left\langle \frac{1}{r^3}\right\rangle = \frac{1}{l(l+\frac{1}{2})(l+1)}\left(\frac{Z}{na_0}\right)^3, \tag{27.6}$$

where $a_0 = \frac{\hbar^2}{(e^2/4\pi\varepsilon_0)m}$ is the **Bohr radius**. We will not derive it and will just take this for granted.

### *27.3.1 Addition of Angular Momentum*

Now we need to evaluate $\langle \vec{L}\cdot\vec{S}\rangle$ and this deserves some attention. We will first claim that the orbital angular momentum, $\vec{L}$, and the spin angular momentum, $\vec{S}$, are both angular momentum. Therefore, an electron has a total angular momentum, $\vec{J}$, due to them. That is

$$\vec{J} = \vec{L} + \vec{S}, \tag{27.7}$$

Fig. 27.2 shows the vector model of the addition of $\vec{L}$ and $\vec{S}$. There are a few things we need to pay attention to. $\vec{J}$ is a constant (in direction and magnitude) because there is no external torque applied to the electron (see Section 8.3.1 in [1]). Therefore, it has a good quantum number, $j$, in this system. Its magnitude and square of magnitude are constant and well-defined by following Eq. (27.1) and thus is $\vec{J}^2 = j(j+1)\hbar^2$ (see also Eq. (27.1)). On the other hand, due to spin-orbit coupling (interaction), $\vec{L}$ and $\vec{S}$ are no longer constants. Their directions change as a function of time (precessing about $\vec{J}$). Luckily, their magnitudes are still constants and thus $l$ and $s$ are still good quantum numbers. Therefore, we have

$$\begin{aligned} \vec{J}^2 &= j(j+1)\hbar^2, \\ \vec{L}^2 &= l(l+1)\hbar^2, \\ \vec{S}^2 &= s(s+1)\hbar^2. \end{aligned} \tag{27.8}$$

We have been using vector notation so far. Eq. 27.8 gives the *possible values* of the square of the magnitudes of the vectors. We know $l$ goes from 0 to $n-1$ and $s$ is $\frac{1}{2}$. How about $j$? We will not derive it. Interested readers can refer to Chapter 3 in [3] for the addition of angular momentum. We just mention here that for any given pair of $l$ and $s$, we have

$$j = |l-s|, |l-s+1|, \cdots, |l+s|. \tag{27.9}$$

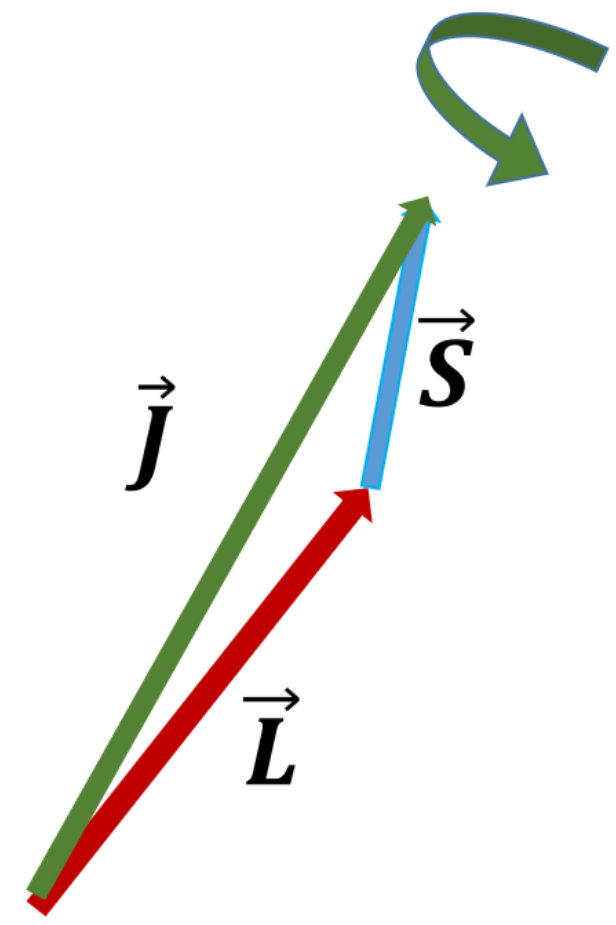

**Fig. 27.2** Addition of the orbital angular momentum, $\vec{L}$, and the spin angular momentum, $\vec{S}$, of an electron

This makes sense, as the maximum magnitude of the sum of two momenta must be the sum of their magnitudes $|l+s|$ and the minimum must be their difference $|l-s|$ (although the quantum numbers, $l$ and $s$, are not exactly the corresponding magnitudes).

To find $\langle \vec{L}\cdot\vec{S}\rangle$, we first derive the following

$$\begin{aligned}
\vec{J} &= \vec{L}+\vec{S}, \\
\vec{J}^2 &= (\vec{L}+\vec{S})^2, \\
&= (\vec{L}+\vec{S})\cdot(\vec{L}+\vec{S}), \\
&= \vec{L}^2 + 2\vec{L}\cdot\vec{S} + \vec{S}^2, \\
\vec{L}\cdot\vec{S} &= \frac{1}{2}(\vec{J}^2 - \vec{L}^2 - \vec{S}^2),
\end{aligned} \tag{27.10}$$

where, in line 3, we remind ourselves that the square of a vector is just the inner product of that vector with itself.

Therefore,

$$
\begin{aligned}
\langle \vec{L}\cdot\vec{S}\rangle &= \langle \frac{1}{2}(\vec{J}^2-\vec{L}^2-\vec{S}^2)\rangle, \\
&= \frac{1}{2}(\langle \vec{J}^2\rangle - \langle \vec{L}^2\rangle - \langle \vec{S}^2\rangle), \\
&= \frac{1}{2}(j(j+1)\hbar^2 - l(l+1)\hbar^2 - s(s+1)\hbar^2), \\
&= \frac{\hbar^2}{2}(j(j+1)-l(l+1)-s(s+1)), \qquad (27.11)
\end{aligned}
$$

where we used Eq. (27.8) in line 3. It is important to clarify which state is used in finding the expectation value. That is, what is the $|\Psi\rangle$ in $\langle \vec{L}\cdot\vec{S}\rangle = \langle\Psi|\vec{L}\cdot\vec{S}|\Psi\rangle$? Obviously, we must be using the eigenstates of $\vec{J}^2$, $\vec{L}^2$, and $\vec{S}^2$ in order to use Eq. (27.8).

Finally, by substituting Eq. (27.6) and Eq. (27.11) into Eq. (27.5), we can calculate the spin-orbit coupling energy as

$$
\begin{aligned}
E_{so} &= \frac{Ze\mu_B}{4\pi\varepsilon_0 mc^2\hbar}\left\langle\frac{1}{r^3}\right\rangle\langle \vec{L}\cdot\vec{S}\rangle, \\
&= \frac{Ze\mu_B}{4\pi\varepsilon_0 mc^2\hbar}\frac{1}{l(l+\frac{1}{2})(l+1)}\left(\frac{Z}{na_0}\right)^3\frac{\hbar^2}{2}(j(j+1)-l(l+1)-s(s+1)), \\
&= \frac{e\hbar\mu_B}{2\times 4\pi\varepsilon_0 mc^2}\frac{1}{l(l+\frac{1}{2})(l+1)}\frac{Z^4}{n^3a_0^3}(j(j+1)-l(l+1)-s(s+1)), \\
&= \frac{\mu_B^2}{4\pi\varepsilon_0 c^2}\frac{1}{l(l+\frac{1}{2})(l+1)}\frac{Z^4}{n^3a_0^3}(j(j+1)-l(l+1)-s(s+1)), \qquad (27.12)
\end{aligned}
$$

where in line 4 we have made the substitution of Bohr magneton, $\mu_B = \frac{e\hbar}{2m}$. We see that as the atomic number $Z$ increases, although $n$ of interest (incompleted shell) will also increase, $E_{so}$ and thus the spin-orbit coupling increases because $Z$ rises with the power of four, and $n$ only rises with the power of three.

The extra energy splittings on top of states $|nlmm_s\rangle$ due to spin-orbit coupling ($E_{so}$) are called the **fine structure**.

## 27.4 Term Symbols

With spin-orbit couplings, $|nlmm_s\rangle$ are no longer the eigenstates of the Hamiltonian. We also need the total angular momentum quantum number to describe the split levels, and **term symbols** are introduced. A term symbol has the form of ${}^{2S+1}L_J$, where $S$, $L$, and $J$ are just the spin, orbital, and total angular momentum quantum numbers, respectively.

For example, for an electron in the $2p$ orbital, due to spin-orbit coupling, the new energy level can be described as $2p\ {}^2P_{1/2}$ or $2p\ {}^2P_{2/3}$. It means these two term symbols (corresponding to two split levels) have $S = 1/2$ (thus $2S+1=2$) and $L = 1$ (thus $P$). Note that, in term symbols, uppercase is used for $l$ (i.e. $P$ instead of $p$).

Since there is only one electron, it is natural that the spin quantum number is $1/2$. Since it is in the $p$ orbital, $l = 1$, the orbital angular momentum quantum number is also 1. However, there are two possible values for $j$. Based on Eq. (27.9), it can be $|l-s| = |1-1/2| = 1/2$ or $|l-s+1| = 3/2 = |l+s|$. Therefore, the two terms are written as $2p\ ^2P_{1/2}$ and $2p\ ^2P_{2/3}$, respectively. Often, it is not necessary to mention $2p$ for convenience. Therefore, we may also write them as $^2P_{1/2}$ and $^2P_{2/3}$. Fig. 27.3 shows the corresponding split levels (fine structure) in a hydrogen atom.

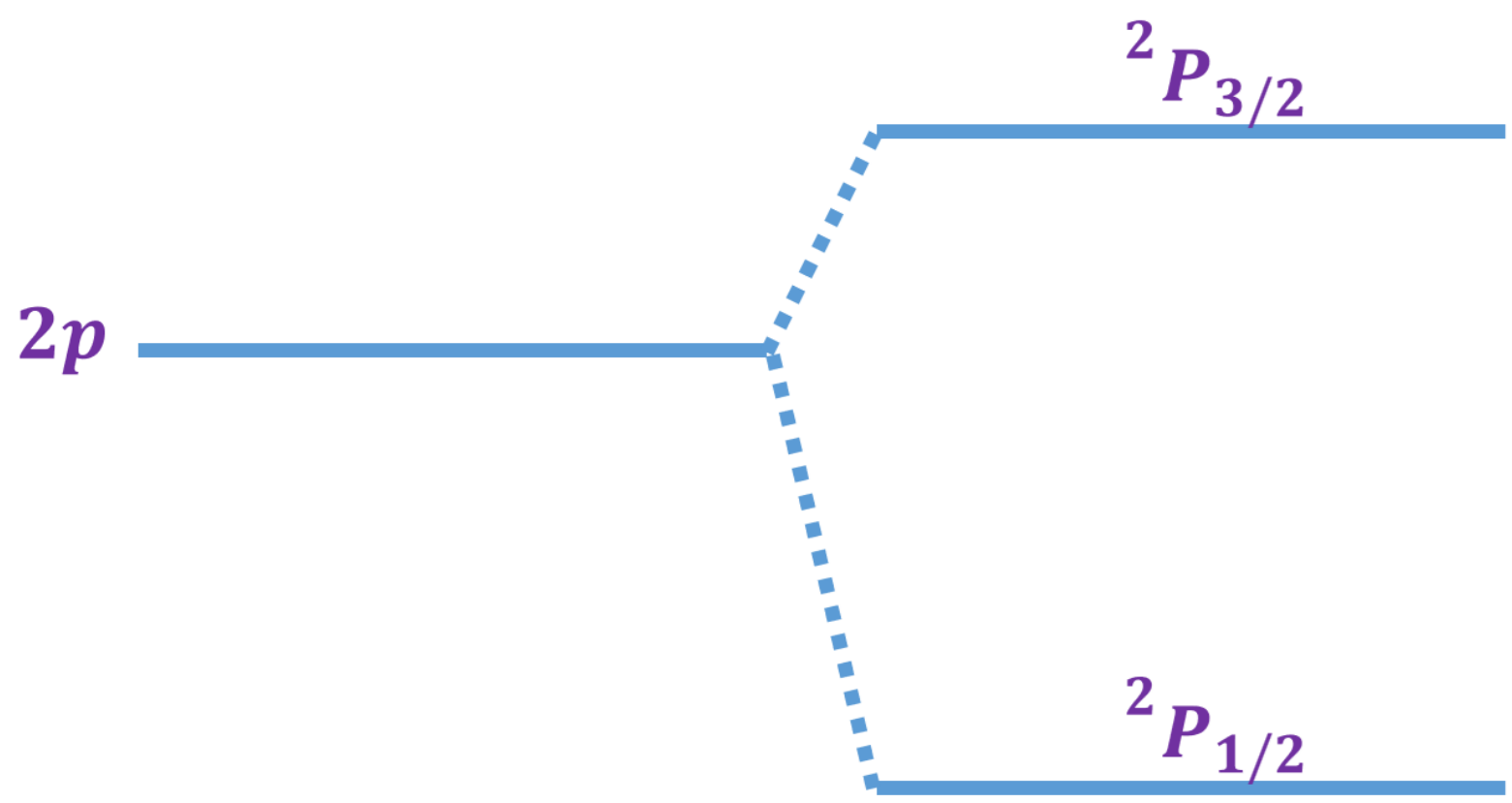

Fig. 27.3: Energy splitting of $2p$ orbital into $^2P_{1/2}$ and $^2P_{2/3}$ due to spin-orbit coupling

Term symbols are very useful and are not limited to one electron. One may have multiple electrons involving multiple orbitals. We can use the same scheme to determine the term symbols (thus the fine structure) of an atom or an ion.

*Example 27.1.* For two electrons in an $l = 1 = p$ orbital. What are the possible term symbols? Assume they have different $n$, so we do not need to worry about any potential violation of the requirement of the total wavefunction to have odd parity.

Firstly, each electron has a spin quantum number of $1/2$. Therefore, with two electrons, by using the same rule of momentum addition in Eq. (27.9), the possible *total* spin quantum numbers, $S$, are

$$\begin{aligned} S &= |1/2-1/2|, |1/2-1/2+1| = |1/2+1/2|, \\ &= 0, 1. \end{aligned} \tag{27.13}$$

For the *total* orbital angular momentum, using the same addition rule, we have

$$
\begin{aligned}
L &= |1-1|, |1-1+1|, |1-1+2| = |1+1|, \\
&= 0, 1, 2, \\
&= S, P, D.
\end{aligned}
\tag{27.14}
$$

For $S = 0$, the maximum and minimum of $J$ are the same for each $L$ because $J = |L - 0| = |L + 0|$ (see Eq. (27.9)). Therefore, the terms are ${}^1S_0$, ${}^1P_1$, and ${}^1D_2$, for $L = 0 = S$, $L = 1 = P$, and $L = 2 = D$, respectively.

For $S = 1$, the maximum and minimum of $J$ are the same for $L = 0$ because $|0-1| = |0+1| = 1$. So $J = 1$ and the term symbol is ${}^3S_1$.

For $S = 1$ and $L = 1$, we have $J = 0, 1, 2$ and the term symbols are ${}^3P_0$, ${}^3P_1$, and ${}^3P_2$.

For $S = 1$ and $L = 2$, we have $J = 1, 2, 3$ and the term symbols are ${}^3D_1$, ${}^3D_2$, and ${}^3D_3$.

Therefore, there are 10 fine-structure levels.

■

## 27.5 Yitterbium Atom Term Symbols and Yitterbium Ion Formation

In Section 27.2, we reviewed the electron configuration of Ytterbium based on the hydrogen model. It is $[\mathrm{Xe}]4f^{14}6s^2$. Now the two electrons in the $6s$ orbital have a ground state term (among others) as $6s\ {}^1S_0$ [5]. This corresponds to $S = 0$ (so that $2S + 1 = 1$), $L = 0$ (note that "S" in the term symbol refers to the "*S*" in "*S*, *P*, *D*, *F*"), and $J = 0$ (the only possible value when $S = L = 0$, see Eq. (27.9)). This means that the two electrons have opposite spins as $S = 0$ (the so-called singlet state).

It also has an excited state in which one of the $6s$ electrons is excited to a $6p$ orbital (i.e. $6s^1 6p^1$). Again, there are many possible term symbols, and one of them is $6s6p\ {}^1P_1$. This means $S = 0$, $L = 1$, and $J = 1$.

Commonly, when there is no confusions, $6s\ {}^1S_0$ and $6s6p\ {}^1P_1$ are written as ${}^1S_0$ and ${}^1P_1$, respectively [6]. Fig. 27.4 shows the relevant terms in the ionization process to produce Ytterbium-171 ion (${}^{171}Yb^+$). It is a **two-color ionization** process. This means that two laser pulses with different wavelengths are used to complete the ionization process. Firstly, a pulse with $\lambda = 398.91nm$ is used to excite one electron from ${}^1S_0$ to ${}^1P_1$. This step is *isotope selective*. That means that the energy spacing between ${}^1S_0$ and ${}^1P_1$ is different in other isotopes of Ytterbium. As a result, we can selectively only ionize Ytterbium-171, which will be used to store a qubit. We will see later that an odd mass number (an odd number of nucleons) is required to enable the so-called **hyperfine structures** required to store the information. Then, another pulse with $\lambda$ =369.53 nm is applied to promote the excited electron to a continuum state (vacuum) to complete the ionization process. Note that this is the same laser we will use for qubit initialization (Fig. 29.3), qubit readout (Fig. 29.1), and (Fig. 30.2) as well, and thus, reduce the complexity requirement of the quantum computer system.

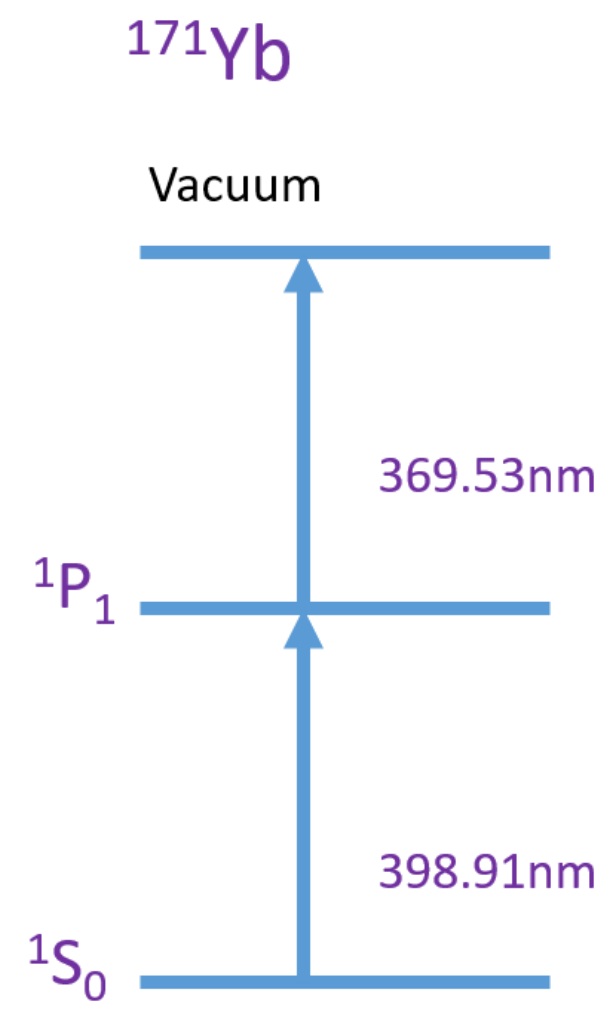

**Fig. 27.4** Two-color ionization process in Ytterbium

*Example 27.2.* Find the energy spacing between $^1S_0$ and $^1P_1$ in terms of frequency and electron-volt (eV).

The corresponding frequency is,

$$\begin{aligned} f &= \frac{c}{\lambda}, \\ &= \frac{3\times 10^8 \mathrm{m/s}}{398.91\times 10^{-9}\mathrm{m}}, \\ &= 752\mathrm{THz}. \end{aligned} \tag{27.15}$$

The corresponding energy in eV is,

$$\begin{aligned} E &= hf, \\ &= 4.14\times 10^{-15}\mathrm{eV}\cdot\mathrm{Hz}^{-1}\times 752\times 10^{12}\mathrm{Hz}, \\ &= 3.11\mathrm{eV}. \end{aligned} \tag{27.16}$$

Here I use $\mathrm{eV}\cdot\mathrm{Hz}^{-1}$ as the unit for Planck's constant. It is easy to memorize as the World Quantum Day is April 14 (4.14).

■

## 27.6 Summary

In this Chapter, we reviewed the electronic structure of atoms based on the hydrogen model. We then mentioned that energy splitting (degenerated energy levels become non-degenerated) occurs when subtle physics is considered. For example, spin-orbit coupling due to the interaction between the electron spin magnetic moment and the perceived magnetic field by the electron results in fine structures. The perceived magnetic field by the electron is due to the nucleus's electric field, seen by the moving electron due to relativity. These split levels can be labeled by term symbols. We then discussed the important term symbols (energy levels) in Ytterbium and the ionization process to create a Ytterbium ion to be used in a trapped ion quantum computer. In the next chapter, we will discuss how the qubit is encoded in a Ytterbium ion through new energy levels created in the hyperfine structures.

## Problems

**27.1. Good Quantum Numbers.**
Show that the eigenstate of an observable labeled by a good quantum number does not change as a function of time (up to a phase).

**27.2. Photon Energy.**
Find the photon energy of the $\lambda$ =369.53 nm laser in Fig. 27.4 in terms of frequency and $electron-volt\ (eV)$.

# Chapter 28
# Trapped Ion Qubit - Qubit Representation and Ion Trap

## 28.1 Introduction

In the previous chapter, we studied some of the basic atomic physics to help us understand the terminology to be used in the trapped ion qubit. In this chapter, we will study how to use trapped ions to store qubit information, before which we need to discuss **hyperfine structure**. Then we discuss the experimental setup to construct a trapped ion quantum computer. **linear Paul trap**, which can be used to trap an array of ions, will be discussed.

### *28.1.1 Learning Outcomes*

Understand how a qubit is stored in a trapped ion; be able to describe the necessary experimental setup to create a trapped ion quantum computer.

### *28.1.2 Teaching Videos*

- `https://youtu.be/nQAujUTRtps`
- `https://youtu.be/mymwdbBBgAQ`

## 28.2 Hyperfine Structure

### *28.2.1 Nuclear Spin*

In Section 27.3, we discussed the origin of fine structure in an atom or an ion. It is due to spin-orbit coupling, which is the interaction of the spin magnetic moment of an electron and the relativistic magnetic field perceived by that electron when it is moving. Spin-orbit coupling splits the otherwise degenerated states predicted by the hydrogen atom model.

Since nucleons (protons and neutrons) also have magnetic moments (due to their spin and orbital angular momenta), it is natural to expect that there might be further energy splitting if they are under a magnetic field. Indeed, this happens and results in the so-called **hyperfine structure**. It is "hyperfine" because compared to fine structures, the energy splitting is much smaller (in the order of GHz instead of hundreds of THz (e.g., Eq. (27.15)) and is more suitable for storing qubit information. Hyperfine structure is due to the interaction between the nuclear magnetic moment and the magnetic field generated by the atomic electrons with spin and orbital angular momenta. We will not dive into the physics too deeply, but we will clarify a few important concepts regarding nucleons and their spins.

First of all, both proton and neutron have a spin of $S_N = 1/2$, and thus they have the corresponding spin angular momentum, $\vec{S_N} = S_N \hbar \hat{z}$ (for simplicity, here we assume the external magnetic field is in $\hat{z}$, see also Eq. (7.7) in [1]). We expect that only charged particles may have a spin magnetic moment. This was discussed in Sections 7.2 and 7.3 [1], in which we understand the relationship between spin angular momentum and spin magnetic moment by making an analogy to that of a charged particle in a circular motion. A neutral particle is expected to have only a spin angular momentum but not a spin magnetic moment.

However, neutron *has a spin magnetic moment*. This is because a neutron is not an elementary particle. It has a structure that is formed by **quarks**. Therefore, it still has a corresponding spin magnetic moment despite being neutral overall.

Protons and neutrons also fill up the nucleus like how the electrons fill up the orbitals. This is explained by the **shell model**. For example, it has orbitals such as $1s$ and $1f$. Note that the potential in the nucleus is not a Coulomb potential. Therefore, it can have $1f$. We will just take this for granted. But like the electrons, its angular momentum numbers are $l = 0$, 1, 2, 3, etc., corresponding to $s$, $p$, $d$, $f$, etc., respectively. Therefore, they also have orbital angular momentum. For a proton, which is charged, this will generate an orbital magnetic moment. For a neutron, which is neutral, it does not have a corresponding orbital magnetic moment (unlike its spin).

Like electrons, we can relate the spin angular momentum of a neutron or a proton to its spin magnetic moment through the **gyromagnetic ratio** (see also Eq. (7.8) in [1]). This is useful for isolated protons and neutrons. However, the nucleus is a complicated system. Therefore, we combine the protons and neutrons in a nucleus to get a net spin. Moreover, in a nucleus, the protons also have orbital magnetic moment. In nuclear physics, *the effect of orbital magnetic moment is combined with the spin magnetic moment* and assumed to be a result of a net **nuclear spin**. *In this book, we define the corresponding angular momentum due to nuclear spin as* $\vec{I}$.

The total magnetic moment of a nucleus is then given by

$$\begin{aligned} \vec{\mu}_I &= g_I \frac{e\hbar}{2M_p} \frac{\vec{I}}{\hbar}, \\ &= g_I \mu_N \frac{\vec{I}}{\hbar}, \end{aligned} \tag{28.1}$$

where $g_I$ and $M_p$ are the **g-factor** and proton mass, respectivley. We also defined **nuclear magneton** as $\mu_N = \frac{e\hbar}{2M_p}$.

Compared to the definition of Bohr magneton, $\mu_B = \frac{e\hbar}{2m}$, which is used for calculating the spin magnetic moment of an electron (also see Eq. (7.9) in [1]), the nuclear magneton is much smaller than the Bohr magneton, as

$$\begin{aligned} \frac{\mu_I}{\mu_B} &= \frac{m}{M_p}, \\ &\approx \frac{1}{1836}. \end{aligned} \tag{28.2}$$

Therefore, the nuclear magnetic moment is much smaller than the electron spin magnetic moment. As a result, hyperfine structure is much more subtle than fine structure.

### *28.2.2 Hyperfine Structure*

As mentioned, hyperfine structure is due to the interaction between the nuclear magnetic moment and the magnetic field created by the atomic electrons, $\vec{B}_e$. The interaction Hamiltonian can be expressed as

$$H_{HFS} = -\vec{B}_e \cdot \vec{\mu}_I \tag{28.3}$$

Readers should compare this to Eq. (7.10) in [1] and Eq. (27.4). The magnetic field created by the atomic electron is complicated. But following the treatment in Chapter 6 in [2], we can use the magnetic field created by an *s*-orbital electron to understand the hyperfine interaction. Since it is an *s* electron, its angular momentum is zero ($l = 0$). Therefore, it only generates a magnetic field due to its spin and $B_e \propto \vec{S}_e$. Here, for clarity, I use $\vec{S}_e$ to represent the spin angular momentum of an electron instead of $\vec{S}$ (in chapter 27). Based also on Eqs. (28.1) and (28.3), we have $H_{HFS} \propto \vec{S}_e \cdot \vec{I}$. Therefore, we set

$$H_{HFS} = A\vec{S}_e \cdot \vec{I}, \tag{28.4}$$

where $A$ is a proportional factor depending on the electron cloud distribution, etc.

When $l \neq 0$, Eq. (28.4) can be replaced by

$$H_{HFS} = A\vec{J} \cdot \vec{I}, \tag{28.5}$$

where it should be convincing that the interaction or the magnetic field must be related to the total angular momentum of the atomic electrons, $\vec{J}$ (Eq. (27.7)), although a proof is not provided.

Note again that, in this book, $\vec{I}$ is the nuclear spin angular momentum. Eq. (28.5) has a similar form to Eq. (27.4). We can also define the total angular momentum of the system (electrons + nucleons), $\vec{F}$ (due to nuclear spin angular momentum $\vec{I}$ and the total angular momentum of the atomic electron, $\vec{J}$), as

$$\vec{F} = \vec{I} + \vec{J}. \tag{28.6}$$

Similar to the case of spin-orbit interaction, the hyperfine interaction does not change the total angular momentum, $\vec{F}$, as this is an internal interaction. However, it changes the directions of $\vec{I}$ and $\vec{J}$. Fig. 28.1 shows that $\vec{I}$ and $\vec{J}$ precess about $\vec{F}$. Therefore, the eigenvalues of the operator corresponding to $\vec{F}$ are good quantum numbers. The quantum numbers are called $F$.

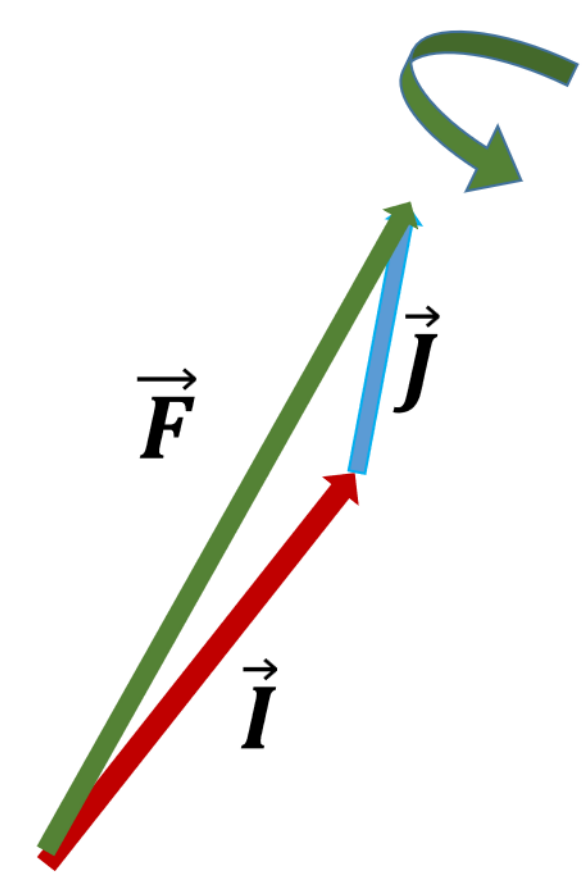

**Fig. 28.1** Addition of the nuclear spin angular momentum, $\vec{I}$, and the total angular momentum, $\vec{J}$, of an electron

*Example 28.1.* For $J = 1/2$ and $I = 1/2$, what are the possible values of $F$?

This is the addition of angular momentum. It still follows Eq. (27.9). Therefore,

$$F = |1/2 - 1/2|, |1/2 - 1/2 + 1| = |1/2 + 1/2|,$$
$$= 0, 1. \tag{28.7}$$

This is similar to what we did in Example 27.1.

One example of this scenario is in a hydrogen atom. It has one proton at the ground state. Thus, $I = 1/2$. It has one electron at $1s$. This corresponds to the ground term of $^2S_{1/2}$. $J = 1/2$ because $L = 0$ and $S = 1/2$. See Example 6.1 in [2].

Therefore, a hydrogen atom has a hyperfine structure containing states $^2S_{1/2},\ F = 0$ and $^2S_{1/2},\ F = 1$.

■

## 28.3 Ytterbium Ion for Trapped Ion Qubit

Now, with all the preparations, we are ready to discuss how a Ytterbium ion can be used to store qubit information. After the two-color ionization process discussed in Section 27.5, its electron configuration becomes $[\mathrm{Xe}]4f^{14}6s^1$. Therefore, we expect that it has a fine structure with a term of $S = 1/2$ and $L = 0$ due to the single electron in $6s$. Therefore, $J = 1/2$ which gives the term $^2S_{1/2}$ (Fig. 28.2). The nucleus of a Ytterbium ion has 70 protons and 101 neutrons. The protons are paired up and have a net spin of zero and also zero *net* angular momentum. The neutrons are also paired up except one of them, resulting in a net spin of $1/2$. As a result, the nuclear spin has $I = 1/2$. Therefore, it also has a further splitting resulting in a hyperfine structure of $^2S_{1/2},\ F = 0$ and $^2S_{1/2},\ F = 1$ (see Example 28.1). We then use $^2S_{1/2},\ F = 0$ as the ground state of the qubit, $|0\rangle$, and $^2S_{1/2},\ F = 1$ as the excited state of the qubit, $|1\rangle$. The energy spacing between $|0\rangle$ and $|1\rangle$ are found to be 12.64 GHz.

*Example 28.2.* Find the energy spacing between $^2S_{1/2},\ F = 0$ and $^2S_{1/2},\ F = 1$ in terms of electron-volt (eV) in Ytterbium ion.

The corresponding energy in eV is,

$$E = hf,$$
$$= 4.14 \times 10^{-15}\mathrm{eV} \cdot \mathrm{Hz}^{-1} \times 12.64 \times 10^{9}\mathrm{Hz},$$
$$= 52\mu\mathrm{eV}. \tag{28.8}$$

We see that this agrees with the typical qubit energy in Fig. 1.2 in [7] and is much smaller than the fine structure in Eq. (27.16).

■

In Fig. 28.2, we show another fine structure term, $^2P_{1/2}$. This is an excited state and, based on the term symbol, we know it has $S = 1/2$, $L = 1 = P$, and $J = 1/2$. Therefore, this corresponds to the $6s$ electron excited to the $6p$ orbital.

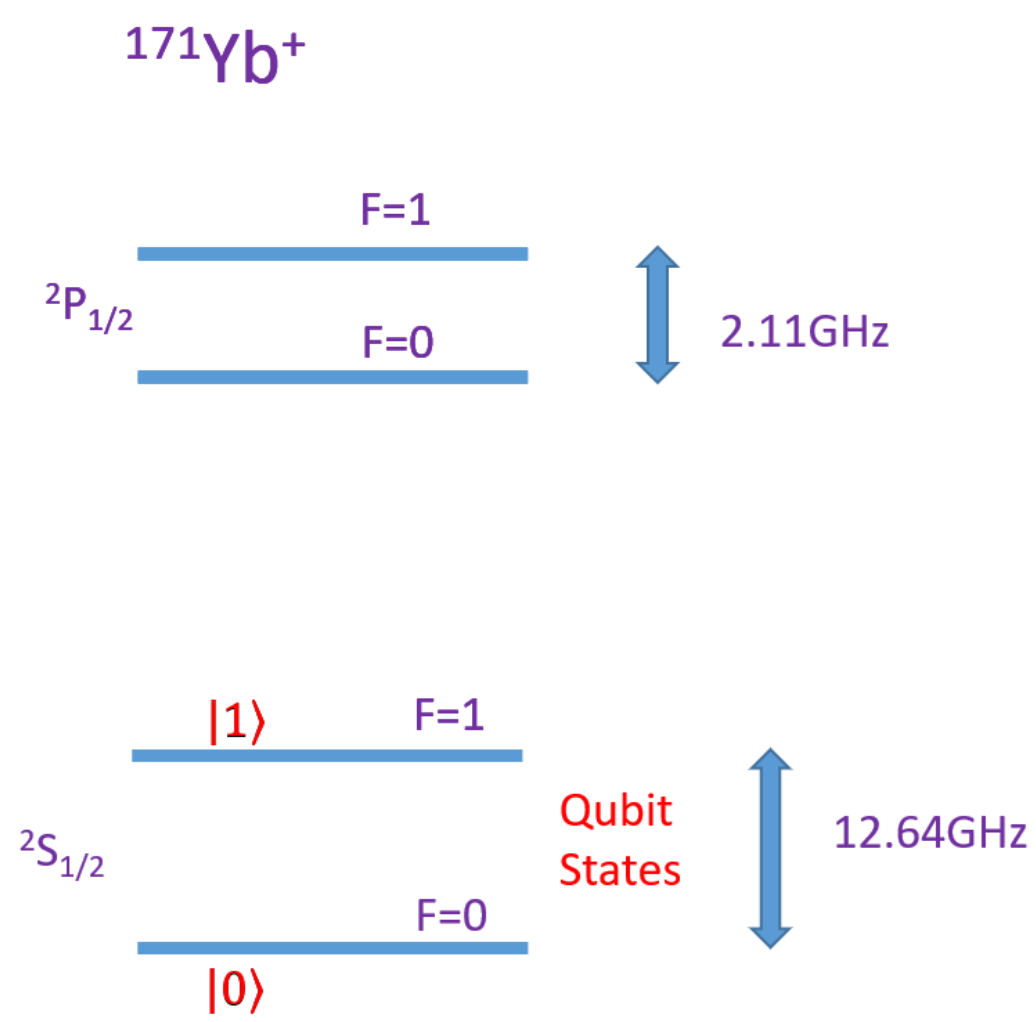

Fig. 28.2: Fine structure and hyperfine structure in Ytterbium ion. Energy is not drawn to scale.

*Example 28.3.* Find the energy spacing between $^2S_{1/2}$ and $^2P_{1/2}$ in terms of electron-volt (eV) if it corresponds to a wavelength of 369nm (e.g., determined by experiment).

The corresponding frequency is,

$$\begin{aligned} f &= \frac{c}{\lambda}, \\ &= \frac{3 \times 10^8 \mathrm{m/s}}{369 \times 10^{-9} \mathrm{m}}, \\ &= 813 \mathrm{THz}. \end{aligned} \tag{28.9}$$

The corresponding energy in eV is,

$$\begin{aligned} E &= hf, \\ &= 4.14 \times 10^{-15} \mathrm{eV} \cdot \mathrm{Hz}^{-1} \times 813 \times 10^{12} \mathrm{Hz}, \\ &= 3.37 \mathrm{eV}. \end{aligned} \tag{28.10}$$

■

This has the same order of magnitude as the energy gap in the fine structure of the Ytterbium atom in Example 27.2, and this makes sense because both involve the excitation of an electron from 6*s* to 6*p*. Of course, this is much larger than the gap between $|0\rangle$ and $|1\rangle$ because it needs a larger energy when it is ionized. Therefore, we have fulfilled one of the **DiVincenzo's criteria** in Section 1.3 in [1] that it is a well-characterized two-qubit system.

Due to the hyperfine structure, $^2P_{1/2}$ also splits into two states, $^2P_{1/2},\ F = 0$ and $^2P_{1/2},\ F = 1$. However, they are not used to store qubit information (as they can decay to $^2S_{1/2}$ in a relatively short time). But they are very important in qubit initialization, qubit readout, and laser cooling.

### *28.3.1 Why Ytterbium-171 Ion?*

At this point, we should have a short summary of the reasons to use Ytterbium-171 ion as the carrier of a qubit. Firstly, it has an odd number of nucleons. Therefore, it has a non-zero nuclear spin to create hyperfine structure to store $|0\rangle$ and $|1\rangle$. Secondly, of course, it is an ion. It can be trapped in an electric field. Thirdly, it has a reasonable ionization energy, and the two-color ionization process allows us to select it among its isotopes. Fourthly, the two hyperfine states that we have chosen are, in first order, magnetic field insensitive, which gives trapped ions a long decoherence time. The decoherence time of a Ytterbium-171 ion qubit is in the order of seconds.

## 28.4 Ion Traps

Now we will discuss how to trap the ions. In silicon electron spin qubit and superconducting qubit quantum computers [1], their qubit carriers are hosted on an integrated chip. In a trapped ion quantum computer, we need to trap the ion in a vacuum using an electric field. A vacuum environment can increase its decoherence time by minimizing its contact with the environment. It also avoids air particles from destabilizing the trapped ions. Moreover, it minimizes the scattering of laser pulses, which are used for qubit manipulation, initialization, and readout.

Naïvely, we will attempt to trap an ion using a DC electric field. In one dimension, we might feel that if we can put $Yb^+$ between two positively charged plates (Fig. 28.3), it will be "trapped" in between. This is true in the lateral direction. However, it is not trapped in the other two directions. For example, it can move vertically (green arrows). We might think that we can add more electrodes such that we will create an energy minimum to trap the ion. Unfortunately, this is not possible. According to **Earnshaw's theorem**, "a charge acted on by electrostatic forces cannot rest in stable equilibrium in an electric field" [2]. We can understand it in this way. To have an energy minimum, it means that all electric field lines in 3D need to point towards the trap location. If not, a positive charge will leave the trap due to any outward electric field line, and thus this point does not have the minimum energy by definition (as a particle should stay at the minimum energy point). However, we know that electric field lines can only terminate at charges. If the space has no charges, then it is impossible for all electric field lines to point towards the trap location. Therefore, we cannot use only electrostatics to trap an ion in space. We need to add AC components.

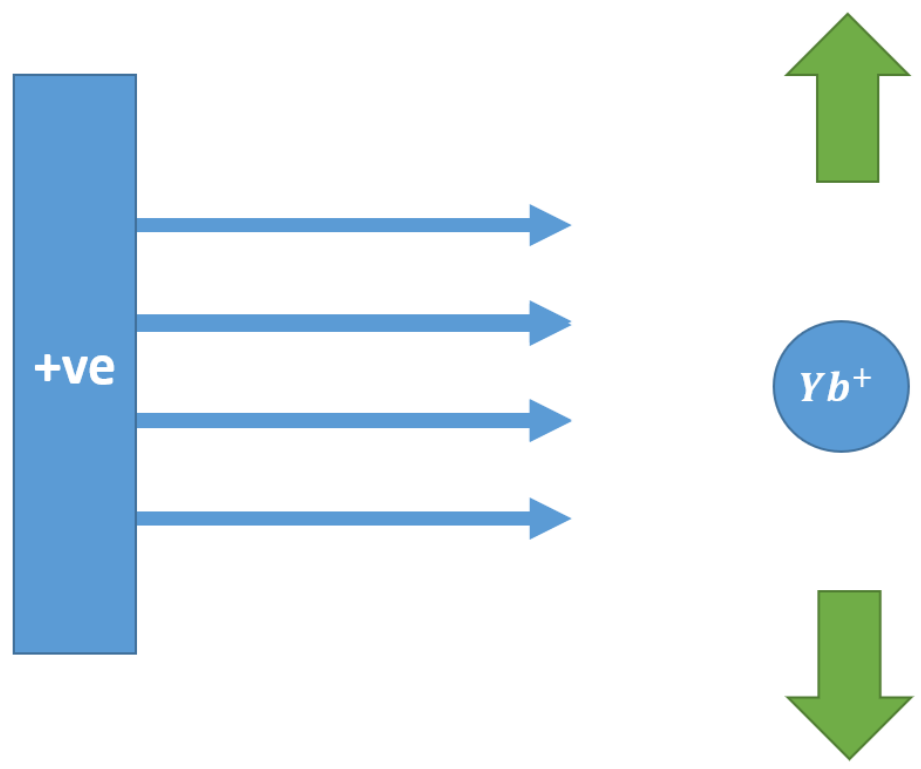

Fig. 28.3: If the Ytterbium ion is put between two positively charged plates, it is "trapped" in the middle horizontally because the middle has the minimum energy laterally. However, it is not stable (not trapped) vertically.

### *28.4.1 Paul Trap*

To confine the ions in 3D space, we also need AC electric fields. One of the design examples is the **linear Paul trap** (Fig. 28.4). A linear Paul trap still uses two positively biased DC electrodes to confine the ions in the lateral direction ($\hat{z}$). Due to mutual repulsion, the Ytterbium ions will form a *lattice of ions*. This lattice and its vibration (phonons) turn out to play a very critical role in trapped ion quantum computers, particularly in 2-qubit gate operations that we will discuss later.

In the orthogonal directions ($\hat{x}$, $\hat{y}$), there are two pairs of electrode rods. The top and bottom electrodes (blue) sandwich the lattice in the $\hat{y}$ direction while the inner and outer electrodes (green) sandwich the lattice in the $\hat{x}$ direction. The rods of the same pair are electrically connected. As an example, the rods may have a diameter of 0.5 mm with a center-to-center spacing of 1.0 mm, and the distance between the electrodes in the $\hat{z}$-direction is 2.6mm [6]. The two pairs are connected through an AC electric field of this form [2]:

$$V_{AC} = V_0 \cos \Omega t, \tag{28.11}$$

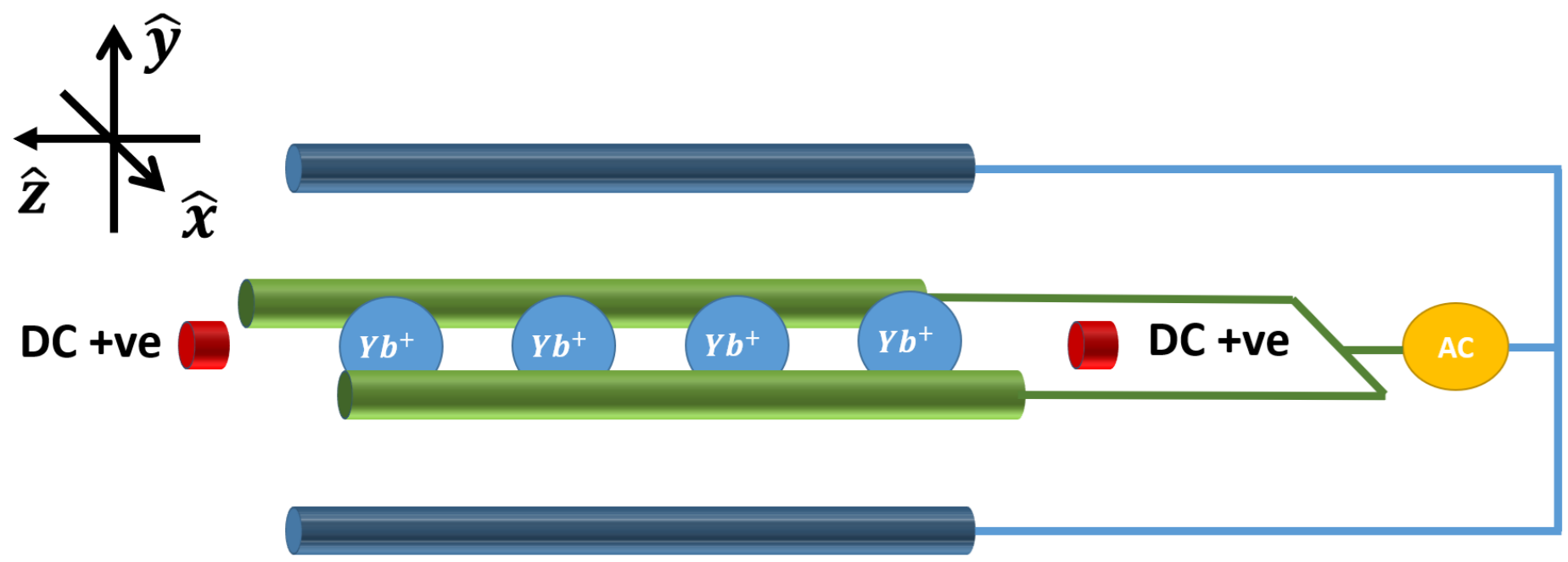

Fig. 28.4: Illustration of a linear Paul trap. The two DC electrodes on the left and right ($\hat{z}$) are biased positively (red). They confine a chain of Ytterbium ions. The top and bottom electrodes (blue, $\hat{y}$) are shorted together. The inner and outer electrodes (green, $\hat{y}$) are also shorted. These two pairs of electrodes are connected through an AC voltage source.

One can solve for the electric potential distribution and thus the equation of motion of the ions. The mathematics is pretty involved, and readers may refer to Chapter 12 in [2]. Here, we will give a brief discussion and highlight the critical aspects.

Firstly, we want to understand intuitively how the ions are held in space. In $\hat{z}$, we agree that the ion will be confined between the positive electrodes due to the DC bias, like in Fig. 28.3. In the $\hat{x}$-$\hat{y}$ plane, we can understand it in this way. If we engineer the system so that a *stable* condition is achieved (i.e. to have the right DC voltage, $V_{DC}$, right $V_{AC}$, right $\Omega$, right ion mass and charge, and right electrode distances), the electric field will change in a way such that it will pushes the ion back to the center whenever it tries to leave.

This is illustrated in Fig. 28.5. Since the orthogonal electrode rod pairs are biased through the AC voltage in Eq. 28.11, they will create a varying electric field pattern that sometimes the ion will be pushed towards the center laterally ($\hat{x}$) and sometimes vertically ($\hat{y}$). The ion will go through a complicated motion, but it will be in resonance with the AC field so that the ion will never be able to leave the center. This is because whenever the ion tries to move in the $\hat{x}$ direction, the electrode pair in the $\hat{x}$ direction is biased positively so that it will push the ion back to the center. When the ion is moving in the $\hat{y}$ direction, the electrode pair in the $\hat{y}$ direction is positively biased and will push the ion back to the center.

Let us further understand how the ion moves under this setup. For example, the equation of motion of an ion in the $\hat{x}$ direction is given by

$$x = x_0 \cos\left(\frac{eV_0}{\Omega^2 M r_o^2}\frac{\Omega t}{\sqrt{2}} + \theta_0\right)\left\{1 + \frac{eV_0}{\Omega^2 M r_o^2}\cos(\Omega t)\right\}, \tag{28.12}$$

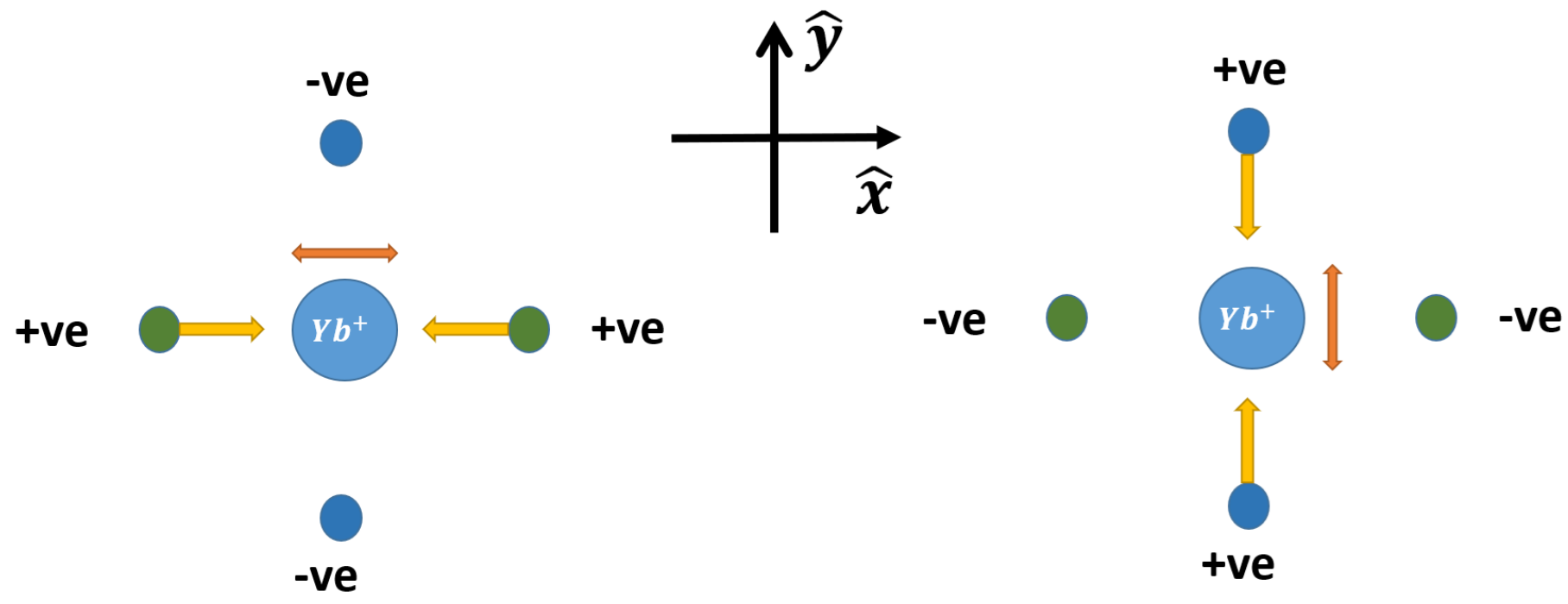

Fig. 28.5: Operations of linear Paul trap. It is designed in a way such that when the ion tries to move in the $\hat{x}$ direction, the electrode pair in the $\hat{x}$ direction is biased positively so that it will push the ion back to the center. Right: similarly, when the ion is moving in the $\hat{y}$ direction, the electrode pair in the $\hat{y}$ direction is positively biased and will push the ion back to the center.

where $M$ is the mass of the ion, $r_o$ is half of the distance between the rods in each pair, $\theta_0$ is the initial phase, and $x_0$ is the amplitude. *Note that this equation is only true under certain assumptions and when a stable condition is achieved.*

Fig. 28.6 shows the $x$ position of a trapped ion under an 8MHz AC signal. A setup to achieve a stable condition was used. We see that it has two features. Firstly, the ion oscillates in $\hat{x}$-direction with an angular frequency equals to $\frac{eV_0}{\Omega^2 M r_o^2}\frac{\Omega t}{\sqrt{2}}$, which has a period of about $4.42\mu$s. This is called the **secular motion**.

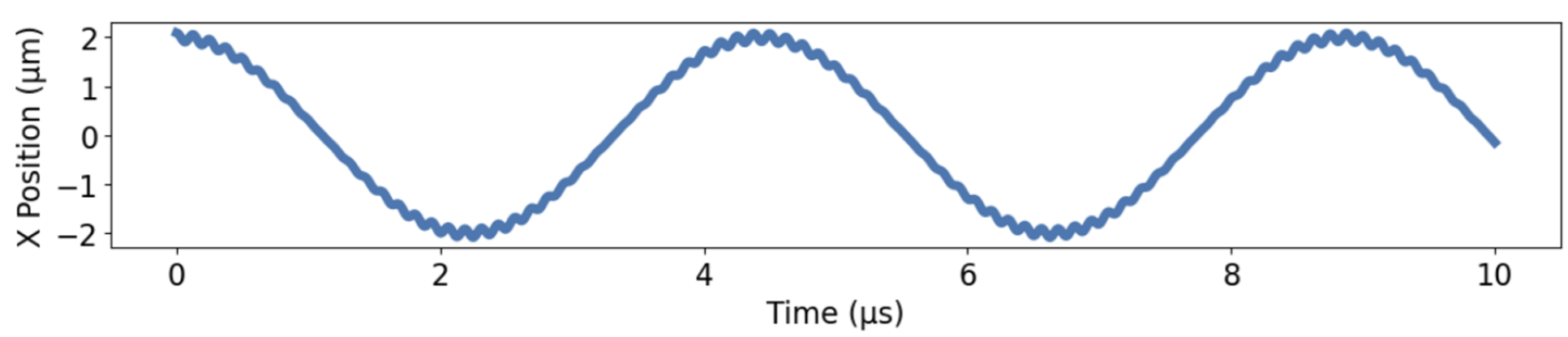

Fig. 28.6: Ion position in the $\hat{x}$ direction as a function of time under an AC field with $\Omega = 2\pi \times 8$MHz

There is another faster oscillation at an angular frequency of $\Omega$ due to the second term in Eq. (28.12). It has a much smaller amplitude, and it is superimposed on the secular motion. This is called the **micromotion**.

Similar motion occurs in the y-direction. If there is a phase difference between the x and y equations of motion, we expect the ion to perform a circular motion (if the phase difference is $\pi/2$) in the $x$-$y$ plane with micromotions as shown in Fig. 28.7.

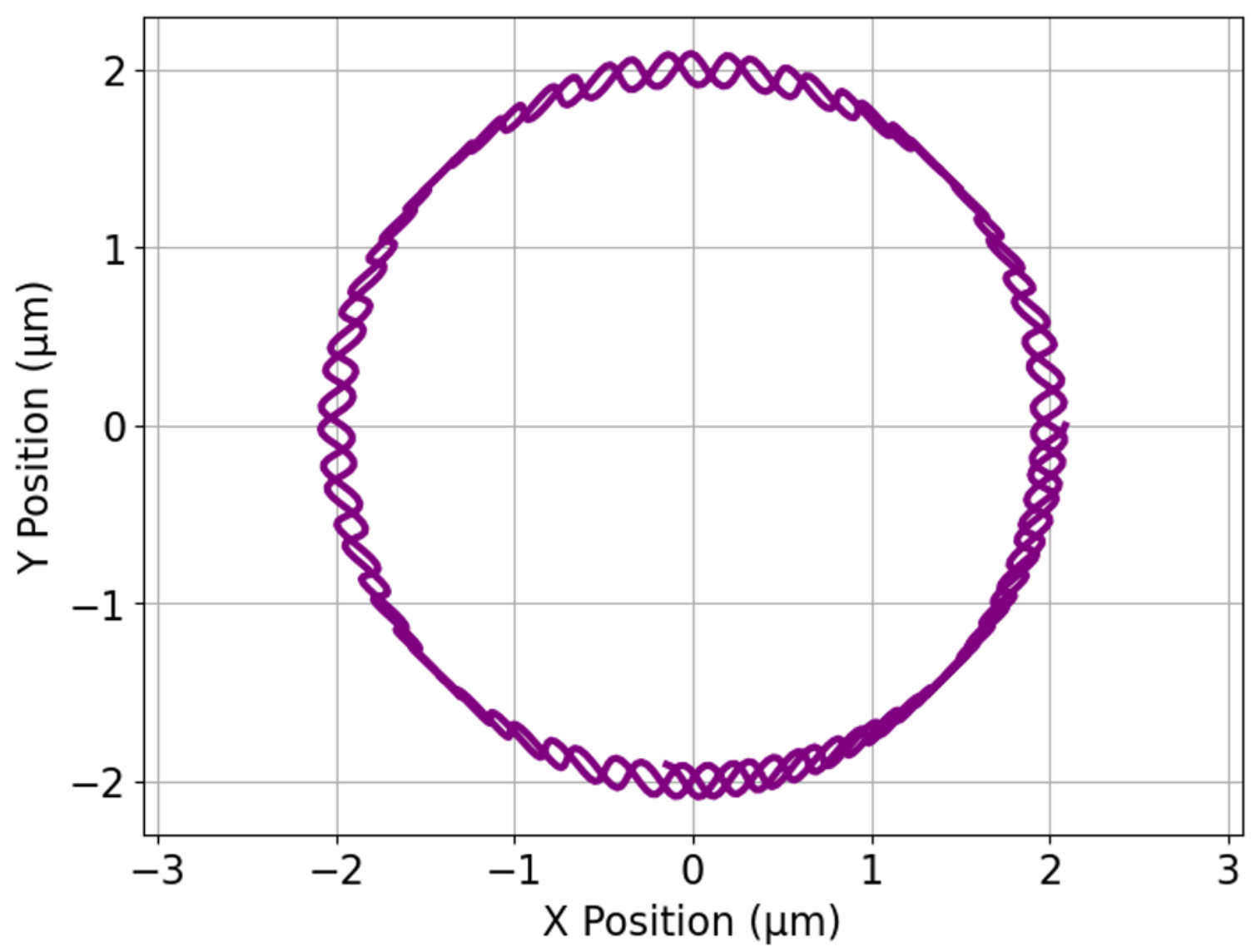

Fig. 28.7: Ion tragetary on the $x$-$y$ plane under an AC field with $\Omega = 2\pi \times 8$MHz after a few period of the secular motion

Finally, I would like to highlight a few aspects. Firstly, the ion also has a simple harmonic motion (secular motion) in the axial direction ($\hat{z}$ direction). This is due to the potential well created by the DC electrodes. However, it does not have a micromotion as there is no AC field. Secondly, it is important to minimize micromotion. Or we should remind ourselves of the micromotion when we consider a trapped ion system. One of the many reasons is that it causes a Doppler shift during its interaction with laser pulses. It will be clear in the next two chapters that the frequencies of the laser pulses are important for qubit initialization, readout, and manipulation (quantum gate). Excessive micromotion will reduce fidelity and increase errors.

With proper engineering, the ions can be stored for several days in the traps [6].

## 28.5 Summary

In this Chapter, we show that due to hyperfine structure, the ground state of a Ytterbium ion is further split into two levels, namely, $^2S_{1/2},\ F = 0$ and $^2S_{1/2},\ F = 1$. We can use them to encode the $|0\rangle$ and $|1\rangle$ states. Hyperfine structure is due to the interaction between nuclear spin and the magnetic field generated by the atomic electrons. It is much smaller than the fine structure. There are also other fine structure and hyperfine structure levels that are critical for trapped ion qubit initialization, readout, and laser cooling, which will be discussed in the following two chapters. We also discussed how to trap ions in a vacuum. While it is forbidden by Physics to trap ions using a pure DC electric field, one may trap them using a combination of DC and AC electric fields. One of the schemes is the famous linear Paul trap. Now, we have shown that the Ytterbium ion has a well-defined two-level system and a long decoherence time. We still need to show that we can perform initialization, readout, one-qubit gate, and entanglement gate manipulations to have it as a suitable quantum computing architecture. These will be covered in the following chapters.

## Problems

**28.1. Micromotion.**
Implement Eq. (28.12) using Python to reproduce Fig. 28.6.

**28.2. Circular motion.**
Now also implement the equation for the *y*-direction. Try various phase differences to reproduce Fig. 28.7.

# Chapter 29
# Trapped Ion Qubit - Qubit Readout, Initialization, and One-Qubit Gate

## 29.1 Introduction

In the previous chapter, we showed that the Ytterbium-127 ion can use its hyperfine structure ($^2S_{1/2},\ F = 0$ and $^2S_{1/2},\ F = 1$) as two well-defined levels to encode a qubit. It also has a long decoherence time. So two of the Divincenzo's criteria are fulfilled. Now, we will discuss how to fulfill or partially fulfill the remaining three Divincenzo's criteria. Firstly, we will discuss how to perform a measurement of the qubit state using a fluorescence technique by illuminating the ions with a laser beam of the same wavelength used in ionization. Then we will study how to use the same laser beam to perform initialization. Finally, we will study how to perform an arbitrary one-qubit gate operation. We will discuss the two-qubit entanglement gate in the next chapter.

### *29.1.1 Learning Outcomes*

Understand how to perform measurement (readout), initialization, and one-qubit gate operation in a trapped ion; appreciate why the same laser beam is used for initialization, readout, and ionization; be able to relate the mathematics of trapped ion one-qubit gate and that of electron spin qubit.

### *29.1.2 Teaching Videos*

- `https://youtu.be/mymwdbBBgAQ`

## 29.2 Qubit State Readout

Qubit readout in a trapped Ytterbium ion is achieved by applying a laser beam with $\lambda = 369.53$ nm to the ion. The energy of the photon corresponds to the energy difference between states ${}^2S_{1/2}$, $F = 1$ (used to encode $|1\rangle$) and ${}^2P_{1/2}$, $F = 0$ (Fig. 29.1). Readers are reminded to revisit Fig. 28.2 to review the fine structure and hyperfine structure of a Ytterbium ion. We should also note that the wavelength is the same as the laser beam used for the second step in the two-color ionization process (Fig. 27.4). This reduces the number of laser sources and simplifies the system.

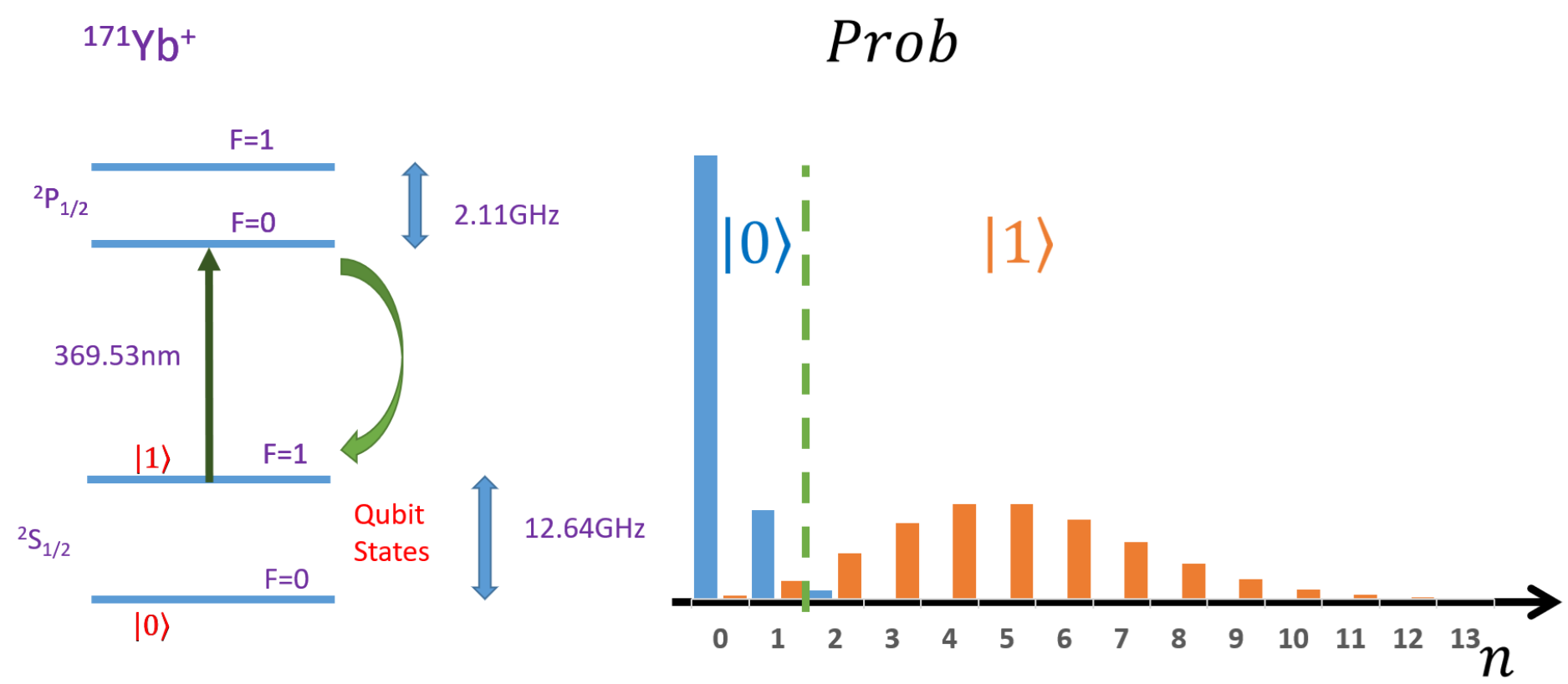

Fig. 29.1: Left: Energy diagram showing the readout process. If the qubit is in state $|1\rangle$, the laser beam will be absorbed and re-emitted. This process is just light scattering. The re-emitted photon (or scattered light) can be detected. Note that the electron will go back to $|1\rangle$ after photon re-emission and will be ready to absorb another photon. If the qubit is in state $|0\rangle$, the laser beam will just go through without being absorbed (i.e., no scattering). Right: The probability of detecting different numbers of scattered photons ($n$) when the qubit is in state $|0\rangle$ or $|1\rangle$. One will calibrate the system to decide a threshold (green dashed line) below (above) which the state is classified as $|0\rangle$ ($|1\rangle$) to minimize the readout error.

If the qubit is in state $|1\rangle$ (i.e., ${}^2S_{1/2}$, $F = 1$), the atom will be *resonantly* excited to ${}^2P_{1/2}$, $F = 0$ by absorbing a photon. After awhile, it will decay *spontaneously* back to ${}^2S_{1/2}$, $F = 1$. In a **spontaneous emission** process, photons are emitted in different directions. The whole process is just the **scattering** of photons. Here, we want to remind ourselves that we are not supposed to be able to see a laser beam if it is not directed at our eyes when there is no scattering. In movies, we usually see how a laser beam propagates in a vacuum. This is impossible and wrong when there is no scattering. The reason we see a beam is due to scattering. The scattered photons will be captured by a camera or a photon-counting photomultiplier tube (PMT) [6].

Another key concept here is that, after the re-emission of a photon, the atom will go back to $^2S_{1/2},\ F = 1$, which will allow it to absorb another photon and go through the scattering process again. This is called the **cycling process** (Fig. 29.2a). Therefore, multiple photons will be scattered when a laser beam is shined on a trapped ion at $|1\rangle$.

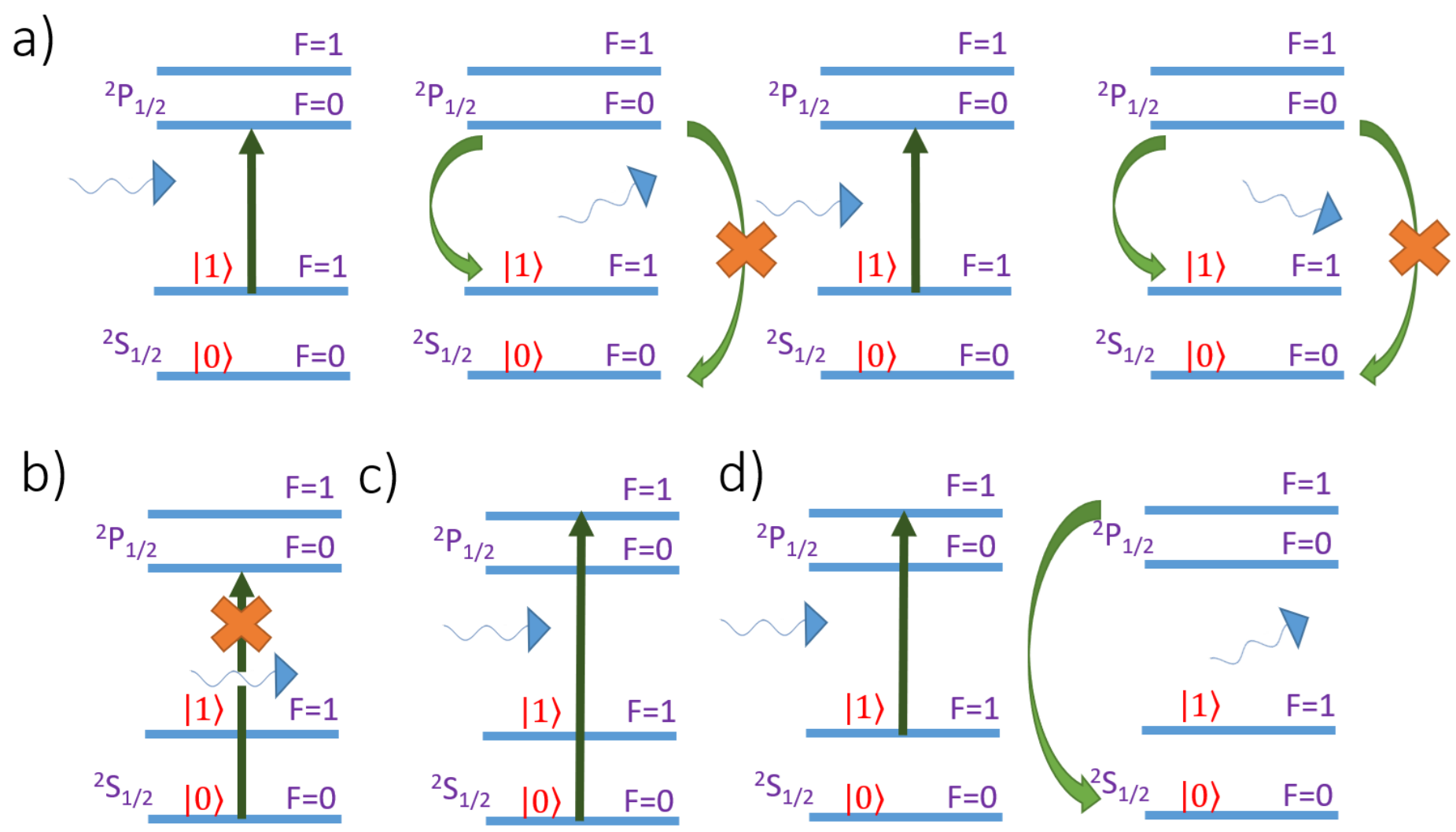

Fig. 29.2: a) shows the cycling process of fluorescence detection in Ytterbium ion when the state is $|1\rangle$. Due to the selection rule, it cannot decay to $^2S_{1/2},\ F = 0$ and thus the cycling process can be maintained. b) and c) show that due to the selection rule, the ion cannot absorb a photon to be excited from $^2S_{1/2},\ F = 0$ to $^2P_{1/2},\ F = 0$ but can be excited to $^2P_{1/2},\ F = 1$. d) shows that $|1\rangle$ can be inadvertently excited to $^2P_{1/2},\ F = 1$ after which it may decay to $^2S_{1/2},\ F = 0$ and thus exists the cycling process.

If the qubit state is at $|0\rangle$ (i.e., $^2S_{1/2},\ F = 0$), the photon does not have enough energy to excite the atom. As shown in Fig. 28.2, it is 12.64 GHz shy (or **detuned**). Therefore, the photon will not be absorbed, and thus there will be no reemission through spontaneous decay. This is the same as a laser beam traveling through a vacuum without scattering, and we would not be able to detect it if it were not directed to our eyes. More importantly, there should be *no* transition between $^2S_{1/2},\ F = 0$ and $^2P_{1/2},\ F = 0$ even if the photon has a high enough energy (short wavelength) because both states have $F = 0$ (Fig. 29.2b). This is due to the **selection rule** in a dipole-induced transition (which is due to the physics when the atom dipole (separation of positive and negative charge centers) interacts with the electric field of the photon). One of the rules states that there shall be no transition from a state with $F = 0$ to another state with $F = 0$. We will not discuss the details, and readers can see Table 5.1 in [2] for the selection rules of $J$, which are similar for $F$. However, it is allowed to be excited to $^2P_{1/2},\ F = 1$, which is 2.11 GHz further away (Figs. 28.2 and 29.2c).

So there is no scattering, and nothing will be captured by the camera or PMT when the state is $|0\rangle$.

As a result, if photons are detected (bright), the qubit is in $|1\rangle$ state. If it is dark, the qubit is in the $|0\rangle$ state. This is called the **fluorescence detection**.

*Example 29.1.* How accurate does the laser beam wavelength need to be in order not to excite a Ytterbium ion at state $|0\rangle$ during the qubit readout process?

We need to make the laser beam narrow enough in spectrum so that it does not have a component that will excite $|0\rangle$. Of course, this also depends on the temperature of the system (including the ions), which can cause broadening of the laser beam and also the Doppler effect (to be discussed in the next chapter).

We will estimate the requirement by comparing the detuned frequency to the frequency between $^2S_{1/2}$ and $^2P_{1/2}$. Based on Example 28.3, the ratio is

$$\frac{(12.64 + 2.11)\text{GHz}}{813\text{THz}} \approx 1.8 \times 10^{-5} \tag{29.1}$$

This is about 18 parts per million (ppm). In reality, it needs to be much lower than that. This example shows us how important it is to have very precise optical and electrical engineering.

■

#### 29.2.0.1 Readout Fidelity

Readout cannot be perfect. Firstly, due to various reasons, even if the qubit is in $|0\rangle$ state, the photons might still be scattered (e.g., by gas molecules as the vacuum is not perfect). As a result, the qubit might appear to be bright occasionally ($n = 1,\ 2$ in Fig. 29.1). Similarly, scattering by the $|1\rangle$ state is a stochastic process. The number of photons to be detected can be very low. Indeed, the scattering process follows **Poisson distribution** [4]. Therefore, the fidelity cannot be 100%. For example, in Fig. 29.1, both states $|0\rangle$ and $|1\rangle$ can result in the detection of $n = 1$ or $n = 2$ scattered photons. We need to calibrate the system to choose a threshold to determine how to classify the state. For example, we may call the state $|0\rangle$ if the number of detected photons is 0 or 1 and call the state $|1\rangle$ if the number of detected photons is 2 or more. This minimizes the error. But we know that if, for example, $n = 1$, while we classify it as $|0\rangle$, there is a possibility that it is actually $|1\rangle$.

Secondly, as shown in Fig. 28.2, the energy spacing between $^2P_{1/2},\ F = 0$ and $^2P_{1/2},\ F = 1$ is only 2.11 GHz. As seen in Example 29.1, this is very small (18 ppm) compared to the laser frequency. It has a higher chance for $|1\rangle$ to be excited *off-resonantly* to $^2P_{1/2},\ F = 1$. And this is a major source of error. Why? This is because if the electron is excited to $^2P_{1/2},\ F = 1$, it might decay to $|0\rangle$ (which is $^2S_{1/2},\ F = 0$) and this will stop the cylcing process in Fig. 29.2a (See Fig. 29.2d). A question to us: why is it not a problem if it is excited to $^2P_{1/2},\ F = 0$ in the cycling process in Fig. 29.2a? Would it not decay to $|0\rangle$? This is again due to the selection rule mentioned that transitions will not occur between two states when they both have $F = 0$.

## 29.3 Qubit Initialization

With the knowledge of trapped ion qubit readout from the previous section, it is very straightforward to design an initialization scheme. Firstly, in order to reuse the same laser equipment, it is desirable to reuse the 369.53 nm laser. Secondly, we can use an "active-reset" like scheme in which we will derive a mechanism such that when the qubit is at $|0\rangle$, there will be no effect and if the qubit is at $|1\rangle$, it will reset the qubit through some processes. This is very similar to the spin qubit initialization scheme described in Fig. 11.5 in [1].

Fig. 29.3 shows the initialization process. The scheme is to shine a laser beam on the ion so that it will excite the ion to a higher energy state if it is at $|1\rangle$, but does nothing if it is at $|0\rangle$. Then we let it decay to $|0\rangle$ spontaneously. We select $^2P_{1/2},\ F = 1$ to be the higher energy state. As a result, we need a laser beam with an energy higher than the 369.53 nm laser by 2.11GHz. This is achieved by passing the laser beam through an electro-optic modulator to increase its frequency before the ion. After the atom is excited to $^2P_{1/2},\ F = 1$, it will decay spontaneously, and it has about $1/3$ of a chance to decay to $|0\rangle$. If it decays back to $|1\rangle$, the process will continue. After a few iterations, most of the $|1\rangle$ will be initialized to $|0\rangle$.

Again, the photons do not have enough energy to excite $|0\rangle$ to $^2P_{1/2},\ F = 1$. Therefore, if the ion is at $|0\rangle$, it will stay at $|0\rangle$.

Why did we not select $^2P_{1/2},\ F = 0$ to be the excited state to avoid using the electro-optic modulator? This is because of the **selection rule** mentioned in the previous section. If $^2P_{1/2},\ F = 0$ is used, it will decay back to $|1\rangle$ only (Fig. 29.2a) and that defeats the purpose of initialization.

*Initialization Fidelity:* We should briefly mention that Fig. 28.2 does not show a complete fine structure and hyperfine structure diagram of the Ytterbium ion. Readers can refer to [4]-[6] for more information. One of the states that matters for initialization is $^2D_{3/2},\ F = 2$. Besides decaying to $|1\rangle$ or $|0\rangle$, $^2P_{1/2},\ F = 1$ also decays to $^2D_{3/2},\ F = 2$, which has a long lifetime and, thus, prevents rapid initialization. Therefore, another laser with $\lambda$ =935.2 nm is needed to pump the atom from $^2D_{3/2},\ F = 2$ to $|0\rangle$.

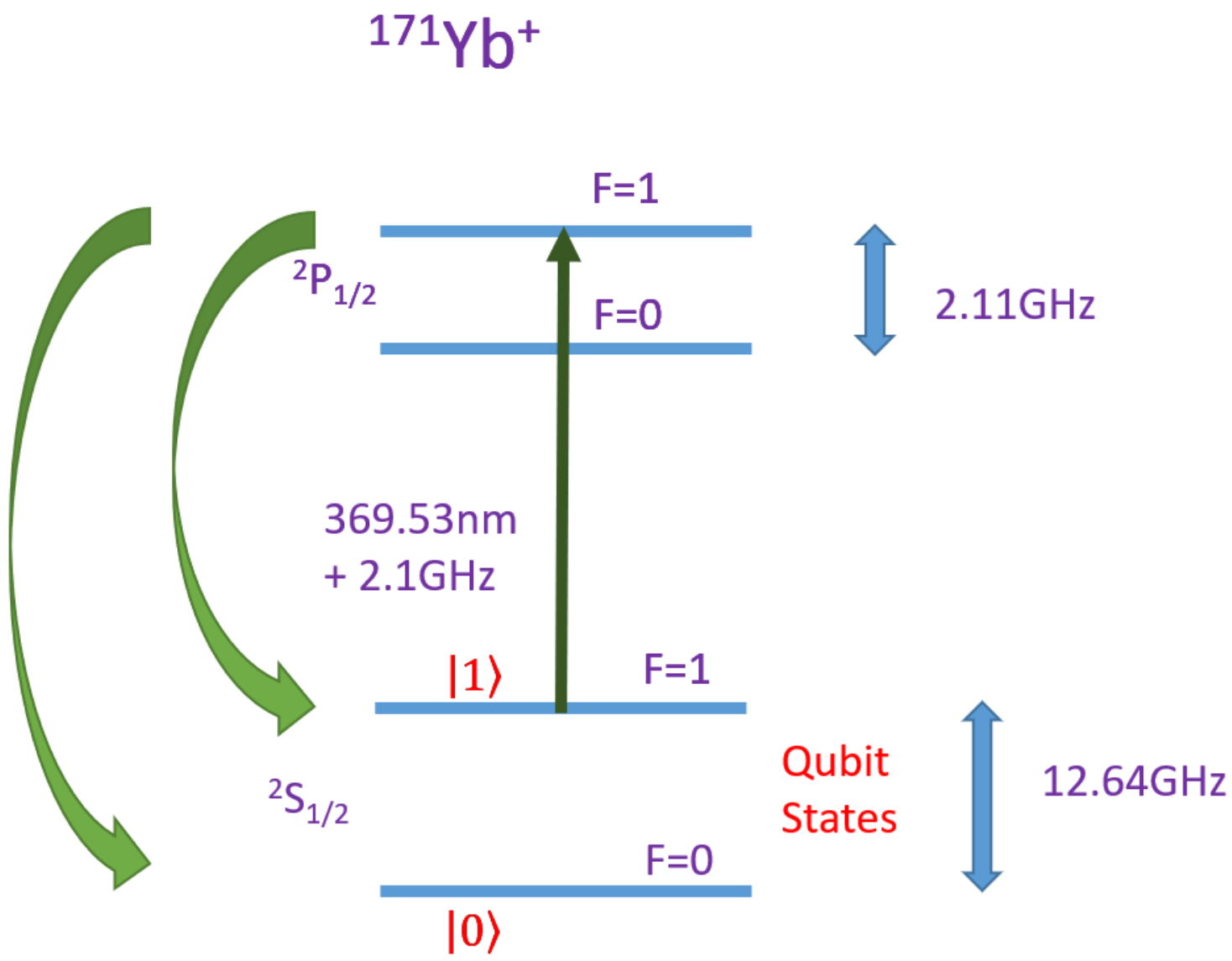

Fig. 29.3: Trapped Ytterbium ion qubit initialization process. The atom may decay to $|1\rangle$ or $|0\rangle$ from the excited state, $^2P_{1/2}$, $F = 1$.

## 29.4 Qubit Frequency and Larmor Precession

Like the spin qubit and the superconducting qubit, the **qubit frequency** of a trapped ion qubit is determined by the energy separation between states $|1\rangle$ and $|0\rangle$. For the Ytterbium ion, it was given in Fig. 28.2 that it has a qubit frequency, $f_L = 12.64$ GHz. In Example 28.2, it is shown that this corresponds to a $52\mu$eV energy separation between the $^2S_{1/2}, F = 0$ and $^2S_{1/2}, F = 1$ states. We should highlight that this energy separation is due to the hyperfine structure as a result of the nuclear magnetic moment, $\vec{\mu}_I$, and the magnetic field created by the electrons, $\vec{B}_e$. Eq. (28.3) is repeated here for convenience:

$$H_{HFS} = -\vec{B}_e \cdot \vec{\mu}_I. \tag{29.2}$$

This has the same form as Eq. (8.1) in [1]. However, it is very important to note that, unlike electron spin, the energy splitting cannot be treated as a result of the magnetic moment parallel or anti-parallel to the magnetic field (see Section 29.4.1 and Fig. 29.4). But since we know the energy difference, we can set the energy at the middle between $^2S_{1/2}$, $F = 0$ and $^2S_{1/2}$, $F = 1$ as the reference (Fig. 29.4). Then the energy of the state $|0\rangle$, $E_0$, and that of $|1\rangle$, $E_1$ are,

$$E_0 = -\hbar\omega_L/2,$$
$$E_1 = \hbar\omega_L/2, \tag{29.3}$$

where $\omega_L$ is the angular **Larmor frequency**. We can write the Hamiltonian directly as (like Eq. (8.4) in [1]),

$$H = \begin{pmatrix} -\hbar\omega_L/2 & 0 \\ 0 & \hbar\omega_L/2 \end{pmatrix}, \tag{29.4}$$

which is consistent with Eqs. (8.4) and (8.11) in [1]. Plugging the Hamiltonian into the Schrödinger equation and solving the equation, we will again observe **Larmor precession** of the qubit on Bloch's sphere. Readers can follow Section 8.3 in [1] to prove this (see Exercise 27.2).

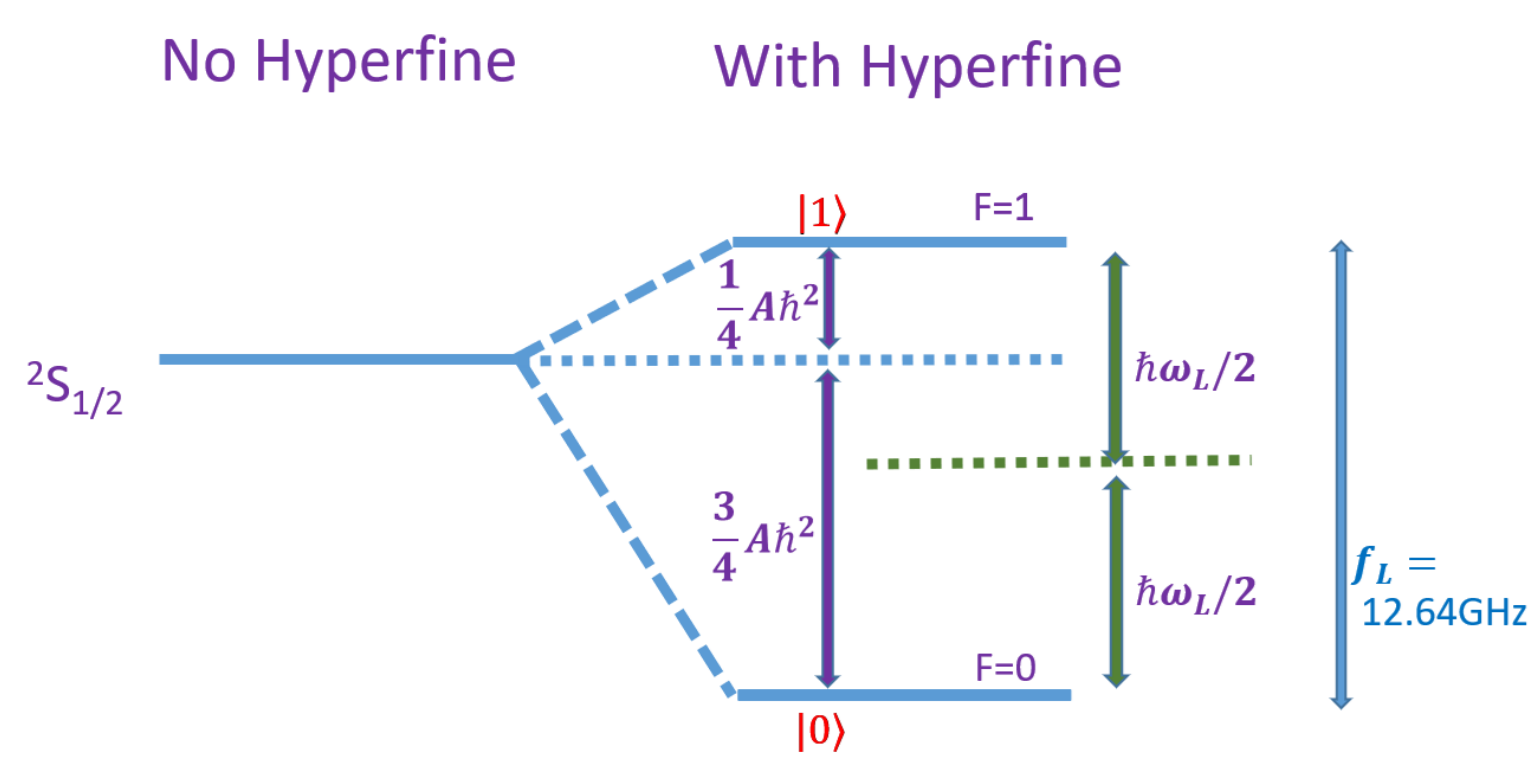

Fig. 29.4: Hyperfine energy splitting of the $^2S_{1/2}$ state

### *29.4.1 Hyperfine Splitting*

In Eq. (28.5), we showed that the hyperfine structure Hamiltonian is given by

$$H_{HFS} = A\vec{I}\cdot\vec{J}. \tag{29.5}$$

Since $\vec{F} = \vec{I} + \vec{J}$, like what we did in Eq. (27.10), we have

$$
\begin{aligned}
\vec{F}^2 &= (\vec{I}+\vec{J})^2, \\
&= (\vec{I}+\vec{J})\cdot(\vec{I}+\vec{J}), \\
&= \vec{I}^2 + 2\vec{I}\cdot\vec{J} + \vec{J}^2, \\
\vec{I}\cdot\vec{J} &= \frac{1}{2}(\vec{F}^2 - \vec{I}^2 - \vec{J}^2).
\end{aligned} \tag{29.6}
$$

As a result,

$$
\begin{aligned}
\langle \vec{I}\cdot\vec{J} \rangle &= \langle \frac{1}{2}(\vec{F}^2 - \vec{I}^2 - \vec{J}^2) \rangle, \\
&= \frac{1}{2}(\langle \vec{F}^2 \rangle - \langle \vec{I}^2 \rangle - \langle \vec{J}^2 \rangle), \\
&= \frac{1}{2}(F(F+1)\hbar^2 - I(I+1)\hbar^2 - J(J+1)\hbar^2), \\
&= \frac{\hbar^2}{2}(F(F+1) - I(I+1) - J(J+1)),
\end{aligned} \tag{29.7}
$$

As discussed in Section 28.3, state ${}^2S_{1/2},\ F=0$ has $I = 1/2$, $J = 1/2$, and $F = 0$. Therefore, the expectation value of its energy *relative to the energy before hyperfine splitting* is,

$$
\begin{aligned}
E_0 &= -\hbar\omega_L/2, \\
&= \frac{A\hbar^2}{2}(F(F+1) - I(I+1) - J(J+1)), \\
&= \frac{A\hbar^2}{2}(0(0+1) - \frac{1}{2}(\frac{1}{2}+1) - \frac{1}{2}(\frac{1}{2}+1)), \\
&= -A\hbar^2\frac{3}{4}.
\end{aligned} \tag{29.8}
$$

Similarly, state ${}^2S_{1/2},\ F=1$ has $I = 1/2$, $J = 1/2$, and $F = 1$ and its energy value *relative to the energy before hyperfine splitting* is,

$$
\begin{aligned}
E_1 &= \hbar\omega_L/2, \\
&= \frac{A\hbar^2}{2}(1(1+1) - \frac{1}{2}(\frac{1}{2}+1) - \frac{1}{2}(\frac{1}{2}+1)), \\
&= A\hbar^2\frac{1}{4}.
\end{aligned} \tag{29.9}
$$

Fig. 29.4 shows the energy splitting due to the hyperfine structure.

## 29.5 One-Qubit Gate

In Section 5.3 in [1], we showed that any one-qubit gate can be decomposed into three rotations about any two of the three axes ($\hat{x}$, $\hat{y}$, and $\hat{z}$). Therefore, we only need to show that we can perform Rabi oscillation to show that it is possible to implement any one-qubit gate.

One possible way we can think of to achieve Rabi oscillation is to use **electric dipole transitions** between the qubit states [2]. This is achieved by letting the atom interact with a laser. The oscillating electric field in the laser interacts with the dipole moment of the atom, resulting in a transition between the states. This means that the qubit energy needs to be in the THz range. Therefore, it cannot be applied to the Ytterbium ion (Fig. 29.4). Secondly, even if the energy matches, to utilize electric dipole transition, $|0\rangle$ and $|1\rangle$ need to have different parities (which will be discussed in Section 29.6 and Eq. (29.29)) and this is not the case for the $|0\rangle$ and $|1\rangle$ states of Ytterbium qubits. Thirdly, if the excited state can be connected to the ground state through an electric dipole transition, the excited state can go through spontaneous emission easily. This is because the excited state can be coupled to the vacuum fluctuation electric field through the electric dipole moment. The qubit will then have a very low decoherence time. Therefore, *any qubit relying on electric dipole transitions for its manipulation has a short decoherence time and thus is not a good qubit*. However, there are some useful mathematics and physics concepts to learn when considering the interaction of atoms with radiation. Thus, we will still study the electric dipole transition of atoms or ions under radiation in Section 29.6.

To reduce spontaneous emission and thus maintain a long decoherence time, we can use a qubit based on an **electric quadrupole transition** [8]. For example, $^{40}Ca^{+}$ can be used as the host of a qubit with $S_{1/2}$ and $D_{5/2}$ as the $|0\rangle$ and $|1\rangle$, respectively. This qubit has a large energy separation (729 nm) and can be manipulated by a laser pulse with the corresponding frequency directly.

Another way to reduce spontaneous emission and, thus, to avoid low decoherence time, is to have a qubit relying on magnetic dipole transitions, like the spin qubit case in Chapter 8 in [1]. While vacuum fluctuation also generates magnetic fields, the magnetic dipole moment interaction with the vacuum fluctuation is generally not the limiting factor for decoherence time. One such example is to use $^{43}Ca^{+}$ ion [9][10]. It uses its hyperfine structure $4^2S_{1/2}, F = 4, M = 1$ and $4^2S_{1/2}, F = 3, M = 1$ as $|0\rangle$ and $|1\rangle$, respectively. The 4 in $4^2S_{1/2}$ refers to the electron in $n = 4$ orbital. Compared to the splitting of Ytterbium ion in Fig. 29.4, we see that the magnetic quantum number associated with $F$, i.e., $M$, is also explicitly mentioned. Nonzero $M$ is chosen so that by applying a constant external magnetic field, one can lift the degeneracy and also fine-tune the qubit frequency. The qubit frequency is 3.1 GHz [9]. This is again too small to interact with lasers (in THz). But it interacts well with the magnetic field in microwave pulses. Transition is obtained when the magnetic dipole moment of the ion interacts with the oscillating magnetic field, like the case of a spin qubit. Readers may refer to [10] to see the details on how to generate microwave pulses on a chip to interact with a trapped ion. This is the so-called **microwave-driven qubit**. It should also be noted that the wavelength of a microwave pulse is much longer than the distance between two trapped ions. Special engineering is required to ensure the selectivity so that a microwave pulse can target a specific qubit.

If a laser is needed but the qubit frequency is in the microwave regime (GHz), **stimulated Raman transition** is used. This is also called the **two-photon process**. The Ytterbium trapped ion qubit, which we have been discussing so far, often uses this process for transition [11]. One of the major advantages of using a laser is that it can address individual qubits easily due to its short wavelength compared to microwave pulses. In a stimulated Raman process, two *coherent* laser pulses of frequency $\omega_1$ and $\omega_2$ are turned on and off at the same time and sent to the trapped ion. The difference in their frequency equals the qubit frequency, $\omega_1 - \omega_2 = \omega_q$. Therefore, it can manipulate the qubit even though the lasers are in the THz regime as long as their difference is in the GHz regime. To have this process successful, there should be a higher energy state in the ion energy system. However, the energy difference between the higher energy state and the qubit states should be larger than $\hbar\omega_1$ and $\hbar\omega_2$ so that it would not be populated during this process by absorbing the photons. A stimulated Raman transition process can be thought of in the following way. The ion will *virtually* absorb one of the photons and emit a new photon stimulated by the second photon. For example, if the ion is in $|0\rangle$ state, it will virtually absorb the photon of $\hbar\omega_1$ and emit a new photon of $\hbar\omega_2$ and get excited to $|1\rangle$ to conserve energy.

## 29.6 Interaction of Atoms or Ions with Radiation

In the previous section, *we emphasize that qubits relying on electric dipole transitions are not good qubits* due to their short decoherence time. However, for pedagogical reasons for engineers, here we discuss the interaction of ions with the radiation field through the electric dipole. If we ignore the fact that it has a short decoherence time, we can appreciate how it can be used to build a one-qubit gate. The Rabi oscillation can be achieved by interacting the trapped ion with an oscillating electric field. Since an electromagnetic (EM) wave, i.e., light, has an oscillating electric field (in addition to an oscillating magnetic field), it is natural to attempt to let a trapped ion interact with an EM wave (i.e., photon) to achieve a qubit operation. Of course, as expected, just like in the case of the spin qubit and the superconducting qubit, the EM wave needs to have the right frequency, right duration, right amplitude, and be applied at the right time as discussed in Section 1.2 in [1]. It turns out that the treatment of trapped ion-photon interaction is very similar to the treatment of the interaction between spin and an oscillating magnetic field in Chapter 9 in [1]. Therefore, we will deliberately reuse some of the mathematical derivations below. Before that, we want to introduce **electric dipole moment** and its interaction with the electric field for readers who are not familiar with the concept.

### *29.6.1 Electric Dipole Moment*

When there is a pair of positive and negative point charges, $+q$ and $-q$, separated by a distance of $r$ (as shown in Fig. 29.5), an electric dipole moment, $\vec{d}$, is formed, with

$$\vec{d} = q\vec{r}, \tag{29.10}$$

where $\vec{r}$ is the displacement from the *negative charge to the positive charge*. Note that, on the other hand, an electric field points from positive charges to negative charges. Readers need to be aware of the difference.

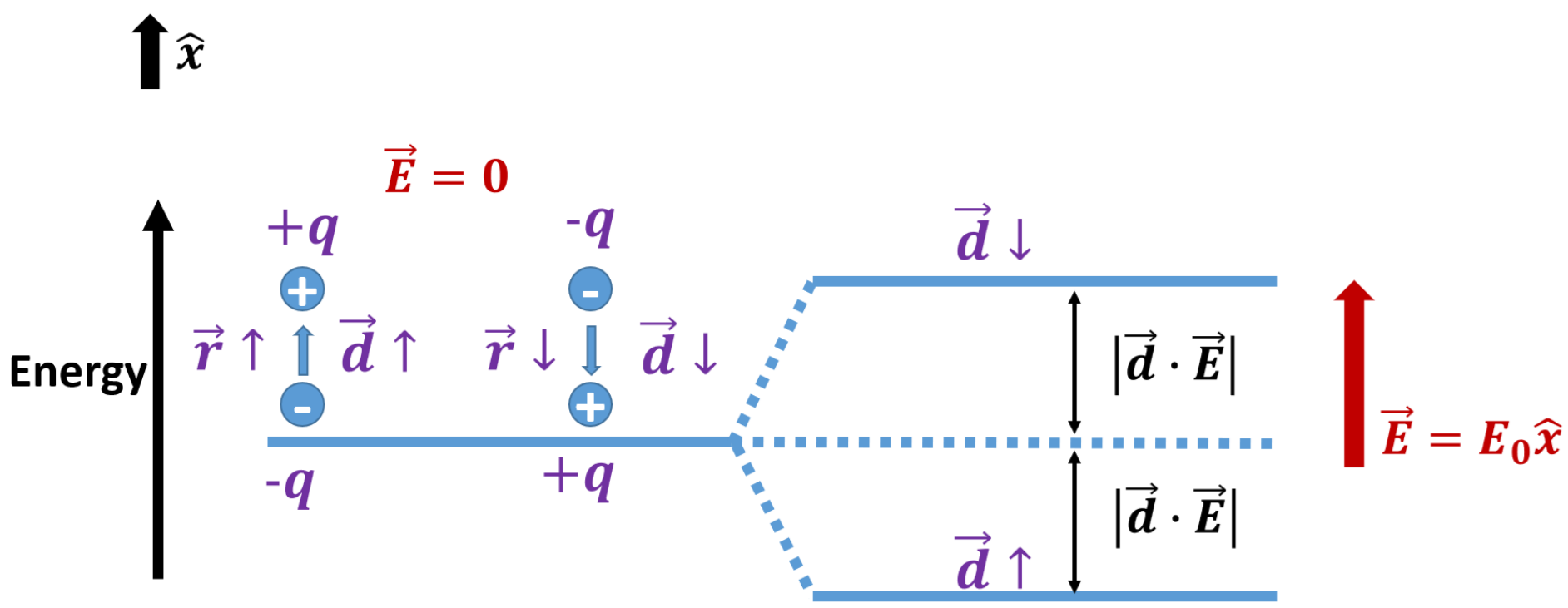

Fig. 29.5: Energy of electric dipole moment under a zero (left) or a finite external electric field (right)

An electric dipole moment will interact with an external electric field (like a magnetic dipole moment interacting with an external magnetic field in Section 7.4 in [1]). This is not difficult to understand. Referring to Fig. 29.5, if the electric field is pointing upwards, the system has a lower energy if the negative charge is closer to the source of the electric field and the positive charge is further away, as the potential energy is lower. This is the case when the dipole moment is pointing upwards. If it is the opposite, we expect a higher energy. When the dipole moment is pointing downwards, eventually, the positive charge, which is closer to the source of the electric field, will be repelled upwards, and the negative charge will be attracted to it to attain a lower potential. Therefore, the energy of the system is lower (higher) if the dipole moment is parallel (anti-parallel) with the external electric field. The energy is given by

$$H = -\vec{E}\cdot\vec{d}, \tag{29.11}$$

which has the same form as Eq. (7.10) in [1] for magnetic moment and magnetic field interaction, $H = -\vec{B} \cdot \vec{\mu}$.

There is a more general equation to calculate the electric dipole moment of a system. For point charges, it is given by,

$$\vec{d} = \sum_{i=1}^{N} q_i \vec{r}_i, \tag{29.12}$$

where there are $N$ point charges with the $i$-th charge having a charge of $q_i$ at location $\vec{r}_i$. Here, the sum of the charges does not need to be zero (neutral), and the origin is conveniently picked to be the center of mass. We have skipped a lot of important discussions regarding the properties and calculations of the electric dipole moment. Readers can refer to elementary EM textbooks for details. What we have reviewed is enough for the purpose of this chapter.

In an atom or an ion, we may treat the nucleus as a point positive charge and the electrons surrounding it as negative charges. We can use Eq. (29.12) to find its electric dipole moment. If we pick the center of mass as the origin, the nucleus, which is composed of many nucleons, is very close to the center of mass as it has a much larger mass than the electrons (see Eq. (28.2)). Therefore, $\vec{r}$ of the nucleus is approximately zero and can be ignored in Eq. (29.12). If we only consider the dipole moment due to one electron, then the dipole moment of interest is

$$\vec{d} = -e\vec{r}, \tag{29.13}$$

where $e$ is the positive elementary charge and $\vec{r}$ is the displacement of the electron from the center of mass. We might also write $\vec{r} = x\hat{x} + y\hat{y} + z\hat{z}$, where $x$, $y$, and $z$ are the $x$, $y$, and $z$ components of $\vec{r}$, respectively.

### *29.6.2 Experimental Setup and Hamiltonian*

In a spin qubit, the well-defined two-level system is achieved by applying a DC magnetic field. After that, a small oscillating magnetic field is added to perform Rabi oscillation (Fig. 9.1 in [1]). Rabi oscillation is due to the interaction between the oscillating magnetic field and the magnetic moment. We do not need a DC electric field in the trapped ion qubit because the two-level system is created already due to the hyperfine structure (see Sections 28.3 and 29.4). Therefore, the electric dipole moment and electric field interaction is **not** used to create the two energy levels in the qubit. We still need a small oscillating electric field to interact with the ion's electric dipole moment to perform Rabi oscillation. Without the DC field, Fig. 9.1 in [1] can be redrawn as Fig. 29.6 for a trapped ion qubit under an oscillating electric field. What we discussed in Section 29.6.1 is just to help us understand the interaction between the oscillating electric field and the dipole moment. Note that the oscillating electric field is due to the EM wave shining on the ion. The interaction Hamiltonian is,

$$\begin{aligned} H_I &= -\vec{d}\cdot\vec{E}_1, \\ &= e\vec{r}\cdot(E_1\cos(\omega_1 t)\hat{x}), \\ &= eE_1\cos(\omega_1 t)\vec{r}\cdot\hat{x}, \\ &= eE_1 x\cos(\omega_1 t), \end{aligned} \tag{29.14}$$

as $\vec{r}\cdot\hat{x} = x$.

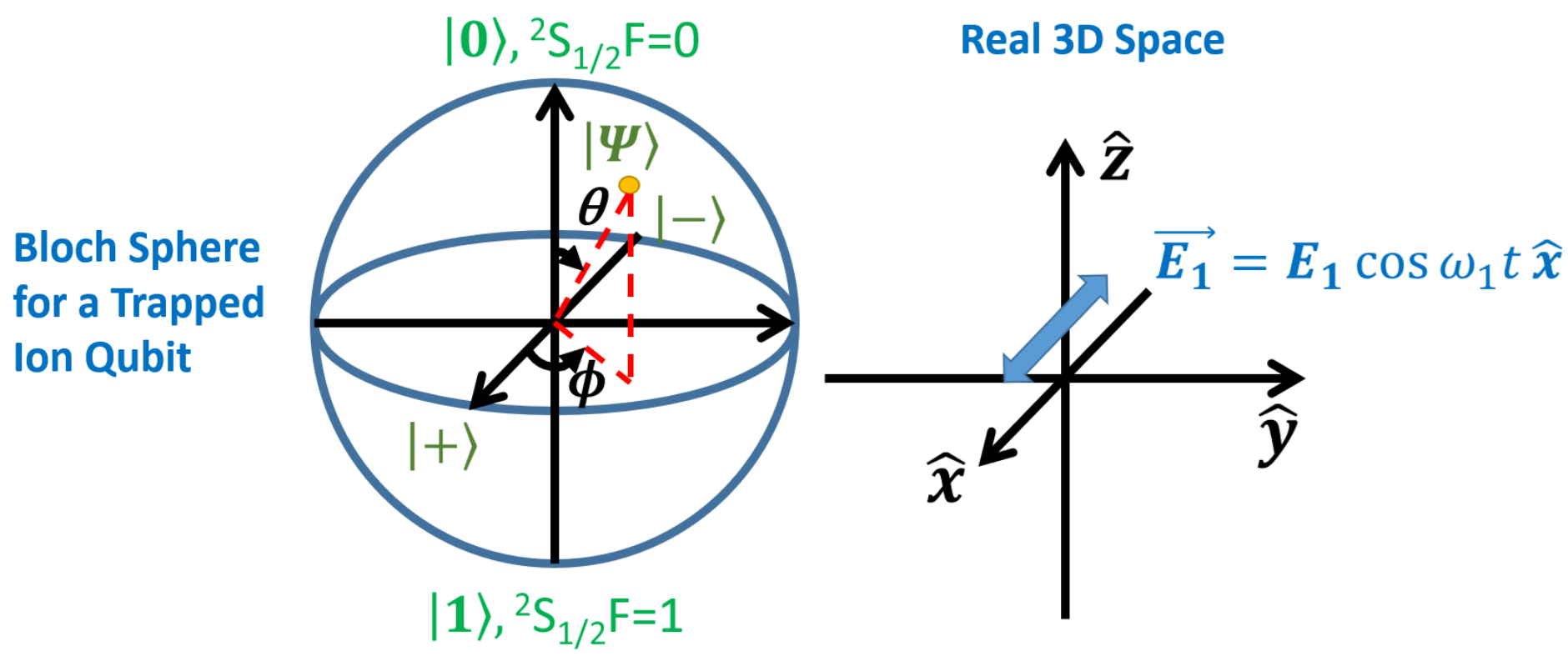

Fig. 29.6: The Bloch sphere representation of a trapped ion qubit (left) and the real 3D space coordinate (right) in which the direction of the oscillating electric field due to an EM wave is shown

*Here, we made a few important but reasonable assumptions*. Firstly, for convenience and simplicity, we assume that the light is a plane wave and it is oscillating in the $\hat{x}$ direction with an angular frequency of $\omega_1$. So, the light is propagating in the $\hat{y}$ or $\hat{z}$ direction as an EM wave is a transverse wave. Secondly, we assume the wavelength is much larger than an atom. We know that the oscillation amplitude of an EM wave is sinusoidal in space at a given time. By assuming that the atom is much smaller than the wavelength, we can treat the atom as being under an oscillating field with a constant amplitude, $E_1$. Thirdly, like what we assumed in the spin qubit case, we assume $E_1$ is small enough such that the interaction Hamiltonian is much smaller than the hyperfine structure Hamiltonian (Eq. (29.2)).

### 29.6.3 Setup of the Schrödinger Equation

The Schrödinger equation corresponding to this system is given by

$$i\hbar\frac{\partial\,|\psi\rangle}{\partial t} = (\boldsymbol{H_0} + \boldsymbol{H_I})\,|\psi\rangle\,, \tag{29.15}$$

with $\boldsymbol{H_0}$ and $\boldsymbol{H_I}$ defined in Eqs. (29.2) and (29.14). Since we assume $E_1$ to be small enough, we can treat $\boldsymbol{H_I}$ as a perturbation to $\boldsymbol{H_0}$. Note that $\boldsymbol{H_0}$ results in the two-level system for the trapped ion with its eigenstates, $|0\rangle$ and $|1\rangle$, in Fig. 29.4. Therefore, we will work on the basis formed by $|0\rangle$ and $|1\rangle$ from now on, and matrices $\boldsymbol{H_0}$ and $\boldsymbol{H_I}$ will be expressed in these two basis states.

It is known that if a perturbation is added to a system, the state of the new system, $\left|\Psi_{perturbed}\right\rangle$, is a linear combination of the eigenstates ($|0\rangle$, $|1\rangle$, $\cdots$ ) of the unperturbed system weighted by the complex exponential of the corresponding normalized eigenenergies ($-E_0t/\hbar$, $-E_1t/\hbar$, $\cdots$ ). Interested readers can refer to time-dependent perturbation in any quantum mechanics textbook, such as [12] for more details. That is

$$\left|\Psi_{perturbed}\right\rangle = c_0(t)e^{-iE_0t/\hbar}\,|0\rangle + c_1(t)e^{-iE_1t/\hbar}\,|1\rangle + \cdots, \tag{29.16}$$

where $c_0(t)$, $c_1(t)$, $\cdots$ are complex coefficients and can be time-dependent. For clarity, we will write them as $c_0$ and $c_1$, but we need to bear in mind that they can be time-dependent.

In the trapped ion case, using the definition in Fig. 29.4 with the reference energy level at the middle of the two levels, the state of the system with the perturbing oscillating electric field can be written as

$$\begin{aligned}\left|\Psi_{perturbed}\right\rangle &= c_0e^{-i\frac{E_0t}{\hbar}}\,|0\rangle + c_1e^{-i\frac{E_1t}{\hbar}}\,|1\rangle\,,\\ &= c_0e^{-i\frac{-\hbar\omega_Lt}{2\hbar}}\,|0\rangle + c_1e^{-i\frac{\hbar\omega_Lt}{2\hbar}}\,|1\rangle\,,\\ &= c_0e^{i\frac{\omega_L}{2}t}\,|0\rangle + c_1e^{-i\frac{\omega_L}{2}t}\,|1\rangle\,.\end{aligned} \tag{29.17}$$

Note that $\left|\Psi_{perturbed}\right\rangle$ is just the $|\Psi\rangle$ in Eq. (29.15). We will now only use $|\Psi\rangle$.

We should also remember that the wavefunction needs to be normalized. Therefore,

$$|c_0|^2 + |c_1|^2 = 1. \tag{29.18}$$

### 29.6.4 Solving the Schrödinger Equation

Now, we will solve Eq. (29.15) by substituting Eq. (29.17). This is a lengthy derivation. If you feel this is too long, you may skip and just trust the answer. If not, I hope you can follow closely as we will practice some very useful skills in quantum mechanics.

We first perform the substitution.

$$
\begin{aligned}
i\hbar\frac{\partial|\psi\rangle}{\partial t} &= (\boldsymbol{H_0}+\boldsymbol{H_I})|\psi\rangle, \\
i\hbar\frac{\partial(c_0e^{i\frac{\omega_L}{2}t}|0\rangle+c_1e^{-i\frac{\omega_L}{2}t}|1\rangle)}{\partial t} &= (\boldsymbol{H_0}+\boldsymbol{H_I})(c_0e^{i\frac{\omega_L}{2}t}|0\rangle+c_1e^{-i\frac{\omega_L}{2}t}|1\rangle) \quad (29.19)
\end{aligned}
$$

Let us first simplify the *left-hand side*.

$$
\begin{aligned}
i\hbar\frac{\partial|\psi\rangle}{\partial t} &= i\hbar\Big(\dot{c_0}e^{i\frac{\omega_L}{2}t}|0\rangle+c_0(i\frac{\omega_L}{2})e^{i\frac{\omega_L}{2}t}|0\rangle \\
&\quad+\dot{c_1}e^{-i\frac{\omega_L}{2}t}|1\rangle+c_1(-i\frac{\omega_L}{2})e^{-i\frac{\omega_L}{2}t}|1\rangle\Big), \qquad (29.20)
\end{aligned}
$$

where we use the chain rule in derivatives. We also use the common notation of time derivative, $\dot{c}=\frac{dc}{dt}$. We will now apply an inner product with $|0\rangle$ to Eq. (29.20). This is equivalent to applying $\langle 0|$ from the left,

$$
\begin{aligned}
&\langle 0|i\hbar\Big(\dot{c_0}e^{i\frac{\omega_L}{2}t}|0\rangle+c_0(i\frac{\omega_L}{2})e^{i\frac{\omega_L}{2}t}|0\rangle \\
&\quad+\dot{c_1}e^{-i\frac{\omega_L}{2}t}|1\rangle+c_1(-i\frac{\omega_L}{2})e^{-i\frac{\omega_L}{2}t}|1\rangle\Big), \\
=\ & i\hbar\Big(\dot{c_0}e^{i\frac{\omega_L}{2}t}\langle 0|0\rangle+c_0(i\frac{\omega_L}{2})e^{i\frac{\omega_L}{2}t}\langle 0|0\rangle \\
&\quad+\dot{c_1}e^{-i\frac{\omega_L}{2}t}\langle 0|1\rangle+c_1(-i\frac{\omega_L}{2})e^{-i\frac{\omega_L}{2}t}\langle 0|1\rangle\Big), \\
=\ & i\hbar\Big(\dot{c_0}e^{i\frac{\omega_L}{2}t}+c_0(i\frac{\omega_L}{2})e^{i\frac{\omega_L}{2}t}\Big), \qquad (29.21)
\end{aligned}
$$

where we have used the fact that $|0\rangle$ and $|1\rangle$ are *orthonormal* in the last step. Therefore, $\langle 0|0\rangle=1$ and $\langle 0|1\rangle=0$.

Now, we will simplify the *right-hand side* of Eq. (29.19) and apply $\langle 0|$ from the left.

$$
\begin{aligned}
&\langle 0|(\boldsymbol{H_0}+\boldsymbol{H_I})(c_0e^{i\frac{\omega_L}{2}t}|0\rangle+c_1e^{-i\frac{\omega_L}{2}t}|1\rangle), \\
=\ & \langle 0|\Big(\boldsymbol{H_0}c_0e^{i\frac{\omega_L}{2}t}|0\rangle+\boldsymbol{H_0}c_1e^{-i\frac{\omega_L}{2}t}|1\rangle \\
&\quad+\boldsymbol{H_I}c_0e^{i\frac{\omega_L}{2}t}|0\rangle+\boldsymbol{H_I}c_1e^{-i\frac{\omega_L}{2}t}|1\rangle\Big), \\
=\ & c_0e^{i\frac{\omega_L}{2}t}\langle 0|\boldsymbol{H_0}|0\rangle+c_1e^{-i\frac{\omega_L}{2}t}\langle 0|\boldsymbol{H_0}|1\rangle \\
&\quad+c_0e^{i\frac{\omega_L}{2}t}\langle 0|\boldsymbol{H_I}|0\rangle+c_1e^{-i\frac{\omega_L}{2}t}\langle 0|\boldsymbol{H_I}|1\rangle. \qquad (29.22)
\end{aligned}
$$

Now, we need to evaluate $\langle 0|\boldsymbol{H_0}|0\rangle$, $\langle 0|\boldsymbol{H_0}|1\rangle$, $\langle 0|\boldsymbol{H_I}|0\rangle$, and $\langle 0|\boldsymbol{H_I}|1\rangle$. Firstly, since $|0\rangle$ and $|1\rangle$ are the eigenstates of the unperturbed system, $\boldsymbol{H_0}$, we have,

$$\begin{aligned}\langle 0|\boldsymbol{H_0}|0\rangle &= \langle 0|\left(-\hbar\omega_L/2\,|0\rangle\right),\\ &= -\hbar\omega_L/2\,\langle 0|0\rangle\,,\\ &= -\hbar\omega_L/2, \end{aligned} \tag{29.23}$$

and

$$\begin{aligned}\langle 0|\boldsymbol{H_0}|1\rangle &= \langle 0|\hbar\omega_L/2\,|1\rangle\,,\\ &= \hbar\omega_L/2\,\langle 0|1\rangle\,,\\ &= 0, \end{aligned} \tag{29.24}$$

where we used the fact that applying the Hamiltonian to its eigenvector results in the eigenvector scaled by the corresponding eigenvalue.

Now, let us find the matrix elements of the interacting Hamiltonian, $\boldsymbol{H_I}$.

*Example 29.2.* Find $\langle 0|\boldsymbol{H_I}|0\rangle$ and $\langle 0|\boldsymbol{H_I}|1\rangle$

To evaluate $\langle 0|\boldsymbol{H_I}|0\rangle$ and $\langle 0|\boldsymbol{H_I}|1\rangle$, we will use $H_I = eE_1 x\cos\left(\omega_1 t\right)$ in Eq. (29.14). We should note that $H_I$ is not written in the basis of $|0\rangle$ and $|1\rangle$ but in the real space of $\vec{r}$ in Eq. (29.14). In Section 14.2.2 in [1], we mentioned that in an 1-D case, position $x$ in real space form a complete basis with basis vector $|x\rangle$ and eigenvalues $x$ with the following completeness property (Eq. (14.8) in [1]),

$$\int_{-\infty}^{+\infty} |x\rangle\,\langle x|\,dx = \boldsymbol{I}. \tag{29.25}$$

In the 3-D space, $\vec{r}$ forms a complete basis also,

$$\int |\vec{r}\rangle\,\langle\vec{r}|\,d^3\vec{r} = \boldsymbol{I}, \tag{29.26}$$

and we will use it to help us compute $\langle 0|\boldsymbol{H_I}|0\rangle$ and $\langle 0|\boldsymbol{H_I}|1\rangle$ by projecting $|0\rangle$ and $|1\rangle$ to the real space, i.e., finding their wavefunctions in real space. The following is a lengthy derivation. I will explain them step-by-step after the equations.

$$
\begin{aligned}
\langle 0|\boldsymbol{H_I}|0\rangle &= \langle 0|\boldsymbol{I}\boldsymbol{H_I}\boldsymbol{I}|0\rangle\,, \\
&= \langle 0|\int \left|\vec{r'}\right\rangle\left\langle\vec{r'}\right|d^3\vec{r'}\boldsymbol{H_I}\int |\vec{r}\rangle\langle\vec{r}|d^3\vec{r}|0\rangle\,, \\
&= \int\int \left\langle 0|\vec{r'}\right\rangle\left\langle\vec{r'}\right|\boldsymbol{H_I}|\vec{r}\rangle\langle\vec{r}|0\rangle\, d^3\vec{r'}d^3\vec{r}, \\
&= \int \langle 0|\vec{r}\rangle\langle\vec{r}|\boldsymbol{H_I}|\vec{r}\rangle\langle\vec{r}|0\rangle\, d^3\vec{r}, \\
&= \int \Psi_0^*(\vec{r})\langle\vec{r}|\boldsymbol{H_I}|\vec{r}\rangle\Psi_0(\vec{r})d^3\vec{r}, \\
&= \int \Psi_0^*(\vec{r})e\vec{r}\cdot(E_1\cos(\omega_1 t)\hat{x})\Psi_0(\vec{r})d^3\vec{r}, \\
&= \int \Psi_0^*(\vec{r})eE_1 x\cos(\omega_1 t)\Psi_0(\vec{r})d^3\vec{r}, \\
&= eE_1\cos(\omega_1 t)\int \Psi_0^*(\vec{r})x\Psi_0(\vec{r})d^3\vec{r}, \\
&= eE_1\cos(\omega_1 t)\langle 0|x|0\rangle\,. \qquad (29.27)
\end{aligned}
$$

In the first line, we inserted identity matrices without affecting the result. In line 2, we used the completeness property in Eq. (29.26). Note that we used two different dummy variables, $\vec{r'}$ and $\vec{r}$, for each of the identity matrices. In line 3, we regroup the terms by recognizing that an integration is just a summation in a limit sense. The terms in the middle will be evaluated for *each pair of $\vec{r'}$ and $\vec{r}$*. Now, we need to apply the physics we know about the interacting Hamiltonian, which is the interaction between the electric dipole moment and the electric field. We know that it is **local**. This means that it depends on the value of the electric field at the dipole's location. In other words, it depends on $\vec{r}$ instead of both $\vec{r}$ and $\vec{r'}$, and, thus, it has a zero value when $\vec{r} \neq \vec{r'}$. Therefore, for a given $\vec{r}$, when we integrate over $\vec{r'}$, only the terms with $\vec{r} = \vec{r'}$ will remain (others are zero). As a result, we obtain line 4 where the integration over $\vec{r'}$ has completed and $|r'\rangle$ is replaced by $|r\rangle$.

In line 5, we used the definition of $\Psi_0(\vec{r}) = \langle\vec{r}|0\rangle$, which is just the wavefunction of state $|0\rangle$ in real space by projecting $|0\rangle$ on $|\vec{r}\rangle$. This is similar to what we did in Eq. (14.43) in [1] in the 1-D case when we projected the eigenstate of a simple harmonic oscillator to the real space to get its wavefunction. We also used the identity $\langle 0|\vec{r}\rangle = \langle\vec{r}|0\rangle^*$ to get $\Psi_0^*(\vec{r}) = \langle 0|\vec{r}\rangle$ (see also Eq. (3.6) in [1]).

$\langle\vec{r}|\boldsymbol{H_I}|\vec{r}\rangle$ means the expectation value of $\boldsymbol{H_I}$ in state $|\vec{r}\rangle$ which is just the energy at location $\vec{r}$, and is given in Eq. (29.14). We perform the corresponding substitution in lines 6 and 7. We factorized the terms independent of $\vec{r}$ in line 8 and defined $\langle 0|x|0\rangle = \int \Psi_0^*(\vec{r})x\Psi_0(\vec{r})d^3\vec{r}$ in line 9.

Similarly, we expect to get

$$
\langle 0|\boldsymbol{H_I}|1\rangle = eE_1\cos(\omega_1 t)\langle 0|x|1\rangle\,. \qquad (29.28)
$$

■

Now, as shown in Example 29.2, we need to know the wavefunctions of $|0\rangle$ and $|1\rangle$ in order to compute the integrals. They are given in atomic physics textbooks (such as [2]). We will not perform the integration. However, for $\langle 0|\boldsymbol{H_I}|0\rangle$,

$$
\begin{aligned}
\langle 0|\boldsymbol{H_I}|0\rangle &= eE_1\cos(\omega_1 t)\int \Psi_0^*(\vec{r})x\Psi_0(\vec{r})d^3\vec{r}, \\
&= eE_1\cos(\omega_1 t)\int \Psi_0^*(\vec{r})\Psi_0(\vec{r})xd^3\vec{r}, \\
&= eE_1\cos(\omega_1 t)\int |\Psi_0(\vec{r})|^2 xd^3\vec{r}. \qquad (29.29)
\end{aligned}
$$

The function $|\Psi_0(\vec{r})|^2$, which is the amplitude squared of the wavefunction, is known to be an even function, i.e., $|\Psi_0(\vec{r})|^2 = |\Psi_0(-\vec{r})|^2$. This is easy to understand because the Columb potential is a central potential (spherically symmetric). Then $|\Psi_0(\vec{r})|^2 x$ is an odd function due to $x$. Therefore, the integral will be 0. That is,

$$
\langle 0|\boldsymbol{H_I}|0\rangle = 0. \qquad (29.30)
$$

For $\langle 0|\boldsymbol{H_I}|1\rangle$, it is usually not 0. From Eq. (29.28), we have,

$$
\begin{aligned}
\langle 0|\boldsymbol{H_I}|1\rangle &= eE_1\cos(\omega_1 t)\langle 0|x|1\rangle, \\
&= e\langle 0|x|1\rangle E_1\cos(\omega_1 t) \\
&= eX_{01}E_1\cos(\omega_1 t), \qquad (29.31)
\end{aligned}
$$

where we defined $X_{01} = \langle 0|x|1\rangle$. Readers are encouraged to compare this to the spin qubit result in Eq. (9.18) in [1].

Now, we go back to continue to evaluate Eq. (29.22), and there are two terms left. By equating it to Eq. (29.21) (the left-hand side), we have

$$
\begin{aligned}
& i\hbar\left(\dot{c_0}e^{i\frac{\omega_L}{2}t} + c_0(i\frac{\omega_L}{2})e^{i\frac{\omega_L}{2}t}\right), \\
= \;& c_0 e^{i\frac{\omega_L}{2}t}\langle 0|\boldsymbol{H_0}|0\rangle + c_1 e^{-i\frac{\omega_L}{2}t}\langle 0|\boldsymbol{H_I}|1\rangle, \\
= \;& c_0 e^{i\frac{\omega_L}{2}t}(-\hbar\omega_L/2) + c_1 e^{-i\frac{\omega_L}{2}t}(eX_{01}E_1\cos(\omega_1 t)). \qquad (29.32)
\end{aligned}
$$

Equating line 1 and line 3 of Eq. (29.32) and recognizing $i\hbar c_0(i\frac{\omega_L}{2})e^{i\frac{\omega_L}{2}t}$ in line 1 and $c_0 e^{i\frac{\omega_L}{2}t}(-\hbar\omega_L/2)$ in line 3 are equal, we have

$$
\begin{aligned}
i\hbar\dot{c_0}e^{i\frac{\omega_L}{2}t} &= c_1 e^{-i\frac{\omega_L}{2}t}(eX_{01}E_1\cos(\omega_1 t)), \\
i\dot{c_0} &= \frac{eX_{01}}{\hbar}E_1\cos(\omega_1 t)e^{-i\omega_L t}c_1. \qquad (29.33)
\end{aligned}
$$

What we have done some far is to perform an inner product with $|0\rangle$ so that we obtain the rate of change of $c_0$ as a function of $c_1$. Now, if we repeat the same process by applying an inner product with $|1\rangle$, we will get an equation relating the rate of change of $c_1$ as a function of $c_0$, which is

$$i\dot{c}_1 = \frac{eX_{01}^*}{\hbar} E_1 \cos(\omega_1 t) e^{i\omega_L t} c_0, \tag{29.34}$$

where $X_{01}^* = \langle 1|x|0\rangle$.

As a reminder, $e > 0$. To solve for $c_0$, we can perform one more time differentiation on Eq. (29.33) and substitute Eq. (29.34) into it. However, we are not interested in the general solution. We are interested in the case when $\omega_1 = \omega_L$.

## 29.7 Rotating Wave Approximation (RWA)

When $\omega_1 = \omega_L$, the system can be said to be in resonance. *That means the horizontal oscillating electric c field oscillates at the same frequency as the Larmor frequency.* By using, $\cos\theta = \frac{e^{i\theta}+e^{-i\theta}}{2}$, Eq. (29.33) becomes

$$\begin{aligned} i\dot{c}_0 &= \frac{eX_{01}}{\hbar} E_1 \cos(\omega_1 t) e^{-i\omega_L t} c_1, \\ &= \frac{eX_{01}}{2\hbar} E_1 (e^{i\omega_1 t} + e^{-i\omega_1 t}) e^{-i\omega_L t} c_1, \\ &= \frac{eX_{01}}{2\hbar} E_1 (e^{i(\omega_1-\omega_L)t} + e^{-i(\omega_1+\omega_L)t}) c_1, \\ &= \frac{eX_{01}}{2\hbar} E_1 (e^{i0t} + e^{-i2\omega_L t}) c_1, \\ &= \frac{eX_{01}}{2\hbar} E_1 (1 + e^{-i2\omega_L t}) c_1, \end{aligned} \tag{29.35}$$

where we have used the fact that $\omega_1 = \omega_L$ in line 4 at resonance. For the term $e^{-i2\omega_L t}$, which is equal to $\cos 2\omega_L t - i\sin 2\omega_L t$, it oscillates very fast compared to the time scale we are interested in. Therefore, it can be ignored. This is because, for any fast oscillation, the average effect is zero on a time scale larger than its period. This is called the **rotating wave approximation (RWA)**. Eq. (29.35) is then simplified to

$$i\dot{c}_0 = \frac{eX_{01}}{2\hbar} E_1 c_1. \tag{29.36}$$

Similarly, by using RWA under resonance, Eq. (29.34) is simplified to

$$i\dot{c}_1 = \frac{eX_{01}^*}{2\hbar} E_1 c_0. \tag{29.37}$$

By taken a further time derivative on Eq. (29.36) and substituting with Eq. (29.37),

$$\begin{aligned}
\frac{d(i\dot{c_0})}{dt} = i\ddot{c_0} &= \frac{eX_{01}}{2\hbar}E_1\dot{c_1}, \\
&= \frac{eX_{01}}{2\hbar}E_1\left(\frac{eX_{01}^*}{2\hbar i}E_1c_0\right), \\
\ddot{c_0} &= -\left(\frac{eE_1|X_{01}|}{2\hbar}\right)^2 c_0, \\
\ddot{c_0} &= -\left(\frac{eE_1X_{01}}{2\hbar}\right)^2 c_0,
\end{aligned} \tag{29.38}$$

where in the last line, $X_{01}$ is real when it is the transition between bound states [2].

Eq. (29.38) is a second-order differential equation for $c_0$. Defining $\omega_R' = \frac{eE_1X_{01}}{2\hbar}$, the equation becomes $\ddot{c_0} = -\omega_R'^2 c_0$ and and the general solution is [13]

$$c_0 = A\cos\omega_R' t + B\sin\omega_R' t. \tag{29.39}$$

To find $c_1$, we will use Eq. (29.36) and substitute Eq. (29.39)

$$\begin{aligned}
c_1 &= \frac{i}{\omega_R'}\dot{c_0} = i(-A\sin\omega_R' t + B\cos\omega_R' t), \\
&= -iA\sin\omega_R' t + iB\cos\omega_R' t.
\end{aligned} \tag{29.40}$$

Note that $c_0$ and $c_1$ are the coefficient of $|0\rangle$ and $|1\rangle$, respectively (Eq. (29.17)). Therefore, the squares of their magnitudes represent the probabilities of finding the ion at $|0\rangle$ and $|1\rangle$, respectively, upon a measurement. Since both of them oscillate at $\omega_R'$, the probabilities of finding the electron at $|0\rangle$ and $|1\rangle$ thus oscillate with time.

## 29.8 Rabi Oscillation and Rabi Frequency

As shown in Eq. (29.39) and Eq. (29.40), the movement of the trapped ion state is complex. We will now inspect a special case to understand how the state moves on the Bloch sphere due to **Rabi oscillation**.

At time $t = 0$, by using Eq. (29.39) and Eq. (29.40), Eq. (29.17) becomes

$$\begin{aligned}
|\Psi(t=0)\rangle &= c_0 e^{i\frac{\omega_L}{2}0}|0\rangle + c_1 e^{-i\frac{\omega_L}{2}0}|1\rangle, \\
&= c_0(t=0)|0\rangle + c_1(t=0)|1\rangle, \\
&= A|0\rangle + iB|1\rangle,
\end{aligned} \tag{29.41}$$

where $A$ and $iB$ are, thus, the coefficients of $|0\rangle$ and $|1\rangle$ at $t = 0$. However, recalling that on the Bloch sphere, the state at $t = 0$ is characterized by an initial polar angle ($\theta_0$) and an azimuthal angle ($\phi_0$) (similar to Eq. (8.10) in [1]). Therefore, $A = \cos\frac{\theta_0}{2}\exp\left\{-i\frac{\phi_0}{2}\right\}$ and $iB = \sin\frac{\theta_0}{2}\exp\left\{i\frac{\phi_0}{2}\right\}$. As a result,

$$
\begin{aligned}
c_0 &= A\cos\omega_R' t + B\sin\omega_R' t,\\
&= \cos\frac{\theta_0}{2}\exp\left\{-i\frac{\phi_0}{2}\right\}\cos\omega_R' t - i\sin\frac{\theta_0}{2}\exp\left\{i\frac{\phi_0}{2}\right\}\sin\omega_R' t,\\
&= e^{-i\frac{\phi_0}{2}}[\cos\frac{\theta_0}{2}\cos\omega_R' t - e^{i\frac{\pi}{2}}\sin\frac{\theta_0}{2}\sin\omega_R' t e^{i\phi_0}],\\
&= e^{-i\frac{\phi_0}{2}}[\cos\frac{\theta_0}{2}\cos\omega_R' t - \sin\frac{\theta_0}{2}\sin\omega_R' t e^{i(\phi_0+\frac{\pi}{2})}],
\end{aligned}
\tag{29.42}
$$

where we used the fact that $i = e^{i\frac{\pi}{2}}$ in line 3. Using Eq. (29.40), we obtain,

$$
\begin{aligned}
c_1 &= -iA\sin\omega_R' t + iB\cos\omega_R' t,\\
&= -i\cos\frac{\theta_0}{2}\exp\left\{-i\frac{\phi_0}{2}\right\}\sin\omega_R' t + \sin\frac{\theta_0}{2}\exp\left\{i\frac{\phi_0}{2}\right\}\cos\omega_R' t,\\
&= e^{i\frac{\phi_0}{2}}[\sin\frac{\theta_0}{2}\cos\omega_R' t + \cos\frac{\theta_0}{2}\sin\omega_R' t e^{-i(\phi_0+\frac{\pi}{2})}],
\end{aligned}
\tag{29.43}
$$

where we used the fact that $-i = e^{-i\frac{\pi}{2}}$. It is still difficult to visualize how the qubit evolves with an arbitrary $\phi_0$. Let us set $\phi_0 = -\pi/2$ (Fig. 29.7). Then $\phi_0 + \pi/2 = 0$ and $e^{-i(\phi_0+\frac{\pi}{2})} = 1$. Eq. (29.42) and Eq. (29.43) are simplified to

$$
\begin{aligned}
c_0 &= e^{-i\frac{\phi_0}{2}}[\cos\frac{\theta_0}{2}\cos\omega_R' t - \sin\frac{\theta_0}{2}\sin\omega_R' t],\\
&= e^{-i\frac{\phi_0}{2}}\cos(\frac{\theta_0}{2}+\omega_R' t),\\
&= e^{-i\frac{\phi_0}{2}}\cos\frac{\theta_0+2\omega_R' t}{2},\\
&= e^{-i\frac{\phi_0}{2}}\cos\frac{\theta_0+\omega_R t}{2},
\end{aligned}
\tag{29.44}
$$

where we use the trigonometry identity $\cos(\alpha+\beta)=\cos\alpha\cos\beta-\sin\alpha\sin\beta$ in the second line. Note that we have already set $\phi_0=\pi/2$. It is kept unsubstituted to show where this initial azimuthal angle is in the equation. Here we define **Rabi Frequency**,

$$
\omega_R = 2\omega_R' = \frac{eE_1X_{01}}{\hbar}. \tag{29.45}
$$

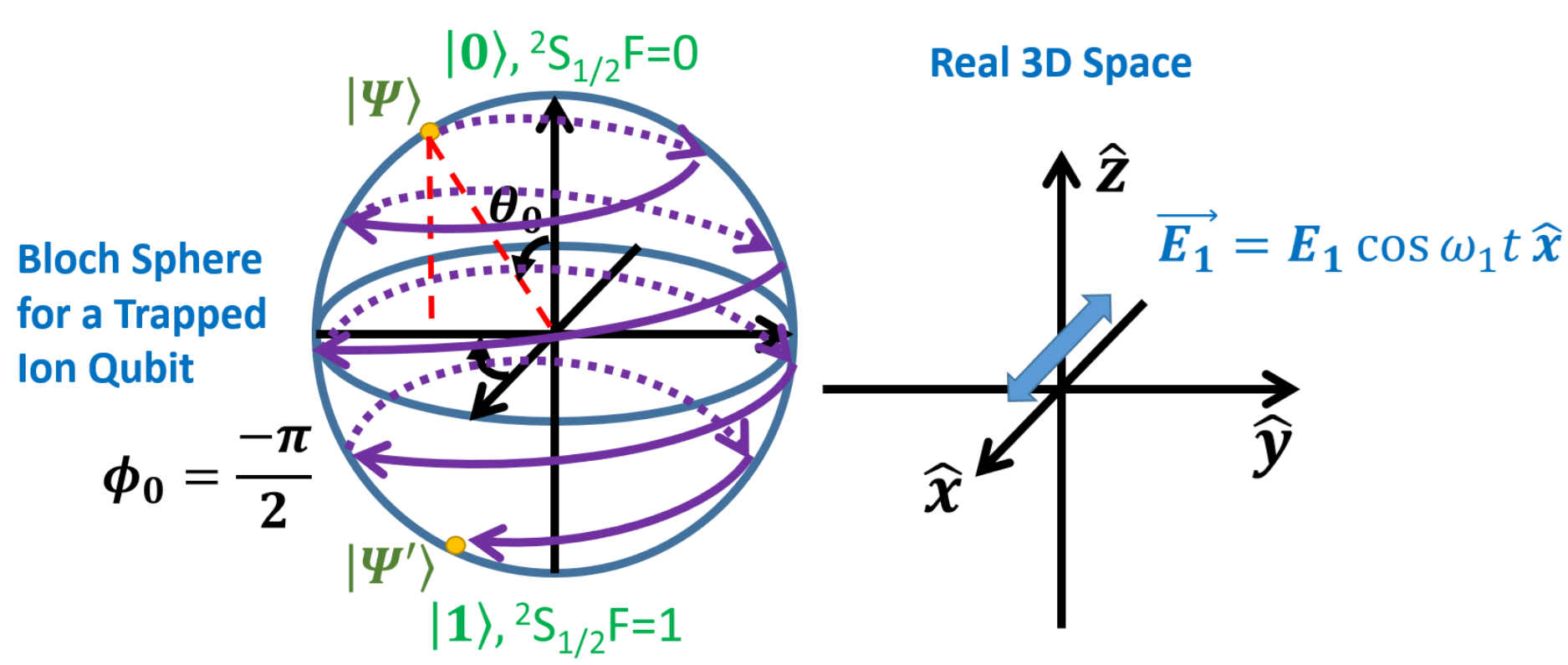

Fig. 29.7: Rabi oscillation of a trapped ion qubit at resonance when the horizontal field is oscillating at Larmor frequency ($\omega_1 = \omega_L$). The initial state $|\Psi\rangle$ has an initial azimuthal angle $\phi_0 = -\pi/2$. The left shows how the state moves on the Bloch sphere due to Larmor precession and Rabi oscillation. The right shows the setup of the experiment.

Similarly,

$$
\begin{aligned}
c_1 &= e^{i\frac{\phi_0}{2}}[\sin\frac{\theta_0}{2}\cos\omega_R' t + \cos\frac{\theta_0}{2}\sin\omega_R' t],\\
&= e^{i\frac{\phi_0}{2}}\sin(\frac{\theta_0}{2}+\omega_R' t),\\
&= e^{i\frac{\phi_0}{2}}\sin\frac{\theta_0+2\omega_R' t}{2},\\
&= e^{i\frac{\phi_0}{2}}\sin\frac{\theta_0+\omega_R t}{2},
\end{aligned}
\tag{29.46}
$$

by using the trigonometry identity $\sin(\alpha+\beta)=\sin\alpha\cos\beta+\cos\alpha\sin\beta$ in the second line.

Now let us substitute Eq. (29.44) and Eq. (29.46) into Eq. (29.17) ,

$$
\begin{aligned}
|\Psi\rangle &= c_0 e^{i\frac{\omega_L}{2}t}|0\rangle + c_1 e^{-i\frac{\omega_L}{2}t}|1\rangle,\\
&= e^{-i\frac{\phi_0}{2}}\cos\frac{\theta_0+\omega_R t}{2}e^{i\frac{\omega_L}{2}t}|0\rangle + e^{i\frac{\phi_0}{2}}\sin\frac{\theta_0+\omega_R t}{2}e^{-i\frac{\omega_L}{2}t}|1\rangle,\\
&= e^{-i\frac{\phi_0-\omega_L t}{2}}\cos\frac{\theta_0+\omega_R t}{2}|0\rangle + e^{i\frac{\phi_0-\omega_L t}{2}}\sin\frac{\theta_0+\omega_R t}{2}|1\rangle.
\end{aligned}
\tag{29.47}
$$

Again, we have already set $\phi_0=\pi/2$ to achieve this result. Like Eq. (8.10) in [1], this equation tells us that the azimuthal angle reduces at a rate of $\omega_L$, i.e., $\phi_0-\omega_L t$. This is the Larmor precession. At the same time, the polar angle also changes as $\theta_0+\omega_R t$, which means it changes at a rate of the Rabi frequency, $\omega_R$ (Fig. 29.7).

Readers are encouraged to read Example 9.2 and Section 9.6 in [1] to understand how the measurement probabilities of $|0\rangle$ and $|1\rangle$ change as a function of time and gain an intuitive understanding of the rotating wave approximation.

## 29.9 Summary

In this Chapter, we showed the mechanisms to perform trapped ion qubit readout and initialization. They use a similar mechanism by exciting state $|1\rangle$ to a higher energy level. In the readout process, it will emit detectable photons through spontaneous emission if the state is $|1\rangle$. $|0\rangle$ will not be excited because of the energy gap. This is called fluorescence detection. In the initialization process, again, only $|1\rangle$ may be excited to a higher energy state and will decay quickly to $|0\rangle$ with a high probability. Both of them are only efficient if the corresponding photon absorption and emission cycle can be maintained. Then we discuss the trapped ion electric dipole moment and photon oscillating electric field interaction. This is used to create Rabi oscillation and thus any arbitrary one-qubit gate. We emphasize that its mathematics and physics have a lot of similarity with those of a spin qubit. Now, we have shown that all Divincenzo criteria have been fulfilled, except that we still need to demonstrate an entanglement gate. This will be discussed in the next chapter.

## Problems

**29.1. Electric Dipole Moment.**
Deduce the electric dipole moment of two opposite charges separated by a displacement $\vec{r}$ using Eq. (29.12).

**29.2. Larmor Precession**
Use the Hamiltonian in Eq. (29.4) and follow the procedure in Section 8.3 in [1] to show that a general qubit state will precess about the z-axis at an angular frequency of $\omega_L$.

**29.3. Solving Schrödinger Equation.**
Prove Eqs. (29.34) and (29.37).

# Chapter 30
# Trapped Ion Qubit - Laser Cooling and Entanglement Gate

## 30.1 Introduction

In this Chapter, we will discuss **Doppler laser cooling** and **resolved sideband cooling**. Like other technologies, trapped ions need to be cooled to a very low temperature to minimize thermal noise and linewidth broadening to enhance decoherence time. More importantly, the ion chain needs to be cooled to its collective minimum energy to enable two-qubit gate operations. The entanglement gate in trapped ion qubits is realized by using the tensor product states of the qubits and the ion chain motion. That is the reason we group laser cooling and entanglement gates in a single chapter.

### *30.1.1 Learning Outcomes*

Understand the principle of Doppler laser cooling and resolved sideband cooling; be able to describe how the Doppler effect is used in laser cooling; understand the quantum state of an ion chain; appreciate how the ion chain motion states are used to enable the entanglement gate in trapped ion qubits.

### *30.1.2 Teaching Videos*

- `https://youtu.be/mymwdbBBgAQ`

## 30.2 Doppler Laser Cooling

At first, it seems counterintuitive that a laser can be used to cool down an ion as a laser is perceived to have a high energy density. Doppler laser cooling is a very smart idea that uses **Doppler effect** to cool an atom or an ion with a laser. In this section, we will just describe qualitatively how Doppler laser cooling is achieved. Readers interested in quantitative treatments may refer to Chapter 9 in [2].

### *30.2.1 Doppler Effect*

When an ambulance approaches you, you hear a siren that has a higher pitch (frequency) than when it is stationary. When it leaves you, you hear a siren that has a lower pitch instead. This is the Doppler effect. The same effect happens to light. Therefore, when a star is moving away from us, its frequency spectrum will shift to a lower frequency, and this is known as the **red shift**. If it is moving towards us, its frequency spectrum will shift to a higher frequency and appear to be more blueish, and this is called the **blue shift**. The amount of frequency shift depends on the relative velocity between the observer and the object emitting the light. The larger the velocity, the more the shift. Fig. 30.1 illustrates the Doppler effect.

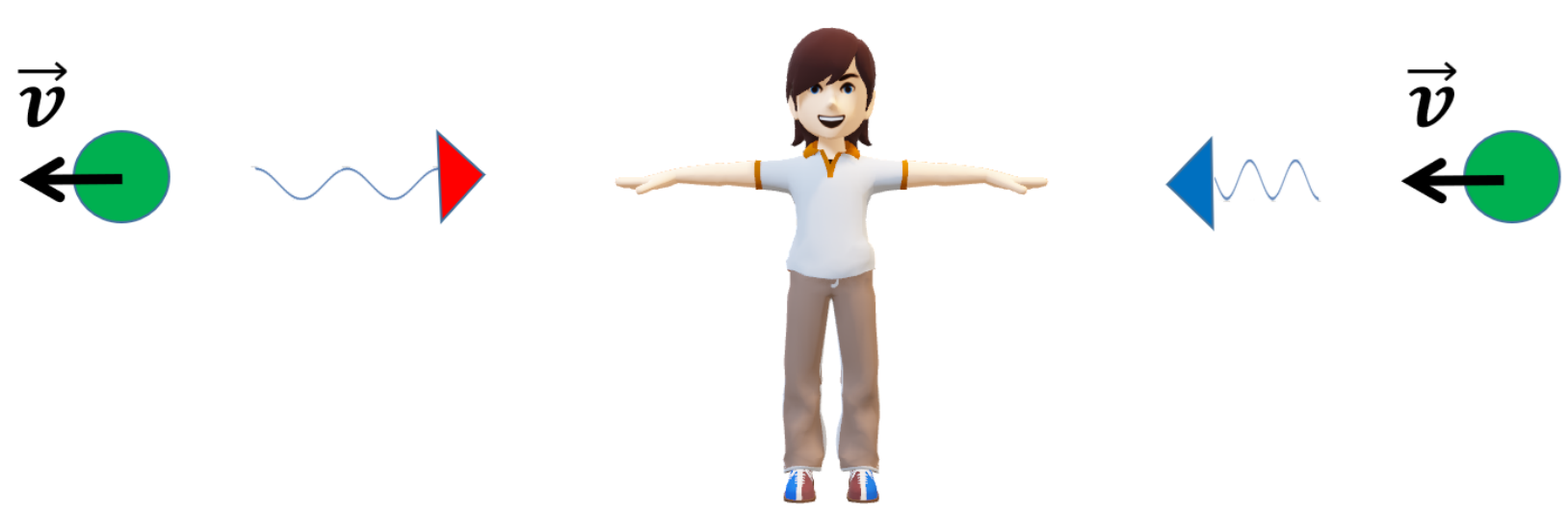

Fig. 30.1: When the left object is leaving the observer in the middle at a speed of $\vec{v}$, the light emitted by that object appears to have a longer wavelength (smaller frequency) to the observer. This is the red shift. For the right object, since it is moving towards the observer, the light emitted by it appears to have a short wavelength (higher frequency) to the observer. This is the blue shift.

### *30.2.2 Simple Doppler Cooling Example*

Let us consider a one-dimensional case. Now, imagine an ion is moving to the right with momentum $p_i$ (Fig. 30.2). Let us assume it is a Ytterbium ion and is at state $|1\rangle$, i.e. ${}^2S_{1/2}$, $F=1$. If it receives a photon with a wavelength of 369.53 nm (note that this is the wavelength it *perceives* after the Doppler effect, and we will discuss this later), it will be excited to ${}^2P_{1/2}$, $F=0$. Let us assume the photon was traveling to the left with a momentum $p_{ph}$. The total momentum before and after the photon absorption or the ion excitation is conserved and is equal to $p_i - p_{ph}$. Note that this process is exactly the same as the photon absorption during qubit readout ( Fig. 29.2a). Therefore, the same laser source can be used.

After a while (bottom row of Fig. 30.2), the excited ion may decay due to spontaneous emission. As mentioned in Section 29.2, photons will be emitted in random directions in spontaneous emission. In this 1-D case, it means it might be emitted to the left or right randomly. After emission, the magnitude of the photon momentum should be the same as before, i.e., $p_{ph}$, because the momentum of the photon is determined by the wavelength ($p_{ph} = h/\lambda$). Due to conservation of momentum, regardless of the emission direction of the photon, the total momentum of the system is still $p_i - p_{ph}$. If the photon is emitted to the right, the ion will slow down with a momentum of $p_i - 2p_{ph}$ so that the total momentum of the system is $p_i - 2p_{ph} + p_{ph} = p_i - p_{ph}$. Therefore, the ion is cooled down as it has a lower thermal velocity. If the emitted photon goes to the left, the ion will continue to move to the right with a momentum of $p_i$ as before, under this setup.

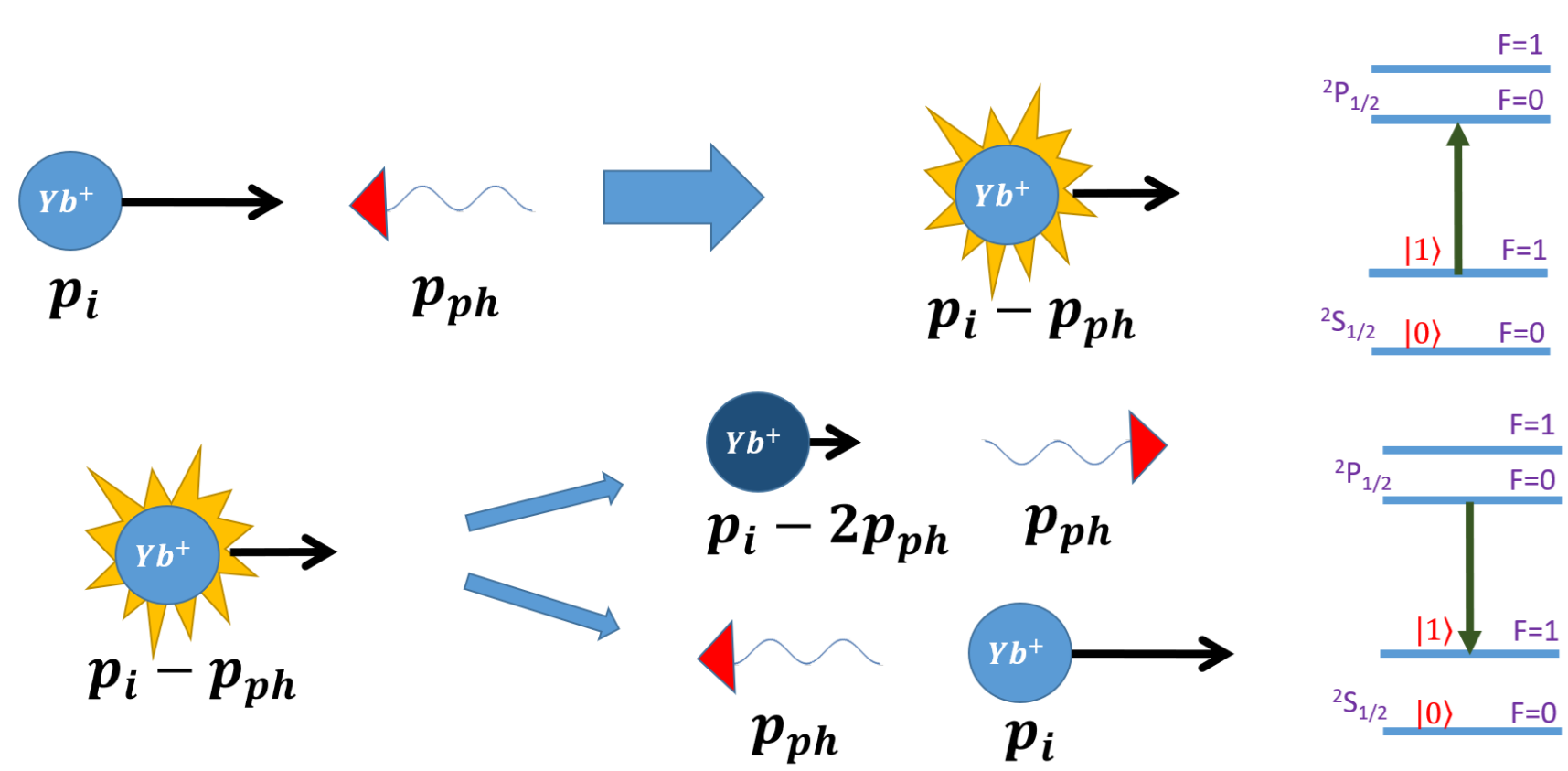

Fig. 30.2: Laser cool process with the ion moving the the right and photon going to the left (top left). Top row: Absorption process. Bottom row: Emission process. The corresponding energy diagrams of the ion and the transition are shown on the far left.

### *30.2.3 Doopler Laser Cooling*

Now, since the ions are moving, there is a Doppler shift. Therefore, the laser beam will be *detuned* to have a longer wavelength (lower frequency) than 369.53 nm. This can be achieved by using an electro-optic modulator as was done during the initialization process (Section 29.3) [6]. Then in Fig. 30.2, the ion that is moving towards the photon will see a blue shift and will absorb the photon to perform the cooling cycle. On the other hand, if the ion is moving in the same direction or away from the photon, it will see less blue shift or even a red shift (even longer wavelength), and there will be no absorption. This is very important because if the ion is moving in the same direction as the photon and if it were able to absorb the photon, it will be heated up (see Problem 30.1). Therefore, due to the Doppler effect, only ions moving towards the detuned laser will be cooled down. In practice, two ion beams will be used in each axis and shot at the ions in both directions. In the 1-D case, it means there will be photons coming from the left and also coming from the right. They are both detuned with a lower frequency than the ion transition frequency. Only the right (left) moving ion will absorb the left (right) moving photons to start the cooling cycle. In 3D, 3 pairs of such laser beams will be used in each axis in the 3D space. After many cycles, they will be cooled down to millikelvin.

The ion needs to go through many absorption and emission cooling cycles to be cooled to millikelvin. As in Fig. 29.2, due to the selection rule, under an ideal situation, the ion will not decay from $^2P_{1/2},\ F=0$ to $^2S_{1/2},\ F=0$ so that it can continue the cooling cycle by absorbing another photon. Of course, due to various reasons, it might decay to $^2S_{1/2},\ F=0$. In that case, it will end the cooling cycle as it cannot absorb the photon anymore, just like how it exited the fluorescence readout cycle.

## 30.3 Resolved Sideband Cooling

Doppler laser cooling cannot cool the trapped ions to a low enough temperature. We need to cool the trapped ions into their **motional ground state**. What does that mean? As shown in Fig. 28.4, the ion chain is trapped in a potential well. When there is a finite temperature, it will vibrate and this is the motion we are referring to. When the motion is small, the system can be approximated as a mechanical simple harmonic oscillator (SHO). We discussed in Chapter 14 in [1] that a mechanical SHO is quantized. The quantization can be noticed at very low temperatures, which is the case in a trapped ion quantum computer. The eigenstates of the SHO are $|N\rangle$, which represents the number of quanta of energy it has, and the corresponding eigenvalue is

$$E_N = \hbar\omega_{chain}(N + \frac{1}{2}), \tag{30.1}$$

where we defined the characteristic angular frequency of the ion chain as $\omega_{chain}$. Therefore, the motional ground state refers to the ground state of this SHO with $N = 0$. The energy is $\frac{\hbar\omega_{chain}}{2}$. Note that its energy quantum is also called **phonon**.

At the same time, we know that the ions can have different energies due to their electronic states. For a Ytterbium ion, it can be at its ground state $|g\rangle = |0\rangle$, which is $^2S_{1/2},\ F = 0$ or at its excited state $|e\rangle = |0\rangle$, which is $^2S_{1/2},\ F = 1$ (see Fig. 28.2). So far, when we discuss the state of the trapped ions, we have been using $|0\rangle$ and $|1\rangle$. We also ignored the motional states $|N\rangle$. From now on (including when we discuss the entanglement gate), we will use $|g\rangle$ and $|e\rangle$ to represent the qubit states to avoid confusion and also consider the motional state, $|N\rangle$. As a result, we have a larger Hilbert space, which is the tensor product of the electronic states (qubit states) and the motional state. For example, $|e, N = 1\rangle$ means that the ion in consideration is at $^2S_{1/2},\ F = 1$ while the ion chain has one quanta of energy. As another example, $|g, N = 2\rangle$ means that the qubit is at $|0\rangle$ ( $^2S_{1/2},\ F = 0$) while the ion chain has two quanta of energy ($N = 2$).

Now we are ready to discuss the resolved sideband cooling. We will shoot a laser with a lower angular frequency, $\omega_{sideband}$, than the qubit angular transition frequency (i.e. the angular Larmor frequency, $\omega_L$, in Eq. (29.3)) by the charcteristic angular frequency of the SHO of the ion chain. That is,

$$\omega_{sideband} = \omega_L - \omega_{chain}. \tag{30.2}$$

Fig. 30.3 illustrates the cooling process. Suppose the chain is at state $|N = 2\rangle$ and the ion is at $|g\rangle$, then the system is at state $|g, N = 2\rangle$. The laser does not have enough energy to excite the qubit to $|e\rangle$ to achieve $|e, N = 2\rangle$ because it is $\hbar\omega_{chain}$ shy of the transition energy. However, the motion of the chain can lose one unit of energy to enable the transition. That means $|g\rangle$ becomes $|e\rangle$ and $|N = 2\rangle$ becomes $|N = 1\rangle$ to conserve the energy. Therefore, in this excitation process, the motional energy of the ion chain SHO is reduced and the overall transition is from $|g, N = 2\rangle$ to $|e, N = 1\rangle$. After a while, the excited ion will decay through spontaneous emission by emitting a photon of $\hbar\omega_L$ (instead of $\hbar\omega_{sideband}$), and the system will become $|g, N = 1\rangle$. Therefore, this cycle of absorption of a red shift sideband photon and then the emission of a photon has successfully cooled down the ion chain by reducing its motional energy from $N = 2$ quanta of energy to $N = 1$ quantum of energy (i.e., taken away one phonon). This process will repeat. For example, in Fig. 30.3, $|g, N = 1\rangle$ will be cooled down to $|g, N = 0\rangle$ after another cycle of sideband cooling, and thus achieve the motional ground state of the ion chain and complete the cooling process.

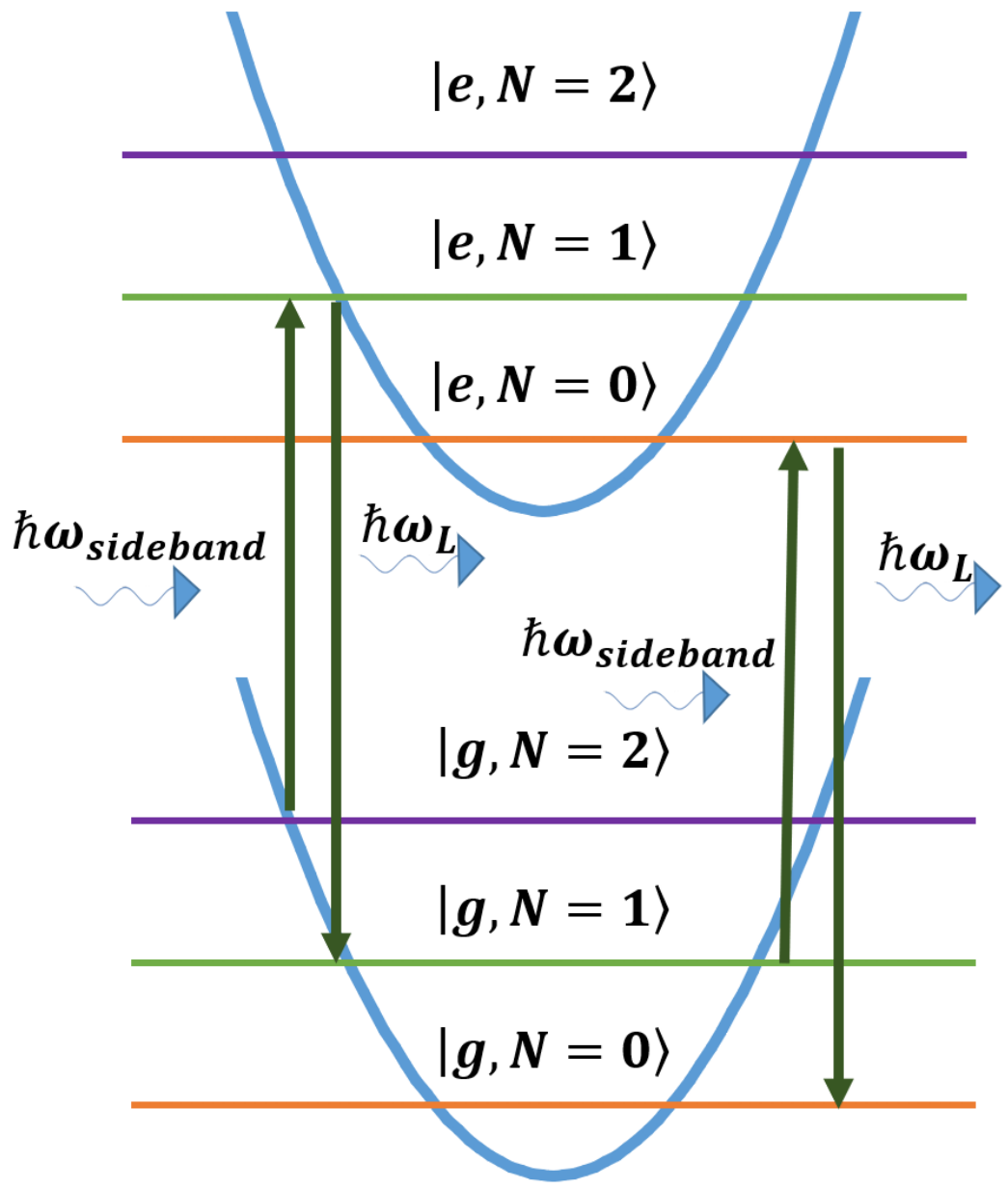

Fig. 30.3: Illustration of resolved sideband cooling. The Hibert space is the tensor product of the ion qubit states ($|g = 0\rangle$ and $|e = 1\rangle$) and the motional states of the ions in the trapping field as a mechanical simple harmonic oscillator ($|N\rangle$). Each horizontal line represents the total energy of the system for the given state. The parabolic lines are just to remind us that the quantization of the SHO is due to the parabolic potential energy.

## 30.4 Cirac–Zoller Gate as an Entanglement Gate

### *30.4.1 Entanglement Gate for Trapped Ions*

The last thing we need to show for trapped ions to be a good quantum computing architecture is that we can implement a two-qubit entanglement gate. There are various types of entanglement gates for trapped ions (e.g. [14] [15]). Here we will discuss the Cirac–Zoller gate, which is a controlled-Z (CZ) gate [14]. As discussed in Section 12.3.1 in [1], a CZ gate (which is a controlled phase shift gate with a phase shift of $\pi$, $\boldsymbol{U}_{CPS,\pi}$) has the following matrix,

$$\boldsymbol{U}_{CPS,\pi} = \begin{pmatrix} 1 & 0 & 0 & 0 \\ 0 & 1 & 0 & 0 \\ 0 & 0 & 1 & 0 \\ 0 & 0 & 0 & -1 \end{pmatrix} \tag{30.3}$$

It can be used to form a CNOT gate by combing with other one-qubit gates because $\boldsymbol{U}_{XOR} = (\boldsymbol{I} \otimes \boldsymbol{H})\boldsymbol{U}_{CPS,\pi}(\boldsymbol{I} \otimes \boldsymbol{H})$ (see Eq. (12.5) in [1]). Therefore, by implementing the Cirac–Zoller gate, together with other one-qubit gates, we can implement or approximate any quantum circuits as mentioned in Section 1.2 in [1].

### 30.4.2 Motional States and Auxiliary States

Like the spin qubit case, we need to borrow states outside of the computational space ($|0\rangle / |1\rangle$) to implement an entanglement gate (see Eq. (12.13) in [1]). This is understandable. For a standalone qubit, it should have a well-defined two-level system. And we expect it to be isolated and not affected by other qubits. Therefore, in order to let them affect each other (to enable an entanglement gate), we need to borrow states from other spaces as a "mediator". Here we borrow two types of states. One is the motional states of the ion chain in the trapping well. We discussed this in detail in Section 30.3. Readers are encouraged to read that section first if they are not familiar with the motional states of the ion chain. In short, in addition to the qubit states, the trapped ion chain also has mechanical vibrational motional states. The motional states are those of a simple harmonica oscillator $|N\rangle$, which means it has $N$ phonons (quanta of vibrational energy, $\hbar\omega_{chain}$).

Since we are working with two trapped ions (not necessarily physically adjacent to each other), by including the motional state, we have 3 entries in the *ket* notation. This is the tensor product of the first ion qubit state, the second ion qubit state, and the ion chain motional state. For example, the system might be in state $|g,e,N=0\rangle = |g\rangle \otimes |e\rangle \otimes |N=0\rangle$. This means that the first ion is at state $|0\rangle = |g\rangle$, the second ion is at state $|1\rangle = |e\rangle$, and the ion chain is in its ground state, $|N=0\rangle$. As another example, $|e,e,N=2\rangle$ means both ions are at their excited states and the chain has two phonons.

The second space we will borrow is an electronic state of the trapped ion outside of the qubit states. We call this $|a\rangle$. This is an excited state and is not a part of the computational space. However, we can excite the ground state $|0\rangle = |g\rangle$ to $|a\rangle$ using a light with a *different polarization*, instead of the one used to excite $|0\rangle = |g\rangle$ to $|1\rangle = |e\rangle$. Fig. 30.4 shows the three electronic states of each trapped ion.

### 30.4.3 Basics for the Cirac–Zoller Gate

Firstly, we will describe rules that are relevant to the implementation of a Cirac–Zoller gate. After that, we will give some insights into its physics and mathematics. Readers may refer to the original paper afterwards for a more advanced treatment [14].

A Cirac–Zoller gate has 3 steps (Fig. 30.4). In **Step 1**, a $\pi$-pulse with an angular frequency $\omega_L - \omega_{chain}$ is shot to the first ion (which hosts the control qubit). So this is red-shifted from the qubit frequency. It is called a $\pi$-pulse because it is intended to flip $|g\rangle$ to $|e\rangle$ and $|e\rangle$ to $|g\rangle$. The pulse has the following effects.

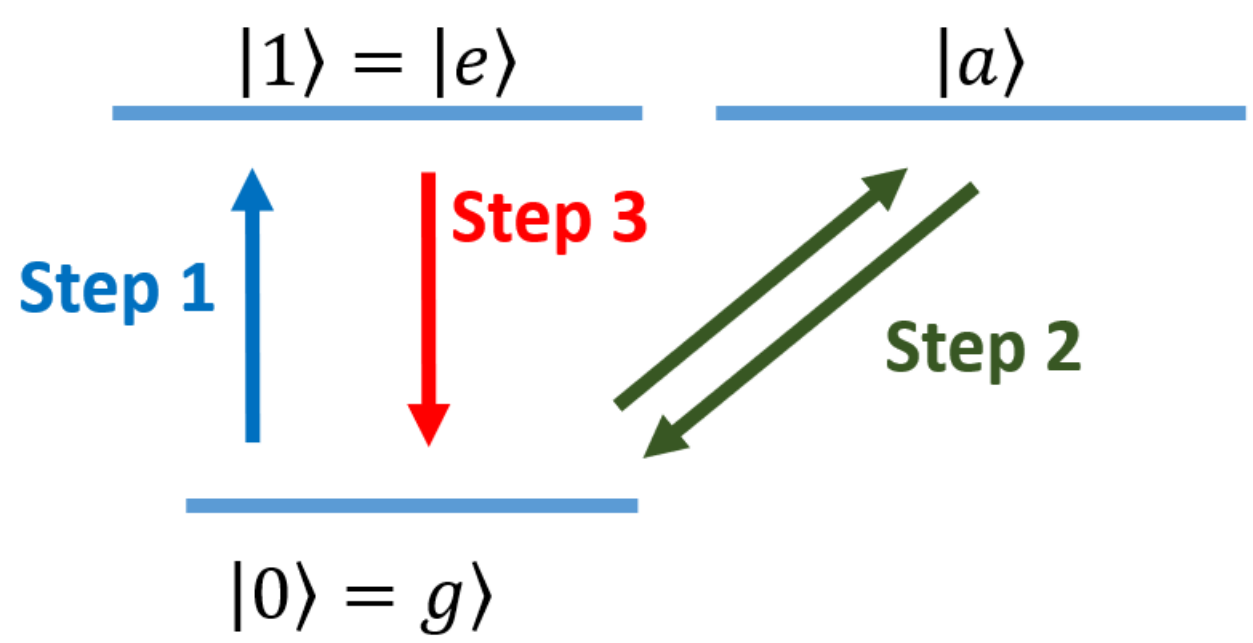

Fig. 30.4: Implementation of the Cirac–Zoller gate in trapped ions. Here, it only shows the transitions in *one* trapped ion. In Step 1, a $\pi$-pulse of a certain polarization is applied. The arrow illustrates $|g, N = 1\rangle \longrightarrow -i|e, N = 0\rangle$. In Step 2, a $2\pi$-pulse of another polarization is applied and an auxiliary state $|a\rangle$ is used. The arrows illustrate $|g, N = 1\rangle \longrightarrow -|g, N = 1\rangle$. In Step 3, the same $\pi$-pulse in Step 1 is used. The arrow illustrates $|e, N = 0\rangle \longrightarrow -i|g, N = 1\rangle$.

$$|g, x, N = 0\rangle \longrightarrow |g, x, N = 0\rangle, \tag{30.4}$$

$$|g, x, N = 1\rangle \longrightarrow -i|e, x, N = 0\rangle, \tag{30.5}$$

$$|e, x, N = 0\rangle \longrightarrow -i|g, x, N = 1\rangle, \tag{30.6}$$

where we label the state of the second qubit as $|x\rangle$ because the laser pulse has no effect on the qubit state of the second ion (as the laser pulse is only shooting at the first ion), and the result does not depend on the qubit state of the second ion. $|x\rangle$ can be $|g\rangle$ or $|e\rangle$.

The frequency of the pulse is not enough to flip the state by itself because its frequency is $\omega_L - \omega_{chain}$ instead of $\omega_L$ (like Eq. (30.2)). However, with the help of the motional state, this is possible. In Eq. (30.5), it says that it will excite the trapped ion from $|g\rangle$ to $|e\rangle$ by borrowing a quantum of motional energy $\hbar\omega_{chain}$. As a result, $|N = 1\rangle$ becomes $|N = 0\rangle$. This makes sense from an energy conservation point of view. After the transition, the photon is absorbed. So the change of photon energy is $-\hbar(\omega_L - \omega_{chain})$. The ion chain energy is also reduced by one phonon with an energy change of $-\hbar\omega_{chain}$. The first qubit has its energy increased by $\hbar\omega_L$. Adding them together, we find that the energy is conserved. That is, the change of energy in this process is $-\hbar(\omega_L - \omega_{chain}) - \hbar\omega_{chain} + \hbar\omega_L = 0$. Additionally, it also adds a phase of $-i$.

This also explains Eq. (30.4) where an excitation is *not* possible because there is no motional energy to be borrowed. As a result, $|g, x, N = 0\rangle$ stays as it is.

*Example 30.1.* Show that Eq. (30.6) make sense from the energy conservation point of view.

Eq. (30.6) tells us that the $\pi$-pulse will flip $|e\rangle$ to the ground state and increase the number of phonons by 1 in the ion chain. It also implicitly means that a photon with a frequency of $\omega_L - \omega_{chain}$ is emitted during this transition. Therefore, after the transition, two photons will come out (as the incident one is *not* absorbed). So the *change* of the photon energy is $\hbar(\omega_L - \omega_{chain})$ due to one extra photon. The change in the ion chain energy is $\hbar\omega_{chain}$ as $|N=0\rangle$ becomes $|N=1\rangle$. The change in the ion state energy is $-\hbar\omega_L$. Therefore, the change of the total energy in this transition is $\hbar(\omega_L - \omega_{chain}) + \hbar\omega_{chain} - \hbar\omega_L = 0$. Therefore, energy is conserved.

■

In **step 2**, a $2\pi$-pulse with an angular frequency $\omega_L - \omega_{chain}$ is shot to the second ion (which hosts the target qubit). It has a *different polarization* than the pulse in step 1 and, thus, it will move the state between $|g\rangle$ and $|a\rangle$ (instead of $|e\rangle$). Note that here we assume that the energy difference between $|g\rangle$ and $|a\rangle$ is the same as that between $|g\rangle$ and $|e\rangle$. So it still has the same transition angular frequency of $\omega_L$. It is a $2\pi$-pulse and thus it is long enough to return to the original state. It obeys the following rules.

$$|x,g,N=0\rangle \longrightarrow |x,g,N=0\rangle\,, \tag{30.7}$$

$$|x,g,N=1\rangle \longrightarrow -|x,g,N=1\rangle\,, \tag{30.8}$$

$$|x,e,N=0\rangle \longrightarrow |x,e,N=0\rangle\,, \tag{30.9}$$

$$|x,e,N=1\rangle \longrightarrow |x,e,N=1\rangle\,. \tag{30.10}$$

Again, the state of the control (first) qubit is irrelevant. So we put it as $|x\rangle$ which can be $|g\rangle$ or $|e\rangle$. Also, note that $|a\rangle$ does not appear in any rule because it is only temporarily used for the transition to obtain the phase change we need and this is a "round-trip" process. Eq. (30.8) tells us that, by applying a $2\pi$-pulse with an angular frequency $\omega_L - \omega_{chain}$, the ion will absorb the photon to $|a\rangle$ and then emitt a photon and go back to $|g\rangle$ with a phase shift of $-1$ (see Fig. 30.4). This is similar to having two consecutive $\pi$-pulses in Eqs. (30.5) and (30.6) although the excited state is $|a\rangle$ and this is applied to the second qubit. But it helps us *appreciate* (not a proof at all) why the phase change is $-1$ as it is $-i \times -i$. *However*, note that we cannot explain the zero phase shift in Eq. (30.9) using the same argument. $|x,e,N=0\rangle$ is insensitive to the pulse just because the polarization of the pulse does not promote $|g\rangle$ to $|e\rangle$. Therefore, there is no interaction between the photon and the ion.

**Step 3** is the same as Step 1. That is, a $\pi$-pulse with an angular frequency $\omega_L - \omega_{chain}$ is shot to the first ion (which hosts the control qubit).

### *30.4.4 The Cirac–Zoller Gate*

With the rules given in Eqs. (30.4) to (30.10), we can show why the three-pulse sequence will make a CZ-gate. To prove that it is a CZ gate, we only need to show that it transforms the basis states in the way a CZ gate will. We also assume that the ion chain has been cooled to its ground state, $|N=0\rangle$, through resolved sideband cooling (Section 30.3).

When the basis state is $|00\rangle$, we expect it to stay as it is (Eq. (30.3)). In the tensor product space we are working on, $|00\rangle = |g,g,N=0\rangle$ after including the ion chain motional state. After Step 1, based on Eq. (30.4), it is still $|g,g,N=0\rangle$. After Step 2, based on Eq. (30.7), it is also $|g,g,N=0\rangle$. After Step 3, again based on Eq. (30.4), it is still $|g,g,N=0\rangle$. Therefore, it is still $|00\rangle$, which is the expected action of a CZ gate on $|00\rangle$.

When the basis state is $|01\rangle$, we expect it to stay as it is (Eq. (30.3)). In the tensor product space we are working on, $|01\rangle = |g,e,N=0\rangle$. After Step 1, based on Eq. (30.4), it is still $|g,e,N=0\rangle$. After Step 2, based on Eq. (30.9), it is also $|g,e,N=0\rangle$. After Step 3, again based on Eq. (30.4), it is still $|g,e,N=0\rangle$. Therefore, it is still $|01\rangle$, which is the expected action of a CZ gate on $|01\rangle$.

When the basis state is $|10\rangle$, we expect it to stay as it is (Eq. (30.3)). In the tensor product space we are working on, $|10\rangle = |e,g,N=0\rangle$. After Step 1, based on Eq. (30.6), it becomes $-i|g,g,N=1\rangle$. After Step 2, based on Eq. (30.8), it becomes $-1\times -i|g,g,N=1\rangle = i|g,g,N=1\rangle$. After Step 3, based on Eq. (30.5), it becomes $-i\times i|e,g,N=0\rangle = |e,g,N=0\rangle$. Therefore, it is still $|10\rangle$, which is the expected action of a CZ gate on $|10\rangle$.

When the basis state is $|11\rangle$, we expect it to have a phase change of $-1$ (Eq. (30.3)). In the tensor product space we are working on, $|11\rangle = |e,e,N=0\rangle$. After Step 1, based on Eq. (30.6), it becomes $-i|g,e,N=1\rangle$. After Step 2, based on Eq. (30.10), it is still $-i|g,e,N=1\rangle$. After Step 3, based on Eq. (30.5), it becomes $-i\times -i|e,e,N=0\rangle = -|e,e,N=0\rangle$. Therefore, it becomes $-|11\rangle$, which is the expected action of a CZ gate on $|11\rangle$.

The following summarizes the transitions we have discussed for the four basis states under the action of the Cirac–Zoller gate.

$$|00\rangle = |g,g,N=0\rangle \to |g,g,N=0\rangle \to |g,g,N=0\rangle \to |g,g,N=0\rangle\,, \tag{30.11}$$
$$|01\rangle = |g,e,N=0\rangle \to |g,e,N=0\rangle \to |g,e,N=0\rangle \to |g,e,N=0\rangle\,, \tag{30.12}$$
$$|10\rangle = |e,g,N=0\rangle \to -i|g,g,N=1\rangle \to i|g,g,N=1\rangle \to |e,g,N=0\rangle\,, \tag{30.13}$$
$$|11\rangle = |e,e,N=0\rangle \to -i|g,e,N=1\rangle \to -i|g,e,N=1\rangle \to -|e,e,N=0\rangle\,. \tag{30.14}$$

Therefore, we have proved that the 3-pulse sequence acts as a CZ gate with the first qubit as the control qubit and the second qubit as the target qubit. We also need to emphasize that the ions do not need to be physically next to each other. In this process, we shoot the laser pulse at two ions and do not require them to be adjacent to each other. Then how are they coupled to each other to have the entanglement action? This is *mediated* through the ion chain motional state.

## 30.5 Summary

In this Chapter, we discussed how to cool the trapped ions to their motional ground states. Firstly, Doppler laser cooling is applied. Then it is followed by resolved sideband cooling. Doppler laser cooling uses the Doppler shift to selectively cool down only atoms approaching the laser beam and avoid heating up atoms departing from it. Resolved sideband cooling couples the electronic states to the ion chain's vibrational state and cools the ions by borrowing a phonon from the ion chain for photon absorption and emission. With resolved sideband cooling, the ion chain can be cooled down to its motional ground state. With this, we can create a Cirac–Zoller gate (an entanglement gate) through a sequence of three pulses. It should be appreciated that the entanglement gate is possible because an auxiliary electronic state and the ion chain motional states are borrowed in the process.

## Problems

**30.1. Doppler Laser Cooling**
Repeat Fig. 30.2, show what will happen if the ion is moving in the same direction as the photon.

**30.2. Cirac–Zoller Gate**
Refer to paper [14], prove Eqs. (30.4) to (30.10). Here are the hints.

- We will take Eq. (1) as given in the paper. But it is easy to understand its meaning through the understanding of creation and annihilation operators in [1]. Here, $|e_0\rangle$ is our $|e\rangle$ and $|e_1\rangle$ is our $|a\rangle$.
- Eq. (2) is just the exponentiation of the Hamiltonian to form the gate for the given amount of time.
- Then we will need to use $\exp iA = \cos A + i \sin A$ for Eq. (2) and note that $A$ is an operator.
- Perform Taylor expansionon on sin and cos. Note that very often, the squared terms are just identity. Then, apply them to the corresponding states. Depending on the states you apply to, some have zero inner product.
- Note that when the pulse is for $q = 0$ ($q = 1$), it has not effect on $q = 1$ ($q = 0$).